*To be submitted*

# Ferroelectric Domain-Wall Motion, Stability, and Frequency Responses under Different Screening-Charge Conservation Laws

Sergei V. Kalinin[1]

Department of Materials Science and Engineering, University of Tennessee, Knoxville, Tennessee 37996, USA

**Abstract**

We analyze ferroelectric domain wall motion and changes in wall structure and geometry arising from the interaction of polarization with screening charge. The universal physical cases are organized by electrical boundary conditions, charge conservation and dynamics, and polarization free energy. We consider exposed surfaces, interfaces beneath a dielectric layer and top electrode, and a ferroelectric adjoining a blocking mixed ionic–electronic conductor. Screening dynamics is described by local exchange, exchange with lateral transport, or strictly conserved redistribution. Analytical derivations and finite-depth calculations connect these cases to screening drag, velocity branches, spatial instability, and oscillatory behavior. For an isolated wall joining differently compensated domains, strict surface conservation and zero far-field current exclude steady translation of the screening profile, while transient displacement, finite-sample arrest, and charge-balanced structural changes remain possible. Charge conservation also constrains the long-distance transport accompanying a moving junction in an adjoining conductor. Large-field simulations show that nonlinear corrugation, transfer toward longer wavelengths, and coupled polarization rotation can emerge in the top-electrode geometry as a consequence of dual screening pathways. A first-order polarization potential introduces the separation of two interphase fronts, while multiaxial polarization allows rotation within the wall and competing intermediate ferroelectric domains. A separate driven structural–screening reduction supports stable periodic behavior. Reduced-variable diagrams summarize the motion and structural regimes, and $BaTiO_3$ and $BiFeO_3$ examples connect them to material thermodynamics. Finally, we derive associated frequency responses for local displacement, polarization dynamics, and far-field electrostatics. Together these results provide a general framework for analysis of the wall dynamics, shape evolution, and internal polarization transitions under different driving and screening conditions.

[1] sergei2@utk.edu

## 1. Introduction

Ferroelectric domain-wall motion is commonly described through the response of polarization to an applied electric field. This description is sufficient when the charge needed to compensate the changing polarization distribution follows the wall rapidly. At a surface or interface, however, displacement of the wall changes the spatial distribution of bound charge. The compensating screening charge must then be created, removed or transported. The wall and its screening distribution become coupled dynamical objects, with generally different characteristic times and spatial scales.

Slow compensation can change the electric field at the surface junction, result in wall bending, and induce morphological and temporal instabilities of moving wall. The polarization can also change its reversal path: a wall may broaden with suppressed polarization, separate into two interphase boundaries enclosing a weakly polarized region, or develop a transverse polarization component. Wall motion and wall structure therefore depend jointly on the electrical geometry, charge conservation and dynamics, and the polarization free-energy landscape.

Classical wall theory connects polarization profiles with gradient energy and strain [1], while nucleation theory describes activated sidewise motion [2]. Later work treated delayed screening [3], electrical forces on charged walls [4, 5], transport through finite-conductivity electrodes [6, 7], and screening-induced shape instability [8]. Surface-layer restoring forces [9], surface-induced broadening [10] and material-specific wall structures [11] provide further parts of the physical description. These studies address different physical realizations. A prescribed surface charge, a mobile conserved surface population, an electrode supplied by contacts and mobile defects inside the ferroelectric obey different balance equations. Their results can be compared only after those distinctions are elucidated and corresponding limiting conditions are analyzed.

Small-amplitude dielectric and impedance measurements provide a complementary probe of domain-wall dynamics. In PZT films, the field-amplitude and frequency dependence of the permittivity reveals contributions from reversible and irreversible wall displacement [12]. The Rayleigh equivalent-circuit element and the frequency-dependent domain-wall pinning element connect such contributions to measured impedance spectra while accounting for the electrical response of the capacitor stack [13, 14]. Wall-related dielectric enhancement can also arise from polarization response within a stationary wall [15], and microwave measurements detect carrier transport at walls that are poorly conducting at dc [16]. A wall-related spectral feature therefore does not by itself identify collective wall translation. Its interpretation requires the driving field, the measured coordinate, and the electrode and screening boundary conditions. Here we calculate the linear displacement response associated with screening dynamics and charge conservation; disorder-induced irreversible Rayleigh motion and conduction within the ferroelectric wall are separate mechanisms. The resulting spectra connect the dynamical cases considered below to frequency-dependent measurements.

Here we organize the problem by electrical boundary conditions and charge dynamics, and explore how the polarization free energy changes both wall motion and internal structure. We consider an isolated front and focus on its interaction with screening charges at the surface junction. This baseline model omits substrate pinning, external mechanical clamping, mobile charge within the ferroelectric bulk, disorder-controlled creep, and nucleation. Polarization and electrostatics through the film thickness are retained because they determine the field acting on the junction and the bending of the wall. Mechanical compatibility is introduced separately for cases where it changes a material prediction; volume charge transport is retained in the adjoining mixed electron-ion conductor (MIEC) where it provides the screening mechanism. This formulation allows the general charge-conservation cases to be compared while preserving the physical boundary conditions of each geometry.

## 2. Experimental evidence and the physical questions it constrains

Here we summarize experimental evidence for the presence and effects of screening charge and its consequences for wall motion. Simultaneous measurement of a wall trajectory and the potential above the surface can resolve the lag between polarization reversal and charge redistribution. The interpretation is clearest when the wall moves under a thermodynamic or mechanical driving force that does not directly inject or drift surface charge. An applied electric field via lateral electrodes acts on both polarization and mobile carriers, so electrically driven measurements require the charge response to the applied field to be included. Temperature-driven experiments also require attention to changes in adsorption, conductivity and diffusivity. Hence, the useful distinction is between the imposed stimulus, the polarization response and the independently evolving screening distribution.

### 2.1. Surface compensation and measured potential

The need to screen spontaneous polarization was understood since the earliest days of ferroelectricity. Känzig discussed near-surface space-charge layers in 1955 [17]. Ivanchik's 1961 treatment of spontaneous-polarization screening, discussed by Dmitriev, belongs to this early foundation [18]. Subsequent theories developed the roles of free carriers, surface states and band bending. Dmitriev compared internal carrier redistribution with surface-state and adsorbate compensation [18], and Ivanchik later examined the equilibrium screening-carrier distribution in a single-domain ferroelectric [19].

Experimental development of this surface-specific picture was limited by the lack of methods able to follow polarization and potential on individual domains. Much of the subsequent work, particularly during the 1960s–1980s, instead explored the coupling of ferroelectric polarization to semiconducting behavior, photoconductivity, and photoinduced charge redistribution. Fridkin's treatments of ferroelectric semiconductors and photoferroelectrics place screening within this broader carrier–polarization problem [20, 21]. Work on the anomalous photovoltaic effect likewise established that carrier generation and transport can be strongly coupled to polar order [22].

The emergence of scanning probe microscopy in the end of XX century made the local surface compensation and its kinetics directly accessible. Kalinin and Bonnell combined electrostatic and surface-potential imaging with local piezoresponse and variable-temperature measurements on $BaTiO_3$ [23, 24]. By following individual domains through thermal changes and observing the potential near moving walls, they separated the rapid polarization response from the slower redistribution of compensating charge. The measured potential could retain the earlier screening pattern after polarization had changed, and its polarity could be opposite to that inferred from unscreened bound charge. Kalinin, Johnson and Bonnell subsequently analyzed the temperature-induced inversion of this contrast through coupled electrostatic and adsorption effects [25]. These observations established that a domain-potential image records the joint state of polarization and compensation, rather than polarization alone.

Hong and colleagues introduced charge-gradient microscopy, in which a grounded conducting probe mechanically removes and collects screening charge while scanning the surface [26]. The resulting current connects domain imaging to a measurable redistribution of the compensating population. Subsequent measurements resolved the removal and recovery of screening without changing the underlying polarization [27], and distinguished contributions from domain interiors and domain walls [28]. These experiments provide a complementary route to the screening kinetics: potential microscopy measures the evolving electrostatic response, whereas charge-gradient microscopy perturbs and collects the mobile surface charge.

Complete compensation of macroscopic bound charge can coexist with a finite surface-potential step because the bound and compensating charges are spatially separated [23, 25].

Potential contrast must therefore be interpreted together with polarization orientation, probe height, and the interfacial dipole contribution. Watanabe, Okano and Masuda demonstrated surface conduction associated with ferroelectric polarization, providing evidence for an electronic compensation route [29]. The broader surface-screening literature shows how electronic, ionic, and adsorbate mechanisms can produce different kinetics even when their static electrostatic effects are similar [30].

### 2.2. Exchange with the chemical environment

Chemical switching demonstrates that changes in the external chemical environment can reverse polarization. Wang and colleagues demonstrated reversible chemical switching of a ferroelectric film [31]. Highland and colleagues measured polarization control by oxygen pressure [32]. Experiments in liquids and ambient environments provide complementary evidence that the available species influence compensation and reversal [33, 34].

For an exposed surface, water adsorption and dissociation provide a concrete microscopic example. Protons or hydroxyl-containing species can be attached, detached or transferred between the surface and its environment, with an accompanying countercharge process. The net surface charge can then change locally. Chemical environment and electrode configuration provide complementary experimental controls. Humidity or oxygen pressure changes the available compensation and reaction rates; an electrode changes the electric field generated by a given charge imbalance. An impermeable dielectric can prevent local chemical exchange even when the electrode above it responds electronically. Conversely, exchange beneath an electrode is possible if the device provides a specific chemical path. Stating that path is necessary when comparing exposed-surface and capacitor measurements [35, 36].

### 2.3. Lateral transport, relaxation and preparation

An important distinction between thermally or chemically induced evolution and field-induced switching is that a biased surface electrode acts directly on mobile charge as well as on polarization. Injection or electrochemical generation at the contact, together with lateral drift and diffusion, can establish a screening distribution ahead of the moving domain boundary. Charge redistribution can then precede wall motion rather than simply relax behind it. Which process leads depends on the electrode geometry, reservoir access, and the relative charge and polarization timescales. Charge injection on $LiNbO_3$ [37], the ionic field effect in local switching [38], and switching control through electrode modification [6, 39] provide experimental evidence of this additional coupling.

Time-resolved charge-injection measurements on $LiNbO_3$ distinguish the evolution of surface potential from the written domain pattern [37]. Gaponenko and colleagues studied the spreading and passivation of injected charge and compared the evolution with a reaction–diffusion description [40]. Humidity-dependent measurements further show how the environment changes the available compensation and switching kinetics [41]. These observations motivate separate local-exchange and lateral-transport terms. They also show why a charge pattern prepared by an electrode pulse cannot automatically be treated as the equilibrium compensation of the domain pattern that remains afterward. Similarly, the initial screening distribution is an experimentally relevant control. A wall displaced from an equilibrated charge pattern leaves a different potential wake from a wall advancing through a nearly uniform or frozen distribution. Surface-domain studies of triglycine sulfate by Likodimos and colleagues provide examples of coupled domain and compensation evolution [42, 43, 44].

### 2.4. Motion, wall shape and strong stimuli

Transport through an electrode, i.e. the rate of screening charge supply, can limit wall propagation. McGilly, Yudin, Tagantsev and colleagues demonstrated domain-wall control through electrode engineering [39]; related theory explicitly includes contact-supplied transport [6, 7]. Such experiments establish the importance of charge delivery, with the contacts supplying a boundary condition that is absent from a closed mobile surface layer.

Under strong switching stimuli, screening may lag far behind polarization reversal. Shur and collaborators reported domain-shape instabilities and dendritic growth in uniaxial ferroelectrics, relating the morphology to ineffective screening and the competition between switching and charge-relaxation times [45]. Artificial surface dielectric layers and altered near-surface conductivity change this competition. Fingering, discrete growth and dense subsidiary domains show that the response can involve nucleation and branching as well as deformation of a pre-existing wall. The corresponding shape-instability theories provide direct precedents for the present corrugation analysis [8, 46]. The isolated-front model describes the onset of a modulation; a branched domain pattern requires its subsequent nonlinear evolution and, where applicable, nucleation.

Oscillatory wall velocities have also been treated with electrical-circuit feedback [47, 48]. Probe-induced intermittency and quasiperiodicity involve localized driving and electrochemical processes [49]. These observations motivate temporal stability calculations. A measured oscillation can arise from circuit feedback, repeated nucleation, a periodically applied stimulus, or an internal instability of a continuously driven wall; the mechanism must be identified from the driving protocol and the coupled signals.

### 2.5. Wall structure and experimental discrimination

A broad wall image can arise from a wider polarization-reversal profile, bending with depth, two adjacent interfaces, polarization rotation or instrumental averaging. Atomic-scale studies of $BiFeO_3$ demonstrate that structural distortions and polarization vary together near walls [50, 51]. Geng and colleagues observed an antiferroelectric transformation in strained $BiFeO_3$ films with dense 180-degree walls and interfacial octahedral constraints [52]. This is evidence for a competing structural state associated with the wall configuration. An antipolar structure has ordered local displacements and must be distinguished from a paraelectric layer.

The useful observables are therefore the wall position, the vector polarization, the width and number of peaks in the polarization gradient, and the time-dependent screening potential. A separated scalar core has two bounding interfaces and an independently evolving separation. A Bloch-like wall has a transverse component localized within the reversal. An intermediate ferroelectric domain has a finite region with its own polarization direction. Temperature-dependent wall measurements provide a further test of changes in local structure [53]. These considerations guide the calculations in Sections 7–9 and the interpretation of existing experimental results.

## 3. Physical cases, equations and boundary conditions

We first specify the common polarization equations and distinguish the electrical geometries and charge laws used below. The behavior of the same isolated wall is analyzed under non-conserved local exchange, exchange combined with lateral transport, and strict conservation of the surface charge population. The adjoining blocking conductor is treated separately because its compensating charge occupies a volume. This formulation fixes the driving, far states, and allowed charge redistribution before comparing wall velocities, changes in structure, and stability.

### 3.1. Geometry, notation and driving

The mean wall advances along $y$, its surface intersection extends along $x$, and $z$ is the outward surface normal. In the finite-film cases the ferroelectric occupies $-h < z < 0$. For the exposed surface, a dielectric exterior occupies $z > 0$. For the electrode case a dielectric of thickness $d$ occupies $0 < z < d$, with electrodes at $z = -h$ and $z = d$. For the MIEC case the ferroelectric and MIEC occupy $z < 0$ and $z > 0$, respectively. A wall that remains a graph is represented by $y = X(x, z, t)$. Figure 1 shows the wall, its surface junction and the directions used to describe displacement and corrugation.

We use $q$ for the Fourier wave number along the direction of advance, $k$ for modulation along the wall, and $Q = (q^2 + k^2)^{1/2}$. The temporal growth variable is $s$: a perturbation is proportional to $\exp(st + iqy + ikx)$. Thus $q$ describes variation across the moving wall, $k$ describes a ripple along its front, and $Q$ is the magnitude of their in-plane wave vector. Positive $\mathrm{Re}s$ means growth; $\mathrm{Im}s$ is an angular oscillation frequency. These wave numbers are distinct from the thermodynamic curvature $K_b$ defined in Section 8.4. Background relative permittivities exclude the explicitly evolved soft polarization. In the MIEC derivation, $\epsilon_m$ denotes the absolute permittivity, and $\epsilon_f = \epsilon_0\sqrt{\epsilon_\parallel \epsilon_z}$ is an absolute effective ferroelectric permittivity.

For the exposed film, the induced bottom potential is zero and nonuniform exterior fields decay at infinity. A remote applied field can be represented by subtracting its background solution. For the electrode stack, $\phi_f(-h) = 0$ and $\phi_d(d) = -V$, so a positive $V$ drives the positive-polarization domain forward. In the ferroelectric half-space, the deep applied field is prescribed; a fixed finite voltage across an infinite thickness would instead give zero drive.

### 3.2. Polarization dynamics and mechanical scope

The uniaxial reference has $\mathbf{P} = p\hat{\mathbf{z}}$ and polarization dynamics is given by

$$f_L(p) = \frac{a}{2}p^2 + \frac{b}{4}p^4 + \frac{c_6}{6}p^6 \tag{1a}$$

$$\Gamma\, \partial_t p = -f_L{}'(p) + \textstyle\sum_j g_j\, \partial_j^2 p + E_z. \tag{1b}$$

For the second-order material $b > 0$, $c_6 = 0$ and $a < 0$. A first-order scalar model uses $b < 0$, $c_6 > 0$. Higher-order and vector potentials are used where required by the material. Electrostatics obeys $\nabla \cdot \mathbf{D} = \rho_{\text{free}}$ and $\mathbf{D} = \epsilon_0 \boldsymbol{\epsilon}_b \mathbf{E} + \mathbf{P}$ in the ferroelectric. There is no mobile bulk charge in the baseline surface-sheet models.

The isolated upper junction interacts with the screening charge distribution, while the wall remains unpinned at the lower interface. The finite-depth equations retain polarization relaxation and electrostatic bending through the film. The baseline natural condition is $\partial_z p = 0$ at both interfaces. A surface energy $f_{\text{surf}} = a_s p^2/2$ instead gives $g_n\, \partial_n p + a_s p = 0$, where $n$ is the outward normal and $\lambda_P = g_n/a_s$ is the extrapolation length [54]. At the lower surface the outward normal points in the negative $z$ direction. Neither condition introduces a lateral pinning force by itself. The model excludes additional substrate-controlled kinetics; calculated depth bending retains its accompanying bulk bound charge.

The baseline scalar calculations omit elastic constraints. Material-specific coherent calculations instead retain compatibility through a stated reduced energy. External mechanical freedom does not mean that every point in a coherent wall can independently adopt its stress-free strain. No coherent correction is inserted into the unconstrained reference without being identified.

### 3.3. Surface charge: sources, currents and conservation

We write the free screening charge density as $\sigma = -c$. Thus $c > 0$ denotes negative free charge, which compensates positive bound polarization charge; $c < 0$ denotes positive free charge. The variable $c$ has units of charge per area and represents the net charge of the surface population. For the analytical model we use the chemical free energy per area $f_s(c) = \chi c^2/2$. This is a quadratic expansion about a reference charge state, with the constant term omitted and the linear chemical offset included in the reservoir potential. The curvature $\chi$ is the inverse differential chemical capacitance. We choose this approximation because it gives a linear charge response and allows the effects of charge conservation, exchange and lateral transport to be compared without specifying a particular adsorbate. It is the local small-deviation approximation to a smooth adsorption free energy, such as the finite-site ionic compensation model of Stephenson and Highland [35]. The electrochemical potential and charge dynamics are

$$\mu = \chi c - \phi_s - \Psi \tag{2a}$$

$$\mathbf{J} = -M\nabla_{\parallel}\mu \tag{2b}$$

$$\partial_t c + \nabla_{\parallel} \cdot \mathbf{J} = -\Lambda\mu. \tag{2c}$$

Here $\mu$ is the electrochemical driving potential per unit compensation charge, and $\Psi$ specifies the reservoir chemical potential relative to the electrical potential reference. The flux $\mathbf{J}$ transports $c$; the corresponding physical electric current is $-\mathbf{J}$. The coefficient $\Lambda$ controls local exchange with the reservoir, while $M$ controls lateral transport. Case I allows local exchange only ($\Lambda > 0$, $M = 0$); Case II also allows lateral transport ($\Lambda > 0$, $M > 0$). In Case III, local exchange is absent ($\Lambda = 0$, $M > 0$) and no charge flows through the lateral boundaries. The total surface charge then remains equal to its initially prepared value, so its spatial average cannot adjust independently to the instantaneous polarization pattern. Charge is frozen when both kinetic coefficients vanish.

A microscopic description instead tracks the nonnegative surface number density $n_i$ of each species with charge $q_i$. The net surface charge is $\sigma = \sum_i q_i n_i$. Each species has electrochemical potential $\mu_i = \partial f_s / \partial n_i + q_i \phi_s$, moves down its electrochemical-potential gradient with flux $\mathbf{J}_i = -L_i \nabla_{\parallel}\mu_i$, and obeys the balance law $\partial_t n_i + \nabla_{\parallel} \cdot \mathbf{J}_i = R_i$. The reaction rate $R_i$ describes creation, removal or exchange of that species. A reservoir may contain many particles, but the number of adsorption sites on the surface is still finite. Creating an electrically neutral pair satisfies $\sum_i q_i R_i = 0$: it changes the carrier populations but does not create net charge. The single variable $c$ retains only their net electrical effect; it cannot describe changes in neutral populations or enforce each species inventory separately.

The quadratic approximation gives a constant chemical capacitance and a linear restoring contribution $\chi c$ to $\mu$. Together with the linear kinetic law, this permits the screening response used below to be calculated mode by mode. For $\chi > 0$, it penalizes charge accumulation but does not impose a maximum surface occupancy or distinguish the chemistry of positive and negative screening species. It is therefore appropriate only over a charge interval in which the chemical curvature changes little. The limit $\chi = 0$ removes this chemical penalty and gives the complete-equilibrium-compensation reference; it does not represent finite-capacity adsorption. Alternatives include dilute-solution mixing entropy, finite-site Langmuir adsorption with separate positive and negative species, and adsorption free energies with interactions between adsorbates [30, 35, 36]. Finite-site entropy makes the chemical potential increase strongly as the surface fills, while adsorbate interactions can introduce cooperative transitions. Such choices preserve the charge-conservation constraints but change the explicit screening response, velocity curves and stability thresholds. The charge conserved field calculations use a nonlinear chemical potential to retain saturation, as specified in S12.

### 3.4. Boundary conditions and initial states

At a surface carrying $\sigma$, the potential is continuous and $D_{\text{out},z} - D_{f,z} = \sigma$. A microscopic dipole sheet would require a separate potential-jump law; it is excluded unless explicitly added.

For a blocking MIEC there is no added surface sheet: $\phi_m = \phi_f$, $D_{m,z} = D_{f,z}$ and $J_{i,z} = J_{e,z} = 0$ at $z = 0$. Blocking particle current does not set the displacement current to zero.

An isolated traveling state must approach specified, time-independent far states as $y \to \pm\infty$. In Cases I–II those states satisfy reservoir equilibrium. In Case III their populations are fixed by preparation and the conserved balance. A finite closed surface has zero current at its lateral edges. A prescribed edge concentration or electrochemical potential introduces a contact and changes the problem. Periodic boundaries along $y$ require a wall pair, whereas periodicity along $x$ simply discretizes transverse perturbations.

**Table 1. Electrical geometry and charge dynamics.**

| Electrical case | Charge location and dynamics | Drive and remote condition | Principal restriction |
|---|---|---|---|
| Exposed film, O-I/O-II/O-III | Mobile surface sheet; exchange and/or lateral current | Grounded bottom; specified remote field and reservoir protocol | Conserved far-charge jump cannot advance without current |
| Dielectric/top electrode, T-I/T-II/T-III | Mobile interfacial sheet; electrode free charge is separate | Fixed electrode voltage; explicit chemical access if exchange is allowed | Ionic inventory remains fixed even when electrode charge changes |
| Blocking MIEC, M | Ionic and electronic volume charge in adjoining medium | Deep ferroelectric field; finite remote potential and concentrations | Isolated line motion requires long-distance charge supply |
| Frozen-charge comparison | Prescribed charge distribution | Same electrical geometry as specified | No redistribution; distinct from conserved mobile charge |

The labels combine these two choices. O means an open, exposed upper surface; T means a dielectric layer terminated by a top electrode. I denotes local exchange with the environment, II denotes exchange together with lateral surface motion, and III denotes lateral motion with strict conservation. Thus O-II is an exposed surface on which charge can both move and enter or leave, whereas T-III is a closed interfacial population beneath an electronically responding electrode. M denotes screening within the adjoining blocking conductor. We use the descriptive names at the start of each case and the abbreviated labels in comparisons.

### 3.5. Dissipation and stability of wall motion

With electrostatics eliminated under the prescribed electrical controls, the appropriate free energy or electric enthalpy obeys

$$\dot{\mathcal{G}} = -\int_f \Gamma\, \dot{p}^2\, dV - \int_S \left(M\left|\nabla_{\parallel}\mu\right|^2 + \Lambda\mu^2\right) dA \leq 0. \tag{3}$$

For the MIEC, the second integral is replaced by the volume sum $\sum_i \int_m L_i\, |\nabla\mu_i|^2 dV$. Reservoir and electrical source work must be treated consistently. Near equilibrium, positive curvature of the free energy in every allowed perturbation direction gives decaying relaxation modes. With the passive kinetics used here, their rates are real and hence the equilibrium wall has no Hopf instability. Sustained oscillation requires a continuing energy supply: a closed system relaxing toward equilibrium cannot repeat the same cycle while losing energy on every cycle. A translating wall can receive energy continuously by converting one domain into the favored domain. Its stability is evaluated in the moving frame, including the delayed charge response.

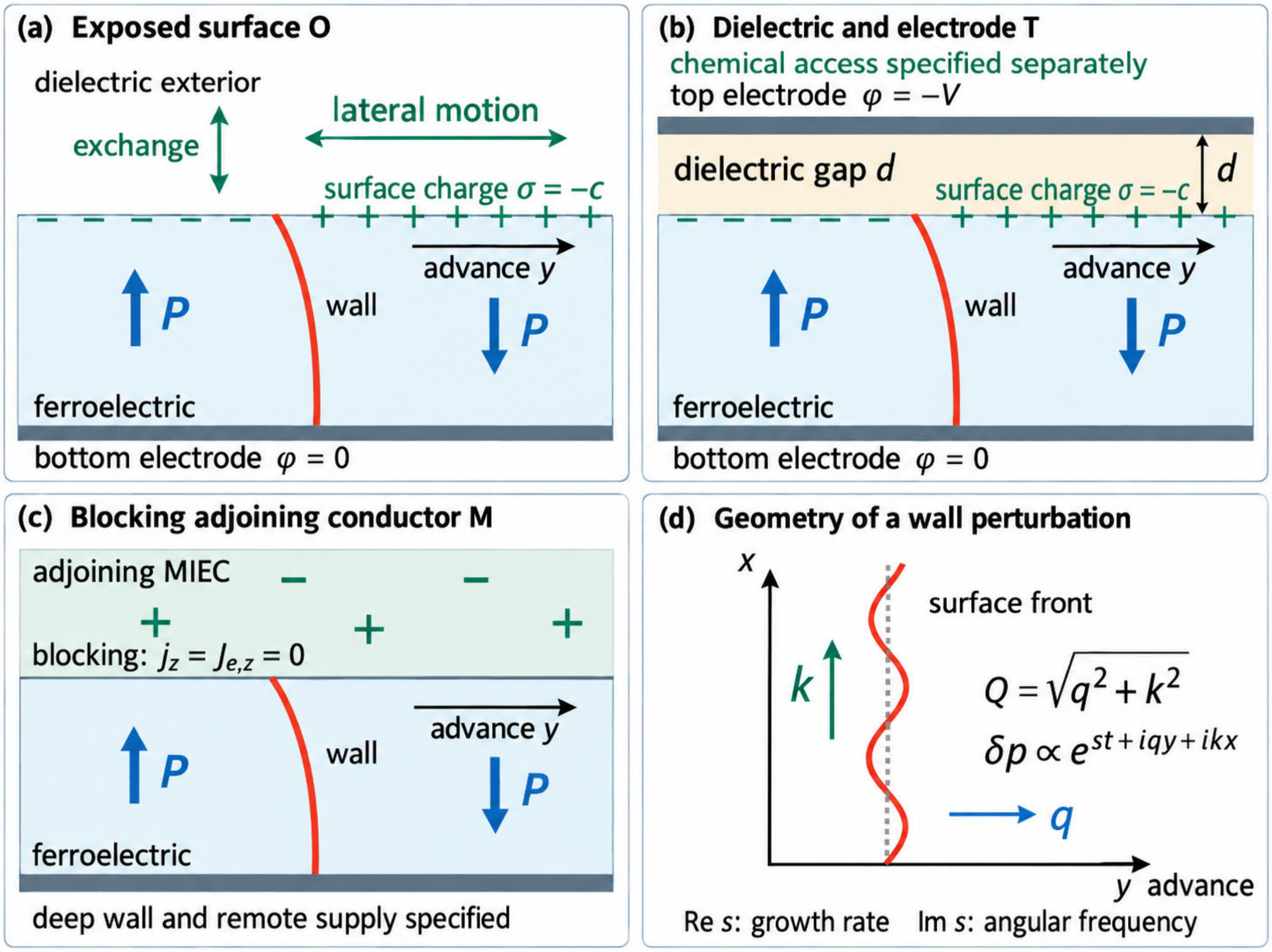

**Figure 1.** Wall geometry, perturbations and electrical cases. The surface displacement is $X(x, 0, t)$; the wall can bend through the film thickness. The lower-right panel defines $q$, $k$, $Q$ and the temporal growth rate $s$. The surface-sheet laws in O and T have independently selectable local exchange and lateral transport. In M, screening resides in the adjoining volume and both species are blocked at the interface. Reservoir access in T is external to the passive dielectric. Arrows indicate allowed processes, not a specified microscopic reaction mechanism.

## 4. Wall motion at an exposed surface

We begin with a wall intersecting an exposed ferroelectric surface. The electrostatic response is set by the finite film and dielectric exterior, while compensation can exchange with the environment, move laterally, or remain strictly conserved. We compare the resulting wall motion and then examine how a first-order polarization potential modifies the internal structure and the coupling to screening.

### 4.1. Electrical response and earlier treatments

The thermodynamic theory of chemical compensation was developed by Stephenson and Highland, who formulated the ionic equilibrium and numerically evaluated polarization phase diagrams [35]. Morozovska, Eliseev and colleagues subsequently developed analytical descriptions of ferroionic states and finite-rate polarization–charge coupling [36]. Static changes in wall width near a surface were analyzed by Eliseev and colleagues [10]. These studies establish the equilibrium charge response and the ways in which a surface modifies the wall profile. Kessler and Balke treated wall deformation in a prescribed surface-charge setting [55]. Cao and Kalinin studied switching under chemical boundary control with surface ions prescribed as uniform and non-evolving in the inspected model [56]. That calculation belongs to a prescribed-charge comparison and does not supply the dynamics of Case II or III. Charge spreading measurements and phenomenological models provide further kinetic context [40].

For a straight, depth-independent polarization, the exposed-film surface potential is $\phi_{s,Q} = G_O(Q)(p_Q - c_Q)$, where

$$G_O(Q) = \{\epsilon_0[\bar{\epsilon}Q\coth(\alpha Q h) + \epsilon_e Q]\}^{-1} \tag{4a}$$

$$\alpha = \sqrt{\epsilon_\parallel/\epsilon_z} \tag{4b}$$

$$\bar{\epsilon} = \sqrt{\epsilon_\parallel \epsilon_z}. \tag{4c}$$

Its zero mode is $G_O(0) = h/(\epsilon_0\epsilon_z)$. Define $A_Q = G_Q + \chi$, $a_Q = G_Q/A_Q$ and $r_Q = (\Lambda + MQ^2)A_Q$. The causal screening response is

$$c_Q(s) = \frac{a_Q r_Q}{s+r_Q} p_Q(s) + \frac{c_Q(0)}{s+r_Q}. \tag{5}$$

Equation (5) describes nonuniform modes, or deviations of the uniform mode after its fixed reservoir offset has been subtracted. The initial-charge term is essential for transients. In Case III, $r_0 = 0$ and $c_0$ remains its initial value. At large wave number $G_O \propto Q^{-1}$; at long wavelength in a finite film it tends to a constant. These two regimes have different transport scaling.

The effective uniform drive depends on which electrical and chemical quantities are controlled. Here an electrical protocol means the imposed voltage or remote field, together with the reservoir electrochemical potential held fixed during a change in that drive. These conditions determine both the field acting on polarization and the equilibrium screening charge. Changing a remote field at a fixed physical reservoir is therefore different from maintaining a prescribed effective polarization bias. With a background bulk field $E_b$, reservoir equilibration gives

$$f_L'(p) + \frac{G_0\chi}{h(G_0+\chi)} p = E_d \tag{6a}$$

$$E_d = E_b + \frac{a_0\Psi}{h}. \tag{6b}$$

For a remote field and fixed physical reservoir, $\Psi = \mu_r - hE_b$. At $\chi = 0$, the equilibrated drive is then $\mu_r/h$: the remote-field change can be compensated at long times. A maintained $E_d$ is a specified control scenario, not an automatic consequence of exposing the film to a field.

### 4.2. Case I: local exchange

The unbiased quartic reference wall is $p_0(\xi) = -P_s\tanh(\xi/w)$, with $P_s^2 = -a/b$, $w = (2g_y/|a|)^{1/2}$ and $I_4 = \int p_0'^2 d\xi = 4P_s^2/(3w)$. Let $W(q) = \int p_0'(\xi)e^{-iq\xi}d\xi = -\Delta P\, f_w(q)$, $\Delta P = 2P_s$, and $f_w = (\pi q w/2)/\sinh(\pi q w/2)$. Projection over the whole wall gives

$$F = \eta v + F_{\mathrm{ch}}(v) \tag{7a}$$

$$\eta = h\Gamma I_4 \tag{7b}$$

$$F_{\mathrm{ch}} = \frac{\Delta P^2}{\pi}\int_0^\infty f_w^2\, G_q a_q \frac{v r_q}{r_q^2+q^2v^2}\, dq. \tag{7c}$$

At weak drive $F = h\Delta P E_d$ and $v = F/(\eta + \zeta_{\mathrm{ch}})$, where

$$\zeta_{\mathrm{ch}} = \frac{\Delta P^2}{\pi}\int_0^\infty \frac{f_w^2 G_q^2}{(\Lambda+Mq^2)(G_q+\chi)^2}\, dq. \tag{8}$$

For local exchange and complete equilibrium compensation, $\zeta_{\mathrm{ch}} = I_4/\Lambda$. For the exposed-film kernel and the tanh profile, the transformation derived in S3 proves an increasing charge drag for Case I. Hence the total rigid-profile force is monotone. When the electrostatic response is approximated by a local capacitance, $G_Q = G_0$, the stronger real-space proof applies to every fixed monotone profile. This approximation neglects the wave-number dependence of the electrostatic interaction. These statements exclude an S-type velocity relation under the specified assumptions; they do not establish stability against every deformation or vector mode.

A deformable wall must satisfy the polarization equation, Poisson equation and surface balance jointly. At zero drive and complete compensation the depth-independent scalar kink is an exact stationary solution of the reference functional. Its longitudinal Hessian is nonnegative, with translation as a zero mode. Positive additional mode stiffnesses provide the starting point for a small-drive continuation after fixing the arbitrary wall position. This is a local existence argument, not a global proof for all fields or incomplete-screening states. S6 specifies the bending sources and the linearized problem.

Finite-depth calculations solve the coupled polarization, Poisson and local-exchange equations across the film, allowing the upper junction to lag the interior wall (Fig. 2). The reference is the quartic model with $P_s = 1$, $a = -1$, $b = 1$, $\Gamma = 1$ and $g_y = g_z = 0.5$. Length is measured in the equilibrium tanh-wall width $w_0 = \sqrt{2g_y/|a|}$, time in $t_0 = \Gamma/|a|$, and velocity in $w_0/t_0$. The film thickness $h$ and the top-to-bottom wall displacement are therefore reported in units of $w_0$. The bias $e$ is the coefficient of $-ep$ in the reduced polarization free energy and acts on polarization without directly driving the surface carriers. The reduced electrostatic coefficients are $\epsilon_f = 10$ and $\epsilon_e = 1$; they include the unit conversion and are not relative dielectric constants. Local exchange has $\Lambda = 0.05$, with $M = 0$ and $\chi = 0$. At the same bias $e = 0.06$, increasing $h$ from 5 to 20 raises the velocity from 0.0404 to 0.0728 and increases the top-to-bottom displacement from 0.095 to 0.985. The thicker wall advances more readily in the interior and bends near its screening-limited surface junction. Numerical methods and refinement checks are given in S12.

The computed branches are monotone over $0 \leq e \leq 0.10$. Their sampled scalar spectra show no growing nontranslational mode for the isotropic gradient coefficients used here. The nearly neutral translation eigenvalue is sensitive to discretization and is excluded from this statement. A complete coarse-grid spectrum at representative points agrees with the nearby-eigenvalue calculation; grid and domain checks are given in S12. The result establishes existence and sampled scalar stability of deformable walls in this model. It does not exclude vector rotation in a different polarization potential.

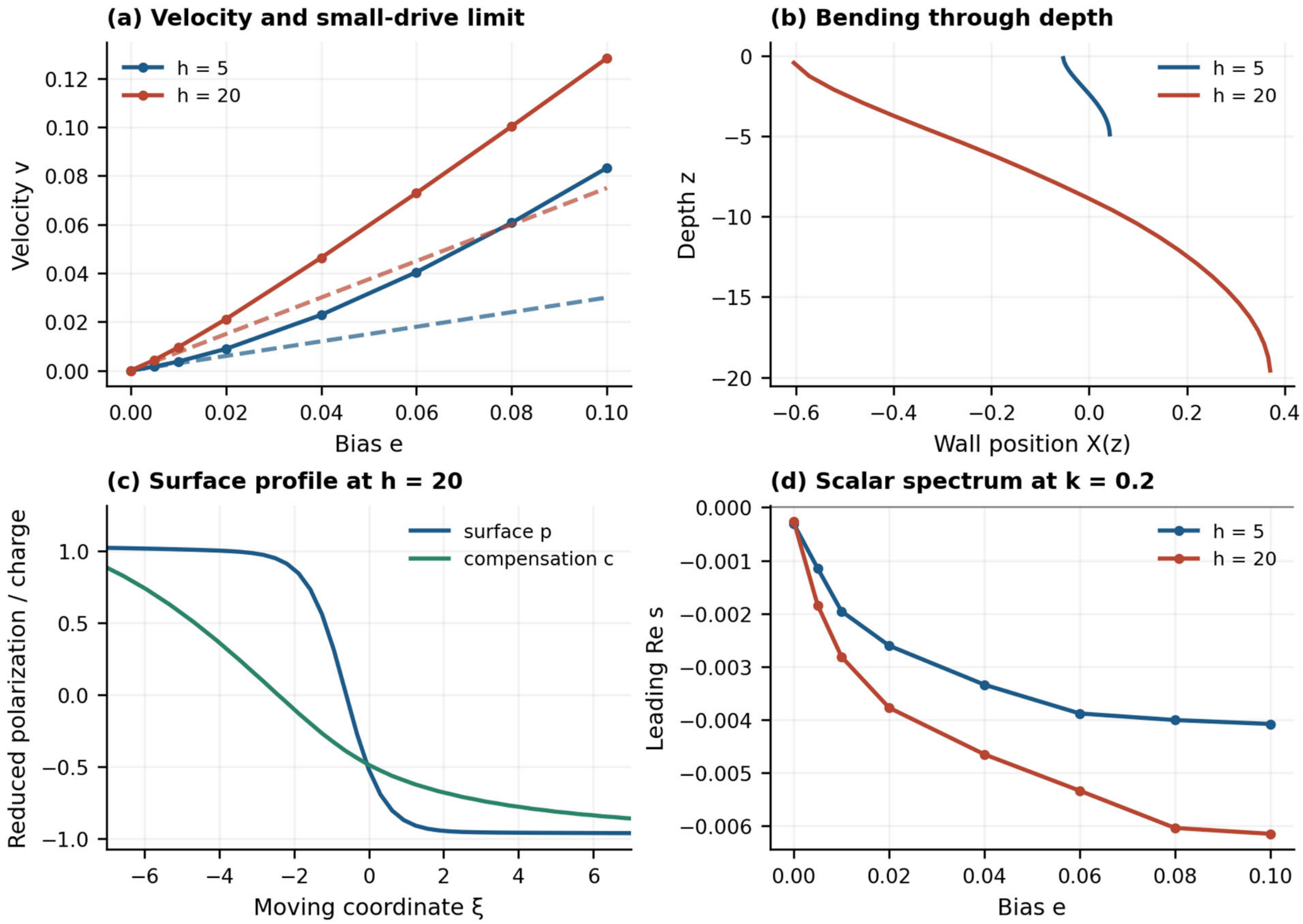

**Figure 2.** Computed open-surface walls with local exchange. (a) Velocity versus thermodynamic bias. Dashed straight lines are the small-drive rigid-wall slopes, not fits. (b) Wall position through the film thickness at $e = 0.06$. (c) Polarization and compensation at the upper surface of the thicker film. (d) Leading computed scalar growth rate at $k = 0.2$. Length, time, polarization and bias use the reduced units specified in S12; $w = 1$ for the reference equilibrium wall.

### 4.3. Case II: exchange combined with lateral transport

In this case charge is supplied locally by the environment and also moves along the surface. When the wall advances, its previous screening distribution produces an electric field that opposes displacement. We call this retarding force charge drag. Lateral motion can carry compensation toward the wall and spread the charge wake; local exchange replenishes the changed far-domain charge. When the relevant lateral wavelengths exceed the electrical lengths, $G_q \simeq G_0$, $r_q = r_0 + Dq^2$, with $r_0 = \Lambda(G_0 + \chi)$ and $D = M(G_0 + \chi)$. The local equation is $\partial_t c = D\,\partial_y^2(c - a_0 p) - r_0(c - a_0 p)$. The derivative acts on the electrochemical deviation, not only on $c$.

For a finite-width tanh profile, the large-speed expansion within this local model is

$$F_{\mathrm{ch}} = \frac{B}{2}\left[1 + \frac{2D/(3w) - r_0 w}{v} + o(v^{-1})\right] \tag{9a}$$

$$B = \Delta P^2 G_0 a_0. \tag{9b}$$

Thus $D/(r_0 w^2) > 3/2$ is sufficient for an interval of decreasing charge drag. A fold of the total force additionally requires $\eta + F_{\mathrm{ch}}' = 0$. The general bound $F_{\mathrm{ch}}' \geq -\zeta_{\mathrm{ch}}/8$ gives a sufficient exclusion condition $\eta > \zeta_{\mathrm{ch}}/8$. The numerical value $3/2$ belongs to this local-capacitance tanh limit; it is not a threshold for an arbitrary film or adsorption law.

If the polarization reverses across a distance much shorter than both charge-response lengths, the charge field sees an abrupt polarization step. Setting $W = -\Delta P$ then gives $F_{\mathrm{ch}} =$ $Bv/(v^2 + 4Dr_0)^{1/2}$, which is monotone. The different results identify the importance of finite wall width and the order of limits. Spatial instability can nevertheless occur on the monotone branch, as shown in Section 8.

At fixed $M > 0$ and vanishing exchange, the low-speed friction and the resulting speed scale as

$$\zeta_{\text{ch}} \sim \frac{B}{2\sqrt{Dr_0}} \tag{10a}$$

$$v \sim \frac{2F\sqrt{Dr_0}}{B}. \tag{10b}$$

In this sharp-step limit, with $\delta = F/B$ and negligible intrinsic friction, $v = 2\sqrt{Dr_0}\,\delta/(1-\delta^2)^{1/2}$ for $0 < \delta < 1$. As exchange becomes slower, the distance $\sqrt{D/r_0}$ over which lateral transport redistributes charge becomes larger and the steady speed tends to zero at fixed force in this regime. At exactly zero exchange, a wall connecting the two differently compensated domains cannot maintain steady translation with zero far current. Its response is then the transient motion analyzed next.

### 4.4. Case III: strictly conserved mobile surface charge

Integrating the charge balance for a putative traveling profile gives the exact identity

$$v(c_R - c_L) = J_R - J_L. \tag{11}$$

With different compensated far charges and zero far currents, nonzero steady velocity is impossible. This does not immobilize the polarization wall and allows for wall deformation. Displacement can be accompanied by an expanding charge-redistribution region and a time-dependent screening wake. At weak drive, the conserved displacement impedance has the long-wave form

$$Z(s) \sim \eta + \frac{B}{2\sqrt{Ds}} \tag{12a}$$

$$X(t) \sim \frac{4F\sqrt{Dt}}{B\sqrt{\pi}}. \tag{12b}$$

The two-sided sharp-interface solution extends this result to finite reduced force: $X = 2\lambda\sqrt{Dt}$, where

$$\frac{F}{B} = \frac{\sqrt{\pi}}{2}\lambda e^{\lambda^2}\operatorname{erfc}(\lambda)\operatorname{erfc}(-\lambda). \tag{13}$$

The square-root law is a finite-film long-wave result. For short wavelengths in the exposed geometry, the relaxation rate can scale as $|q|$ rather than $q^2$. In a finite closed specimen, the global domain fraction and conserved charge eventually produce a restoring force or drive the system to another equilibrium. Stability during spreading or arrest is a nonautonomous initial-value problem; the dispersion relation of a constant-speed wall cannot be assigned to it.

Figure 3 was evaluated by time integration of the coupled polarization and conserved surface-charge equations, with Poisson's equation enforced at every step. We first relaxed a compensated wall on a closed surface, then applied a constant thermodynamic bias $e = 0.01$ and followed its displacement to $t = 5000$. The surface length is $L = 40$, the film thickness is $h = 1$, and the transport coefficients are $M = 10$ and $\Lambda = 0$, with zero charge flux at the lateral edges. The reference units are those used for Fig. 2. The wall leaves a spreading charge wake and approaches a final displacement of 2.00075, close to the leading finite-inventory prediction of 2.00000. Arrest follows from redistribution of a fixed total charge and the resulting restoring field as the domain fractions change; no pinning force is introduced. The conservative time-stepping scheme, charge-balance check and mesh refinement are documented in S12.

The square-root displacement gives the infinite-domain long-wave asymptote. The finite calculation approaches this trend before crossing over to arrest; increasing the interval to $L = 200$ extends the spreading regime (Fig. 3b). Intrinsic wall friction and finite-size effects remain visible, so the curves are not fitted as exact square-root trajectories. Perturbations of the position at $k = 0, 0.15, 0.30$, evolved on the time-dependent base state, decay for the stated preparation and stiffness. This is a direct transient stability test for those perturbations, not a steady-front dispersion relation. The calculation is well away from a polar spinodal; a separate near-spinodal continuation remains necessary if far-domain susceptibility becomes large.

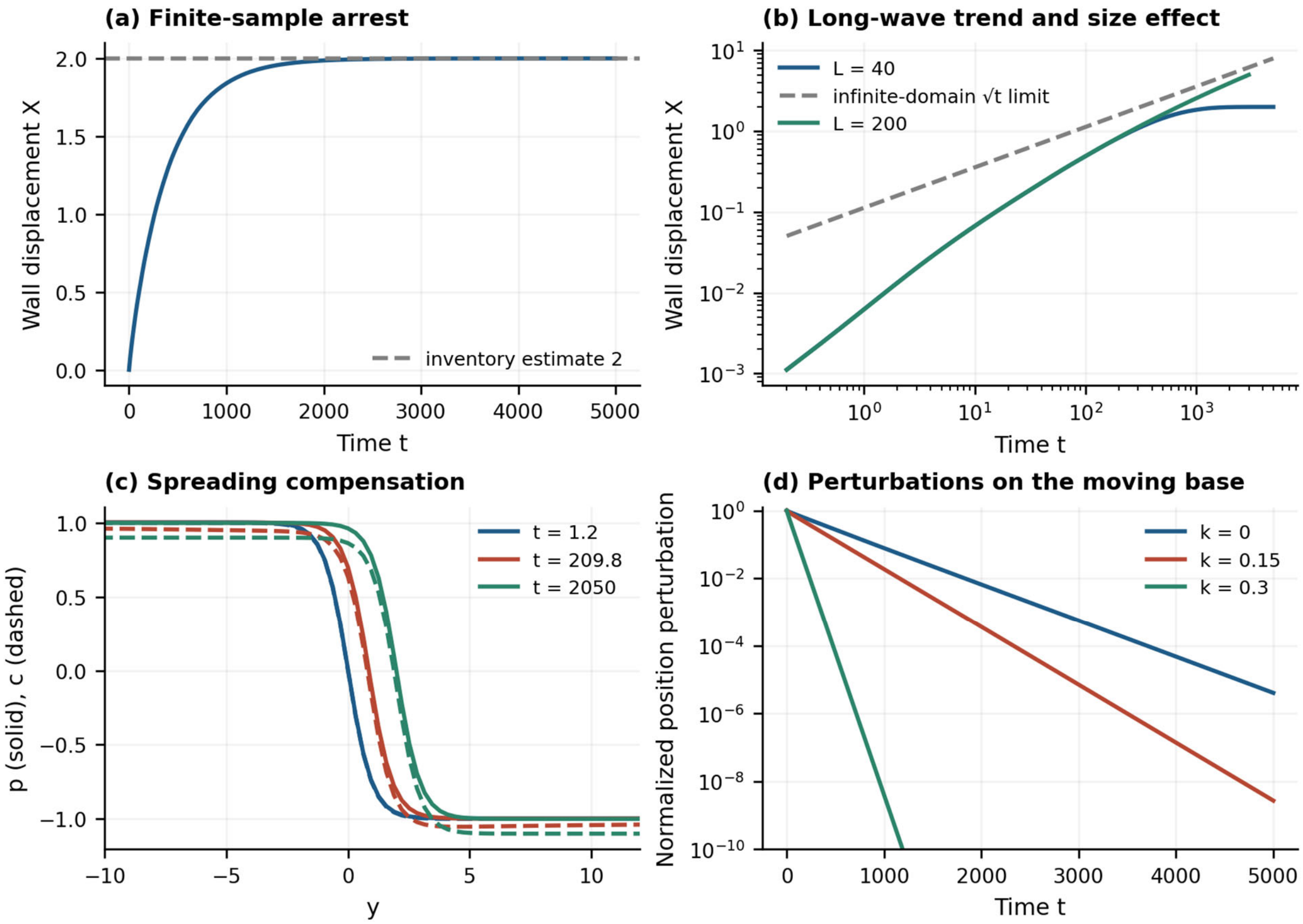

**Figure 3.** Conserved finite-depth evolution. (a) Wall displacement and finite-inventory arrest. (b) Early and intermediate displacement compared with the long-wave square-root asymptote. (c) Surface polarization and compensation during relaxation. (d) Displacement perturbations on the evolving base, normalized to their initial amplitude. The lateral boundaries have zero particle current; no reservoir supplies compensating charge.

### 4.5. First-order versus second-order ferroelectrics

We further analyze this dynamics for the material with the first order transition. The exposed electrostatic response and the conservation identity are unchanged. The polarization profile, tension and friction change, and a new internal coordinate appears near coexistence. The exact sixth-order profile and its finite friction integral are given in Section 7. Away from splitting, replacing $W_4$ by $W_6$ and $I_4$ by $I_6$ in Eqs. (7a–c) and (8) gives the prescribed-profile first-order velocity law. A growing geometric width does not imply vanishing friction because the gradients remain concentrated at two interfaces.

For local capacitance, the Case I monotonicity proof remains valid for a fixed monotone sixth-order profile. For the full exposed kernel, the tanh-specific proof must not be extended to a two-peak form factor without checking its hypotheses. Lateral transport can resolve the two interfaces and couple their translation to their separation. A fold calculated at fixed separation is not automatically a fold after the separation is allowed to relax.

Under strict conservation, common translation still requires a net charge transfer. Symmetric expansion of the central layer removes equal amounts of positive and negative polarization and has zero net charge demand. In the separated-interface limit, the conserved screening contribution to the width friction is finite and proportional to the separation, whereas the quasistatic translation friction diverges. Thus the first-order extension adds a slow charge-balanced structural process even when a steadily advancing screened wall is unavailable. A paraelectric interpretation requires a stable weak-polarization state under the actual electrical and mechanical constraints; first-order bulk thermodynamics alone is insufficient.

## 5. Wall motion beneath a dielectric layer and top electrode

The top electrode provides an additional screening mechanism. When a wall advances beyond its slowly moving ionic compensation, electronic charge on the electrode reduces the resulting field. Its effectiveness depends on the lateral scale of the charge variation, the dielectric gap, the dielectric contrast, and the film thickness. The interfacial ionic population nevertheless retains its own exchange or conservation law. We therefore compare the same three charge-dynamics cases in this electrical geometry, distinguishing the response of the electrode from the dynamics of the buried screening population.

### 5.1. Electrical response, voltage control and previous studies

The finite stack changes the electrical field generated by a given polarization–charge mismatch. Classical surface-layer theories treated domain formation and restoring forces in electroded specimens [9, 57]. Eliseev and colleagues derived screening-retarded wall velocities for a capacitor with an interfacial dielectric layer [3]. Ferroionic models subsequently coupled polarization to a potential-dependent ionic equilibrium and finite local charge relaxation [36, 58]. The comparisons below retain the charge law and the region over which the electrical force acts in each treatment.

For the same polarization and charge conventions as Section 4,

$$G_T(Q) = \{\epsilon_0[\bar{\epsilon}Q\coth(\alpha Q h) + \epsilon_d Q\coth(Qd)]\}^{-1}. \tag{14}$$

Let $C_f = \epsilon_0\epsilon_z/h$, $C_d = \epsilon_0\epsilon_d/d$ and $b_V = C_d/(C_f + C_d)$. Then $G_T(0) = (C_f + C_d)^{-1}$ and $\phi_{s,0} = G_0(p_0 - c_0) - b_V V$. With $\Psi$ defined by Eq. (2a), reservoir equilibration gives

$$a_{\text{eff}} = a + \frac{G_0\chi}{h(G_0+\chi)} \tag{15a}$$

$$E_d = \frac{\chi b_V V + G_0\Psi}{h(G_0+\chi)}. \tag{15b}$$

For a fixed uniform compensation $c_b$, use instead $a_{\text{fr}} = a + G_0/h$ and $E_{\text{fr}} = (b_V V + G_0 c_b)/h$. These are different electrical/chemical ensembles. A local quadratic contribution already absorbed in $a_{\text{eff}}$ must be subtracted from the remaining nonlocal energy to avoid counting it twice.

### 5.2. Case I and the Eliseev screening-relaxation calculation

For passive local electrochemical exchange, Eqs. (5), (7a–c) and (8) apply with $G_T$. In the local-capacitance approximation, the surface potential at each lateral position is determined by the charge mismatch at that same position, with $G_Q$ replaced by its long-wavelength value $G_0$. This is the limit in which the relevant lateral length scales are large compared with the electrostatic thickness scales of the stack. Neglecting the remaining wave-number dependence gives the following real-space drag for any fixed monotone wall profile:

$$F_{\mathrm{ch}}(v) = \frac{H_0}{2}\int\int p_0'(\xi)p_0'(\eta)\exp[-r_0|\xi-\eta|/v]d\xi d\eta \tag{16a}$$

$$H_0 = \frac{G_0^2}{G_0+\chi} \tag{16b}$$

$$v > 0. \tag{16c}$$

Every integrand increases with $v$, so the force is monotone. This is a statement about the specified passive local model and whole-wall projection.

The historical calculation instead uses the relaxation law $\tau\dot{c} = p_0(y - vt) - c$, derived under its conductivity-based reduction [3]. Its traveling solution is

$$c(\xi) = \frac{1}{v\tau}\int_{\xi}^{\infty} e^{-(u-\xi)/(v\tau)}\, p_0(u)\, du \tag{17a}$$

$$c_q = \frac{p_{0,q}}{1 - iqv\tau}. \tag{17b}$$

Inserting $p_0 - c$ into the capacitor Green function gives the electric field produced by the delayed screening charge. The field at the wall–surface intersection weights each Fourier mode by $\alpha|q|\coth(\alpha|q|h)$, whereas the thickness-averaged field weights it by $1/h$. In the model of Eliseev and colleagues [3], the speed of this surface intersection is set by the local electric field acting there. A force balance for the entire wall instead projects the field through the film onto the translating polarization profile. These are different approximations to wall motion and can give different slopes of the field–velocity curve. Their difference becomes substantial when the delayed field is concentrated near the surface and the wall bends with depth. S4 derives both fields and the corresponding velocity relations.

Removing the top electrode does not remove screening delay. It changes the electrical Green function and the applied-field protocol. The exposed response must then be solved with its own far-field conditions; a capacitor velocity formula cannot simply be retained unchanged.

### 5.3. Case II: exchange and lateral transport

For a wall narrow relative to both charge-response lengths, define $B = H_0 \Delta P^2$ and $S_v = (v^2 + 4Dr_0)^{1/2}$. The local single-step result is

$$F = \eta v + Bv/S_v \tag{18a}$$

$$F'(v) = \eta + 4BDr_0/S_v^3 > 0. \tag{18b}$$

Its transverse stiffness can nevertheless decrease enough to destabilize a straight front. The exact linear dispersion within this local sharp-interface reduction is

$$\mathcal{C}(s,k) = \eta s + Tk^2 + B\left[\frac{s + v^2/(2D)}{\sqrt{v^2 + 4D(r_0 + Dk^2 + s)}} - \frac{v^2}{2DS_v}\right] = 0, \tag{19}$$

where $T = hg_x I$ and the square root has positive real part. The effective long-wave tension is $T_{\mathrm{eff}} = T - BDv^2/S_v^3$. Equation (19) supplies a stationary corrugation instability on a positive-slope velocity branch and excludes a Hopf crossing in this particular unresolved-interface limit. The proof and wavelength formulas are given in S7.

The $r_0 \to 0^+$ supplied branch has diverging response lengths and retains its reservoir-selected far states. Its pointwise force limit is not the same as the solution prepared with $r_0 = 0$ and equal far charges. This is the same balance-law issue as in the exposed film. It remains present even though the electrode can exchange electronic charge with its voltage source.

### 5.4. Case III: conserved interfacial populations

For each conserved species, $v[n_i] = [J_i]$. Zero endpoint currents forbid a steadily translating profile joining distinct screening populations. A different running state is possible in the linear local model when the far compensation is identical on both sides. For $v > 0$ and a sharp descending polarization step,

$$c - c_b = -a_0 \Delta P \, e^{-v\xi/D} \Theta(\xi) \quad (20a)$$

$$\int (c - c_b) d\xi = -\frac{a_0 \Delta P D}{v}. \quad (20b)$$

The pulse is a deficit in the compensation convention $c = -\sigma$. Its excess inventory must be present initially or balanced elsewhere. Its drag is $B/2$. Different velocities correspond to different pulse inventories, so $F = \eta v + B/2$ is a family of prepared states, not a universal velocity curve at fixed inventory. Finite closed samples generally require transient redistribution and changing domain fractions.

The evaluated Eliseev junction-field relation exhibits a negative-slope interval in a sufficiently thick film for the chosen screening relaxation time (Fig. 4a). Reducing the film thickness narrows and then removes that interval. The thickness-averaged force closure gives a different velocity curve (Fig. 4b), showing that the field sampled at the surface junction cannot generally replace the force integrated through the wall. Gap-dependent comparisons and numerical values are retained in Fig. S1 and S12. These curves evaluate the historical equations [3] with explicitly stated reduced parameters.

The coupled finite-depth model with bounded adsorption, local exchange and lateral transport instead has a monotone velocity branch over the computed driving interval (Fig. 4c). At $e = 0.06$, its velocity is approximately 0.0473. Here the imposed thermodynamic bias acts on polarization, while the surface charge responds to its electrochemical potential through passive exchange and transport. The bounded chemical law limits the available compensation. The stability of the moving wall is evaluated from the coupled polarization and charge equations: reducing the gradient stiffness along the front produces corrugation even though the velocity continues to increase with driving. The nonlinear scalar and vector examples are shown in Fig. 8.

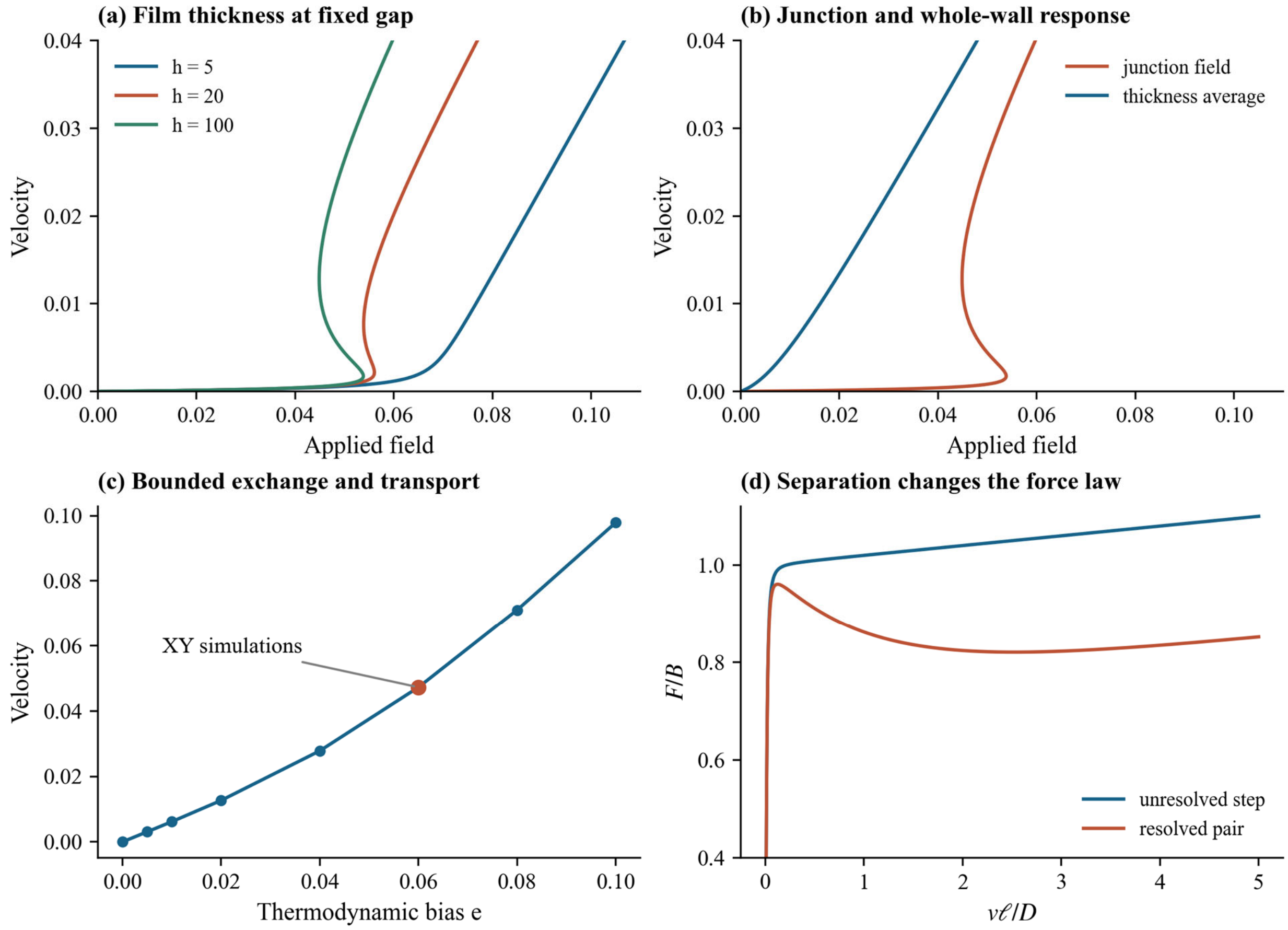

**Figure 4.** Electrode screening and the role of wall structure. (a) Eliseev junction-field velocity curves for different film thicknesses at fixed gap. (b) Junction-field and thickness-averaged closures for the same stack. (c) Coupled finite-depth velocity branch with bounded adsorption, exchange, and lateral transport; the marked driving is used for the xy simulations. (d) Local mixed-transport force laws for an unresolved step and a resolved pair at fixed separation. Panels (a,b) use the historical screening-relaxation law; (c) uses a passive electrochemical law and thermodynamic driving; (d) isolates the effect of resolving the two interfaces. Full comparisons and parameters are given in S12 and S15.

### 5.5. First-order walls and interphase separation

Near first-order coexistence, a 180-degree wall can separate into two interphase boundaries enclosing a weakly polarized layer. Each boundary changes the polarization by approximately $P_s$, and their separation $\ell$ becomes an additional dynamical variable. If the screening distribution resolves the two boundaries, each leaves a charge wake that also acts on the other. The resulting electrical forces can therefore alter both the common translation speed and the thickness of the intervening layer. This coupling is absent when the wall is treated as one abrupt polarization step. With $a_\pm = (S_v \pm v)/(2D)$, the local exchange-plus-transport model gives

$$F_{\mathrm{ch}}^{(2)} = \frac{Bv}{4S_v}\left[2 + e^{-a_-\ell} + e^{-a_+\ell}\right] \tag{21a}$$

$$F_\ell^{\mathrm{ch}} = \frac{Bv}{8S_v}\left[e^{-a_-\ell} - e^{-a_+\ell}\right]. \tag{21b}$$

The center and separation equations are $F = \eta_X v + F_{\mathrm{ch}}^{(2)}$ and $\eta_\ell \dot{\ell} = -U'(\ell) + F_\ell^{\mathrm{ch}}$. Both must be satisfied before a steady pair is claimed. In the regime $r_0\ell^2/D \ll 1$ and $v\ell/D = O(1)$, the fixed-separation drag becomes $B\left[3 + e^{-v\ell/D}\right]/4$ and can decrease with speed. This provides a resolved-pair mechanism absent from the monotone unresolved-step result. Along a freely adjusting branch, $d\ell/dv$ also contributes to $dF/dv$.

For local exchange alone, the corresponding narrow-pair drag is $B\left[1 + e^{-r_0\ell/v}\right]/4$, and the charge wake contributes an opening force $Be^{-r_0\ell/v}/8$. A cooperative delay-system calculation excludes growing center/separation modes when the retained binding stiffness is positive. The existence of a first-order separation coordinate therefore does not guarantee oscillation.

For the equal-background conserved running family, the pair drag is $B\left[1 + e^{-v\ell/D}\right]/4$ and the charge force on the separation is compressive, $-Be^{-v\ell/D}/8$. Its reference potential is the frozen-background potential, and its pulse inventory remains $-2a_0P_sD/v$. The sign difference from the reservoir-fed opening force is a consequence of different charge preparations. The complete pair response and its stability determinant are given in S5 and S7.

## 6. Wall motion adjoining a blocking mixed ionic–electronic conductor

An adjoining conducting medium screens polarization through mobile charge distributed over a finite depth. A solid mixed ionic–electronic conductor carries both ionic and electronic current. An aqueous electrolyte provides a related diffuse-screening geometry but normally requires positive and negative ions rather than the ionic/electronic pair used below. A semiconductor can often be described by drift, diffusion and Poisson equations, but its electrons, holes, fixed dopants and recombination law must be specified [18, 59]. The displayed blocking model applies directly when the retained species are separately conserved and neither crosses the interface. Faradaic transfer, carrier recombination or nonblocking contacts change those equations and boundary conditions.

### 6.1. Physical model and relation to earlier work

The MIEC occupies the half-space above an insulating ferroelectric. Screening is carried by ionic and electronic volume densities, with no particle transfer through their common boundary. Earlier studies provide the diffuse-charge transport framework, MIEC electromechanical response and equilibrium semiconductor screening [59, 60, 61, 62, 63]. Genenko and colleagues treated migration of charged defects inside the ferroelectric, including self-consistent numerical reconstruction of its domains [64, 65]. Those calculations retain their own charge location and boundary conditions but do not provide the solutions of an external blocking-MIEC junction.

For species $j$ of charge $q_j$, concentration $n_j$, susceptibility $\chi_j = \partial n_j/\partial\mu_j$ and diffusivity $D_j = L_j/\chi_j$, the baseline equations are

$$\partial_t n_j = -\nabla\cdot\mathbf{J}_j \tag{22a}$$

$$\mathbf{J}_j = -L_j\nabla\mu_j \tag{22b}$$

$$\mu_j = \partial_{n_j} f_{\text{chem}} + q_j\phi_m \tag{22c}$$

$$-\epsilon_m\nabla^2\phi_m = \textstyle\sum_j q_j\, n_j + \rho_{\text{fix}}. \tag{22d}$$

The interface is blocking for both species. At large depth, the concentrations and potential approach mutually consistent reference values. A half-space has an unbounded inventory at nonzero carrier density, but transport from distant regions still requires a finite conductivity and an electrical potential drop.

### 6.2. Static and dynamic screening response

The nonlinear planar equilibrium satisfies $\epsilon_m\phi'^2/2 = -\int_0^{\phi}\rho\,(u)du$ . This gives the interface charge–potential relation by quadrature for a specified carrier equation of state. The homogeneous linear response requires small relative concentration changes; for dilute carriers, $|q_j\phi| \ll k_BT$ is the corresponding condition. It must not be assumed solely because the electric field is treated quasistatically.

Define $\kappa_j^2 = q_j^2\chi_j/\epsilon_m$, $\kappa^2 = \sum_j \kappa_j^2$ and $\sigma_m = \epsilon_m \sum_j D_j \kappa_j^2$. For equal ionic and electronic diffusivities $D_m$, the blocking half-space admittance is

$$Y_m(Q,s) = \epsilon_m Q p \frac{s+D_m\kappa^2}{sp+D_m\kappa^2 Q} \tag{23a}$$

$$p = \sqrt{Q^2+\kappa^2+s/D_m}. \tag{23b}$$

Here $D_{m,z} = Y_m\phi_s$. The equal-diffusivity assumption is used only for the displayed closed expression. S8 gives the algebraic two-species solution for unequal diffusivities. The interface response to polarization is

$$H(Q,s) = \left[\epsilon_f Q + Y_m(Q,s)\right]^{-1}. \tag{24}$$

The static limit is $Y_m(Q,0) = \epsilon_m\sqrt{Q^2+\kappa^2}$; the high-frequency limit is $\epsilon_m Q$. The harmonic depth mode is necessary to enforce blocking and must be retained along with the space-charge modes. Unequal diffusivities also introduce a composition mode whose timescale is distinct from charge relaxation.

### 6.3. Charge conservation at the isolated moving junction

An isolated wall advancing at speed $v$ changes the integrated screening charge by $|\Delta P v|$ per unit junction length and time. In the long-distance Ohmic sector of a half-plane, the associated spreading current gives

$$|\Delta\phi| \simeq \frac{|\Delta P v|}{\pi\sigma_m}\ln\frac{R}{a_{\text{in}}} \tag{25a}$$

$$P_{\text{diss}}/L_x \simeq \frac{(\Delta P v)^2}{\pi\sigma_m}\ln\frac{R}{a_{\text{in}}}. \tag{25b}$$

The inner length $a_{\text{in}}$ describes the near-junction region and $R$ the actual outer supply scale. Taking $R \to \infty$ at finite $v$ and finite remote potential makes the steady screening problem singular. Thus a bounded stationary screening cloud for this isolated line cannot be maintained solely by a finite potential reference at infinity. This is a transport constraint, not an absence of available carriers and not a violation of interfacial blocking.

Transient wall motion, incomplete screening, depth bending and charge-balanced rearrangements remain possible. A translated periodic pattern of fixed domain fractions has no changing mean charge and avoids the isolated-front supply term. A finite contact distance or finite junction length changes the outer problem and must be specified before assigning a stationary force–velocity relation. The outer scale of a transient cannot in general be replaced by a universal $\sqrt{D_m t}$ because charge and composition relaxation need not have the same dynamics.

### 6.4. Second-order wall motion, bending and conditional velocity branches

Far below the interface, a neutral scalar wall has the exact local quartic traveling solution given in S2. Its bulk velocity satisfies

$$\Gamma v_{\text{bulk}} \int p'^2 d\xi = f_E(p_R) - f_E(p_L) \tag{26a}$$

$$f_E = f_L - E_0 p. \tag{26b}$$

A localized junction force cannot select a separate finite drag correction for an infinitely deep rigid wall: both the deep-field force and intrinsic wall drag grow with its depth. The relevant surface response is therefore a finite-depth deformation of a wall whose deep motion is specified, or the motion of a finite specimen with a stated outer supply geometry.

For an admissible finite or charge-balanced mode, the passive response has a positive relaxation representation. A single retained spatial and relaxation mode produces the conditional force law

$$F = \eta v + \frac{Av}{1+(v/v_c)^2} \tag{27a}$$

$$v_c = r/q_0 \tag{27b}$$

$$A > 0. \tag{27c}$$

With $\mathcal{A} = A/\eta$ and $u = v/v_c$, folds exist if and only if $\mathcal{A} > 8$, at $u_\pm^2 = \left[\mathcal{A} - 2 \pm \sqrt{\mathcal{A}(\mathcal{A} - 8)}\right]/2$. This is an exact result for the one-mode model. It is not a steady-state prediction for the unregularized isolated half-space, and a decreasing near-field drag cannot eliminate the positive outer supply dissipation.

### 6.5. First-order wall structure at a blocking interface

The blocking conditions, two-species response and long-distance charge transport remain unchanged. The sixth-order wall changes the source form factor and intrinsic friction, and near coexistence it adds a separation mode. Common advance of the two interfaces still carries the net polarization jump and therefore retains the isolated-junction supply constraint. Symmetric separation changes have no net source at zero bias and can couple to finite-wave-number screening without the same logarithmic translation term.

The first-order scalar model permits a bound interphase pair below coexistence, logarithmic separation at coexistence, and expansion of the lower-energy weakly polarized phase above coexistence, as long as the actual constrained potential supports that phase. Charge relaxation alters the separation kinetics and can couple the mode to temporal dynamics. The passive finite-mode calculations in S9 demonstrate possible oscillatory onset on a supplied moving base; they do not establish sustained oscillation of the strict isolated-infinity problem.

For multiaxial materials, a rotated or structurally reconstructed core can compete with the scalar paraelectric route. The mode that becomes soft first must be determined from the same electrical, mechanical and structural model used to construct the base wall. A first-order bulk transition supplies additional possible states and modes; it does not by itself select a velocity fold, a corrugation or a Hopf bifurcation.

## 7. Changes in wall structure and the multiaxial extension

A wall can accommodate a separation from its compensating charge by changing the polarization-reversal profile. The simplest response is broadening with reduced polarization in the center. If the local free energy also supports a weakly polarized state, the wall can separate into two interphase boundaries enclosing that state. A distinct core requires an interval long compared with the relevant interphase thickness, polarization close to a locally stable central branch, and two identifiable bounding interfaces. At nonzero field the central branch is generally weakly polarized rather than exactly nonpolar. A structural assignment also requires the remaining order parameters of that phase.

A quartic second-order material below its transition has $f_L''(0) = a < 0$. A static extended $p \simeq 0$ layer is therefore unavailable in its homogeneous local scalar potential. Heating toward the transition broadens the wall as $w \propto |a|^{-1/2}$ while $P_s$ and the bulk curvature decrease. A transient low-polarization region may persist because screening temporarily provides restoring stiffness, but its eventual stability must be assessed after the charge has relaxed. Stabilizing a true slab requires an additional local change of the effective free energy, a boundary constraint or another coupled field.

The scalar interphase structure and wetting mechanism were developed in the classical work of Ishibashi and Suzuki and of Dorfman and colleagues [66, 67, 68]; Lajzerowicz provides an earlier wetting precedent [69]. The present calculation asks how screening changes the motion and separation of those interfaces. Electron microscopy supplies evidence that distinct structural states can occur near complex domain configurations: atomic distortions in $BiFeO_3$ walls [51] and wall-associated antipolar structures in strained films [52] illustrate why polarization and structural order must be measured together. Those observations motivate a coupled structural model rather than identifying every weak-polarization core as paraelectric.

### 7.1. Exact sixth-order wall and the separation coordinate

At zero effective field, let $S = P_s^2$ be the stable polar root of $a + bS + c_6 S^2 = 0$, define $d_6 = 2S + 3b/(2c_6)$, and set $\rho = d_6/S$. Then

$$f_L(p) - f_L(P_s) = \frac{c_6}{6}(p^2 - S)^2(p^2 + d_6) \tag{28a}$$

$$\Delta f = f_L(0) - f_L(P_s) = \frac{c_6 S^3 \rho}{6}. \tag{28b}$$

For $\rho > 0$, the exact decreasing local scalar wall is

$$p_6(\xi) = -P_s \frac{\sqrt{\rho}\sinh(\kappa_w \xi)}{\sqrt{1+\rho\cosh^2(\kappa_w \xi)}} \tag{29a}$$

$$\kappa_w = S\sqrt{\frac{c_6(1+\rho)}{3g_y}}. \tag{29b}$$

The regime $0 < \rho < 1/2$ has a locally stable central state and two gradient maxima. Define $\ell$ by the two locations where $p^2 = S/2$. Then

$$\ell = \frac{2}{\kappa_w}\operatorname{asinh}\sqrt{\frac{1+\rho}{\rho}} \tag{30a}$$

$$I_6 = \frac{S^2}{4}\sqrt{\frac{c_6}{3g_y}}\left[(2-\rho)\sqrt{1+\rho} + \rho(4+\rho)\operatorname{asinh}\left(\rho^{-1/2}\right)\right]. \tag{30b}$$

The width between $p = \pm P_s/2$ is a different quantity, $w_{1/2} = 2\kappa_w^{-1}\operatorname{asinh}\sqrt{(1+\rho)/(3\rho)}$. This distinction is retained in every figure. At coexistence $\rho \to 0^+$, $\ell \sim \kappa_t^{-1}\ln(4/\rho)$ while $I_6 \to S_t^2\sqrt{c_6/(3g_y)}/2$. The two interfaces move apart while their gradients, total tension and translational friction remain finite.

For large separation, the energy per unit junction length is

$$U(\ell) = 2h\gamma_{F0} + h\Delta f\,\ell + hA_w e^{-\kappa_t \ell} + \cdots \tag{31a}$$

$$A_w = 8\gamma_{F0} \tag{31b}$$

$$\gamma_{F0} = g_y S_t \kappa_t / 4. \tag{31c}$$

This convention is tied to the definition of $\ell$ above. For $\Delta f > 0$, the weakly polarized phase has a higher bulk free-energy density than the surrounding polar domains. Increasing its thickness therefore costs $h\Delta f\,\ell$ per unit wall length. This contribution tends to bring the two interfaces together, while their short-range repulsion favors greater separation. Their balance gives a finite layer thickness. At coexistence the bulk free-energy difference vanishes, and the remaining repulsion leads to logarithmic separation in the intrinsic overdamped limit. When the weakly polarized phase has lower free energy, it expands into the metastable polar domains. Exact isolated interphase-front velocities and the screening contribution to the separation dynamics are derived in S2 and S5.

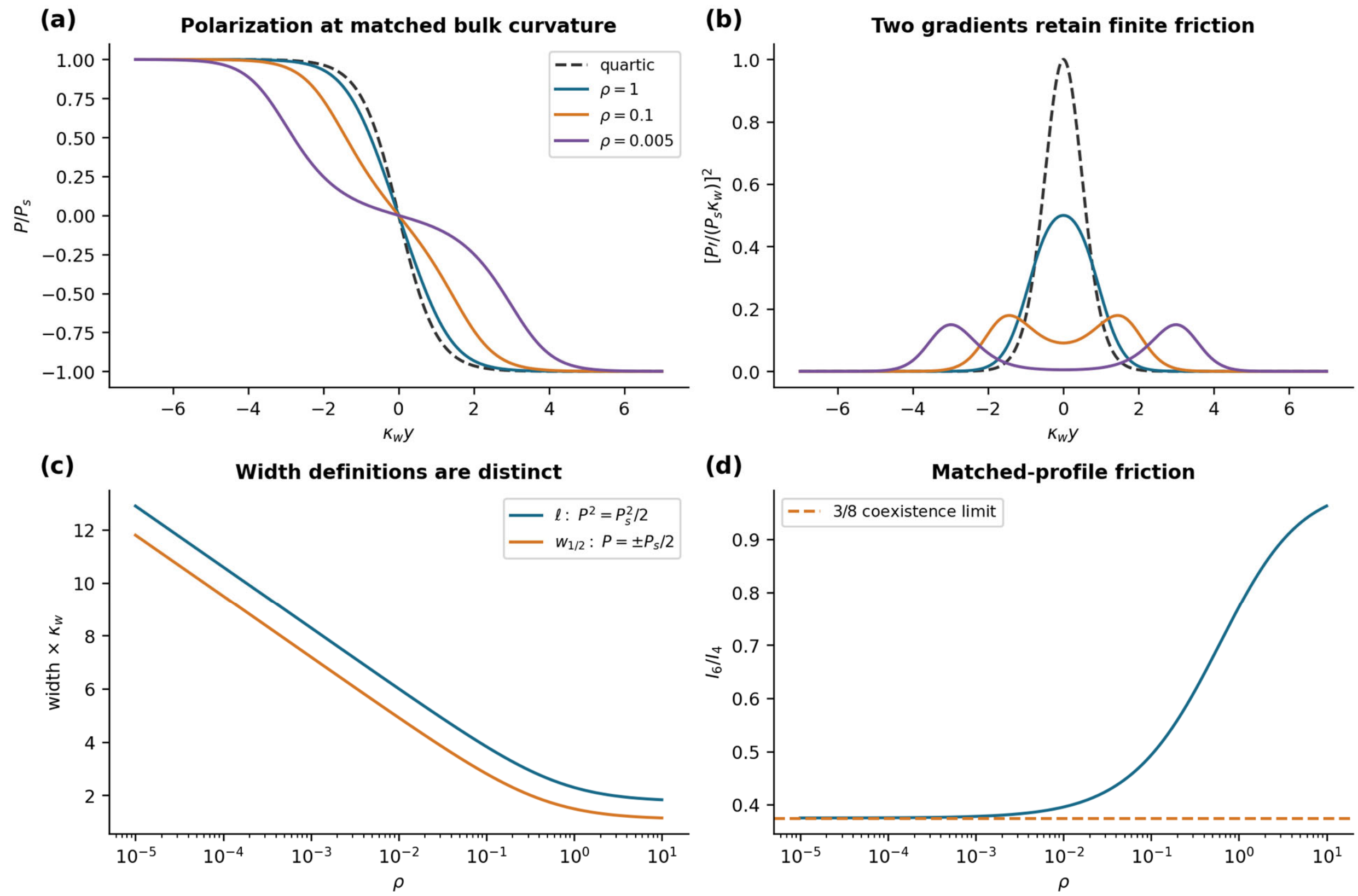

**Figure 5.** Exact scalar structure at matched $P_s$ and far-domain curvature. (a) Quartic and sixth-order polarization profiles. (b) The corresponding gradient densities; a separated first-order wall develops two peaks. (c) The two explicitly defined widths. (d) The finite friction ratio $I_6/I_4$, approaching $3/8$ at matched $P_s$, bulk curvature and $g_y$. The inverse ratio $8/3$ is the corresponding intrinsic small-drive mobility ratio, not a comparison of two arbitrary materials at the same temperature.

### 7.2. Localized polarization rotation

In a multiaxial ferroelectric the polarization can reverse by turning within the wall instead of passing through zero magnitude. A component parallel to the wall plane produces a Bloch-like core when the corresponding geometry is uncharged. Rotation can therefore lower the energy of a strongly suppressed scalar core, but it competes with the anisotropy that favors the bulk polarization directions. Classical theories describe non-Ising walls and their structural transitions [70, 71]. Marton, Rychetsky and Hlinka calculated compatible wall profiles in the different phases of $BaTiO_3$ [11], and Taherinejad and colleagues examined Bloch-like rhombohedral walls [72]. Yudin and colleagues established mechanisms for chirality and bistability [73, 74]. Strain-induced changes, Ising lines and mixed wall structures further show that rotation can acquire spatial structure along a wall [75, 76, 77]. For a moving surface junction the same competition is evaluated about the screened moving profile. A localized transverse component and an expanding intermediate domain are separate outcomes, because only the latter requires a stable bulk variant in the intervening region.

An explicit dimensionless reference is

$$\mathscr{F} = \int \left\{ \frac{p'^2 + a_\perp'^2}{2} + \frac{\left(p^2 + a_\perp^2 - 1\right)^2}{4} + \frac{\kappa_\perp a_\perp^2}{2} \right\} dy. \quad (32)$$

The Ising wall $p = \tanh(y/\sqrt{2})$ has a localized transverse eigenvalue $\kappa_\perp - 1/2$. For $0 < \kappa_\perp < 1/2$, exact rotated solutions are $p = \tanh(\sqrt{\kappa_\perp} y)$ and $a_\perp = \pm\sqrt{1 - 2\kappa_\perp}\text{sech}(\sqrt{\kappa_\perp} y)$. Their transverse polarization is confined to the wall. This potential does not have a stable macroscopic in-plane variant for $\kappa_\perp > 0$; it therefore cannot demonstrate growth of an in-plane domain.

### 7.3. Finite intermediate ferroelectric domains and bistability

A separate potential with stable axial variants is $f = -A(p^2 + a_\perp^2)/2 + B(p^4 + a_\perp^4)/4 + Cp^2 a_\perp^2/2$, with $C > B > 0$. It admits both out-of-plane and in-plane minima. Whether a finite intermediate domain grows depends on its bulk energy, bounding-wall energies, charge balance, closure geometry and strain compatibility. The absence of a linear rotation instability does not exclude a lower-energy finite-amplitude path.

For a symmetry-preserving amplitude $A_\perp$, the local expansion $\mathcal{U} = rA_\perp^2/2 + uA_\perp^4/4 + wA_\perp^6/6$ gives a continuous onset for $u > 0$. If $u < 0, w > 0$, the nonzero branch appears at $r = u^2/(4w)$, has equal equilibrium energy at $r = 3u^2/(16w)$, and the unrotated state loses local stability at $r = 0$. A conjugate transverse field unfolds this symmetry: a small induced transverse component is then not a spontaneous instability. Equal-energy lines of this equilibrium amplitude model must not be used as coexistence lines of an arbitrary driven system.

The structural calculations distinguish relaxation of a broadened wall from formation of a new central phase (Fig. 6). In the quartic scalar reference, the prepared broad core contracts to an ordinary wall. A sixth-order potential supports a bound weakly polarized layer below coexistence and an expanding layer above it. The screening population is conserved in these examples. Symmetric expansion consumes equal parts of the oppositely polarized domains and can therefore change the wall structure without the net charge transfer needed for common translation. The parameter sets and width histories are given in S12.

Vector continuation gives the separate rotation result. For the quartic radial potential, the computed Ising-to-Bloch threshold is $\kappa_\perp = 0.49986$, consistent with the exact value $1/2$. Sixth-order radial potentials matched in $P_s$ and far-domain curvature have much smaller linear rotation thresholds: 0.04317 at $\rho = 0.3$ and 0.005746 at $\rho = 0.1$. Stable finite-amplitude rotated walls persist beyond these linear thresholds, giving coexistence of locally stable Ising and rotated structures over part of the sampled interval. These are equilibrium reference branches under complete compensation. They show that first-order core softening can produce a discontinuous structural choice; they do not assign a universal rotation threshold to all moving walls.

A final time-dependent calculation uses a potential with an actual in-plane minimum, $f = -p^2/2 + p^4/4 - (1 - \delta)a_\perp^2/2 + a_\perp^4/4 + Cp^2 a_\perp^2/2$, with $C = 2$ and $\delta = -0.06$. A prepared rotated region expands, reaching a width of approximately 15.5 at $t = 300$. Thus localized rotation, a bound nonpolar core, an expanding nonpolar layer and an expanding rotated domain all occur, but under different free-energy conditions. No periodic forcing is used. The simulations and equilibrium continuation retain their own stated constraints; a full moving vector wall with nonlocal elastic relaxation is a remaining material-specific extension.

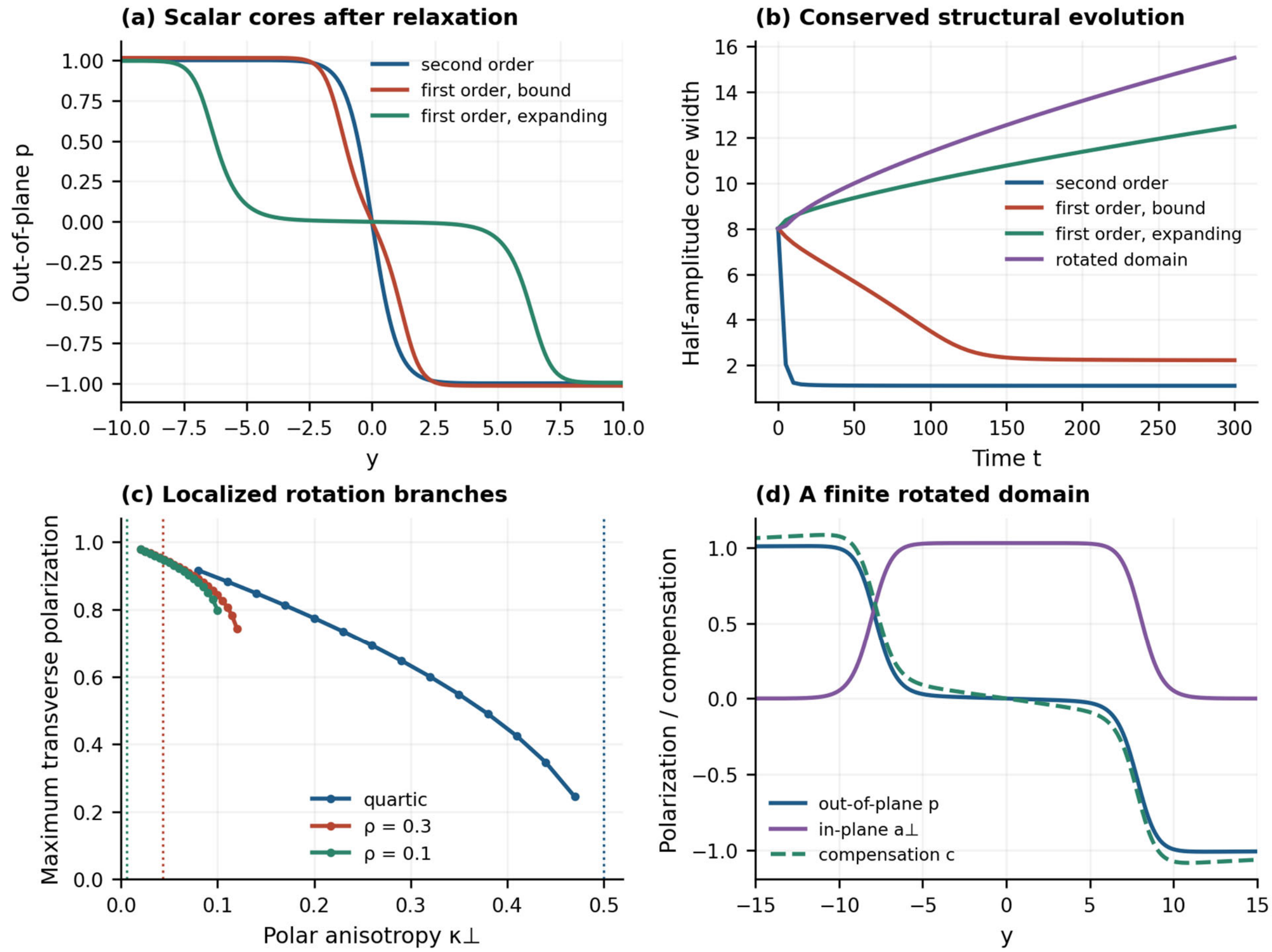

**Figure 6.** Calculated structural mechanisms. (a) Final scalar profiles after relaxation of the same broad initial core. (b) Core width versus time for conserved screening. (c) Transverse polarization on continued equilibrium vector-wall branches; solid segments are locally stable calculated branches, and vertical markers indicate loss of Ising stability. (d) The expanding rotated domain retains a nearly uniform in-plane interior, unlike a localized Bloch core. The time-dependent panels use the long-wave electrostatic model; the continuation panel uses a fully compensated charge-free planar reference. See S12 for potentials, preparation and convergence.

## 8. Stability of wall motion and comparative analytical diagrams

We now compare the stability of wall motion and internal structure. Displacement of the front, rotation of polarization, and separation of two interphase boundaries provide distinct perturbations and can couple through screening. A translating wall may lose stability even when its velocity increases with driving. When charge conservation prevents a stationary traveling profile, perturbations must instead be followed on the evolving transient. We first examine spatial instability and its nonlinear development, then the conditions for oscillatory behavior.

The retained fields are displacement $u(x,z,t)$, transverse polarization amplitude $A_\perp(x,z,t)$, and, where a core is resolved, separation $\ell(x,z,t)$. For $p = p_0(y - vt - u)$ and a tangential component $P_x = P_s A_\perp \psi(y - vt)$, the leading charge sources are

$$\delta\rho_b = p_0{}' \, \partial_z u - P_s \psi \, \partial_x A_\perp \tag{33a}$$

$$\delta\sigma_b = -p_0{}' u|_{z=0}. \tag{33b}$$

A uniform tangential Bloch component can be uncharged; a spatial modulation of that component generally is not. Coupling coefficients must be obtained from these sources and the appropriate Green function. They cannot be assigned independently merely to obtain an instability.

### 8.1. Displacement stability and velocity folds

We first approximate the wall by a fixed polarization profile that is uniform through the film thickness. The wall may translate and its surface position may vary along $x$, but its internal polarization profile and its position relative to depth are held fixed. Thus this displacement calculation describes a corrugated wall moving as a whole; bending through the thickness and changes in internal structure require the coupled field calculation or additional wall variables. For this prescribed-profile approximation, define

$$Z(Q,z) = G_Q - \frac{G_Q a_Q r_Q}{z+r_Q} \tag{34a}$$

$$\mathcal{C}(s,k) = \eta s + Tk^2 + \int_{-\infty}^{\infty} \frac{|W(q)|^2}{2\pi}\left[Z(Q,s-iqv) - Z(|q|,-iqv)\right]dq. \tag{34b}$$

The integral is evaluated causally and symmetrically; for complex $s$ it must remain an analytic function rather than an arbitrary real-part operation. The subtraction comes from varying both the polarization source and the force projection and guarantees $\mathcal{C}(0,0) = 0$. For the surface models, $\partial_s \mathcal{C}(0,0) = dF/dv$. A negative total force slope gives a real growing uniform mode within this reduction. A positive slope does not establish transverse or internal stability.

At rest, the residual static kernel is $K_{\mathrm{eq}}(Q) = \chi G_Q/(G_Q + \chi)$. Its transverse graph stiffness is $Tk^2 + \int |W|^2\left[K_{\mathrm{eq}}(Q) - K_{\mathrm{eq}}(|q|)\right]dq/(2\pi)$. Complete compensation makes this electrical correction vanish. Incomplete compensation can soften the wall. For the exposed surface, the $Q$ term in the exterior admittance can generate nonanalytic long-wave behavior, so a polynomial expansion in $k^2$ is not universally justified.

### 8.2. Corrugation and wavelength selection

We next consider a wall whose thickness is small compared with the screening lengths, so that the charge distribution responds to a single abrupt polarization step. Surface charge can both exchange locally with a reservoir and move laterally, while electrostatics is approximated by the local capacitance $G_0$. The model therefore describes displacement and corrugation without resolving the internal polarization profile. The reduced speed is $u = v/\left(2\sqrt{Dr_0}\right)$, and $\theta = T\sqrt{r_0}/\left(B\sqrt{D}\right)$ compares the wall-tension penalty with the destabilizing effect of delayed screening. Equation (19) gives

$$\theta_c(u) = \frac{u^2}{2\left(1+u^2\right)^{3/2}} \tag{35a}$$

$$\max\theta_c = \frac{1}{3\sqrt{3}}. \tag{35b}$$

The straight front is unstable to long waves when $\theta < \theta_c$. With $R = BDv^2/(TS_v^3) > 1$ and $s_c = \left[\sqrt{1+8R} - 1\right]/2$, the unstable band is $0 < k < k_c$, where $k_c^2 = S_v^2(s_c^2 - 1)/(4D^2)$. The onset wavelength diverges at a long-wave threshold in an infinite system. Near threshold, for $Q_k = k\sqrt{D/r_0}$, $s/r_0 \simeq \left(a_2 Q_k^2 - a_4 Q_k^4\right)/c_\lambda$, with $a_2 = \theta_c - \theta$ and $a_4 = 3u^2/\left[8(1+u^2)^{5/2}\right]$. Thus $Q_{\max}^2 = a_2/(2a_4)$ and $s_{\max} = r_0 a_2^2/(4a_4 c_\lambda)$. The positive coefficient $c_\lambda$ is given in S7. These are initial growth scales, not a nonlinear selected period.

A coupled displacement–rotation reduction illustrates a second route. For

$$K(Q_k) = \begin{pmatrix} b_s Q_k^2 & i c_s Q_k \\ -i c_s Q_k & \mu + d_s Q_k^2 \end{pmatrix} \tag{36a}$$

$$b_s, d_s > 0, \tag{36b}$$

the uniform unrotated wall with $\mu > 0$ loses mixed-mode long-wave stability when $b_s\mu < c_s^2$. At $b_s = d_s = D_s$, the fastest wave number is $Q_{\max}^2 = c_s^2/(4D_s^2) - \mu^2/(4c_s^2)$ and the maximum rate is $(c_s^2 - D_s\mu)^2/(4D_s c_s^2)$. These are conditional projected coefficients; a symmetry may set $c_s = 0$. S9 provides the eigenvalues and the finite-wave-number alternative.

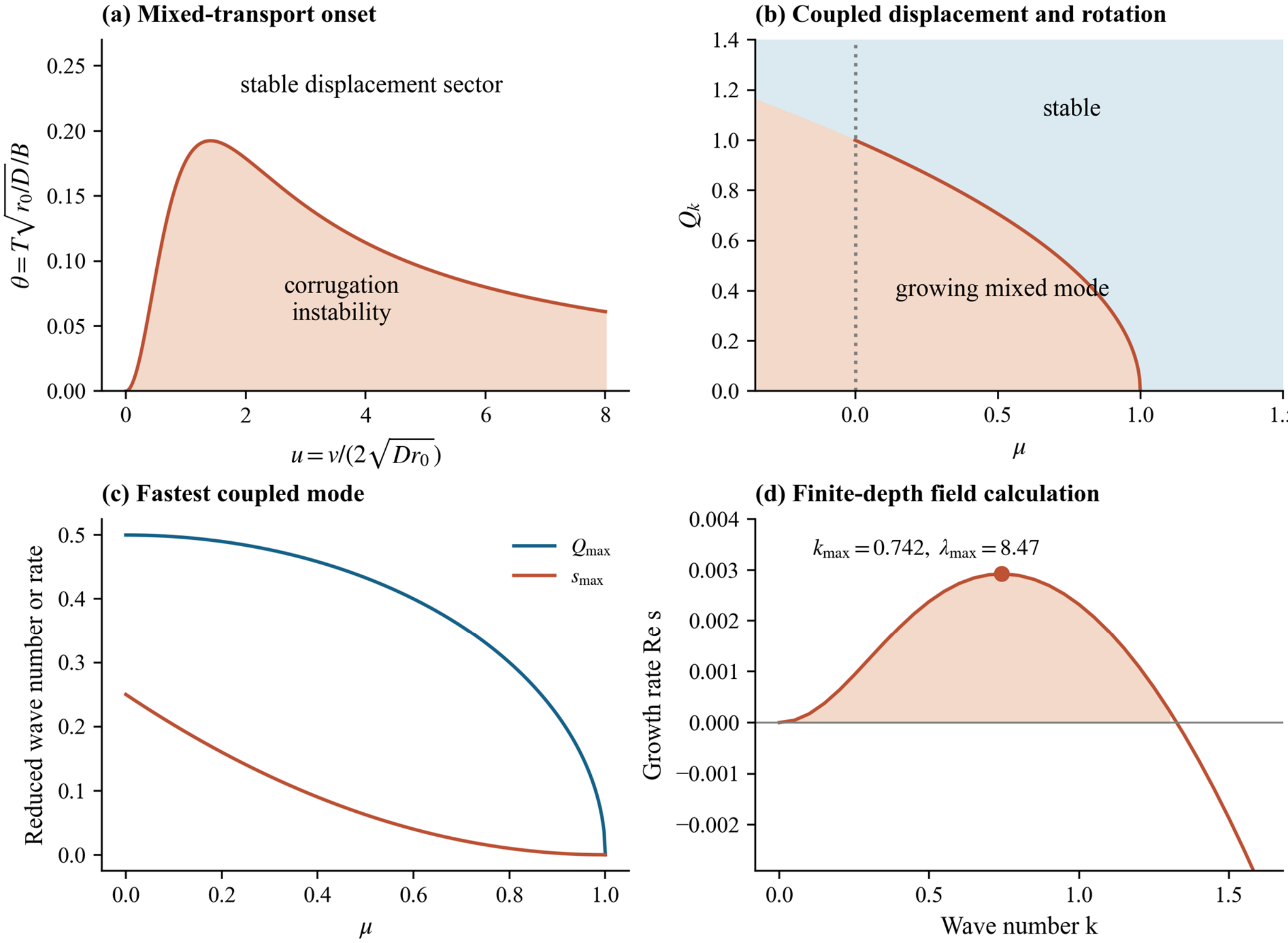

**Figure 7.** Spatial instability of a moving wall. (a) Corrugation threshold in the local mixed-transport displacement model. (b,c) Growing-mode region and fastest-mode scales in the coupled displacement–rotation reduction with equal diagonal gradient coefficients and unit coupling. (d) Finite-depth scalar spectrum for the top-electrode exchange-plus-transport branch. The initial fastest wavelength is approximately 8.47 in the simulation units. The analytical panels and field calculation retain their separate constitutive assumptions; complete numerical spectra are given in Fig. S3.

The finite-depth exchange-plus-transport branch provides a field-level example of corrugation on a positive-slope velocity curve (Fig. 7d). At $e = 0.06$, lowering only the gradient coefficient along the front leaves the straight base state and its velocity unchanged, but creates a finite band of growing real modes. The fastest mode has $k \simeq 0.742$, corresponding to wavelength 8.47 and an e-folding time of about 342 in the simulation units. The short-wavelength gradient penalty closes the unstable band. Full spectra and refinement checks are given in Fig. S3 and S12.

The eigenvalue calculation includes polarization deformation across the wall and through the film, together with the evolving surface charge. It establishes the initial growth band. The following time-dependent calculations examine how that modulation develops after the linear stage. The nonlinear field calculations extend this instability into the $xy$ plane while retaining the finite film depth and self-consistent charge response (Fig. 8). Both lateral dimensions are twice those of the earlier runs, and both evolution intervals are twice as long. The scalar wall develops a substantial corrugation. At intermediate times its spectrum is dominated by the wavelength close to the linear maximum, $\lambda \simeq 8.47$. During the later evolution, the component at $\lambda \simeq 16.94$ overtakes it, while smaller-scale structure remains. The result shows the beginning of coarsening beyond the linear growth stage; the initial fastest wavelength does not uniquely determine the later front morphology.

Allowing tangential polarization changes both the internal wall structure and the spatial evolution. The vector wall develops broad regions of opposite rotation and loses the near-repetition of the two halves of the larger field. At the final recorded time, the maximum tangential polarization is approximately $0.94P_s$. In the part of the surface profile with $|P_z| < 0.15P_s$, the median polarization magnitude is approximately $0.91P_s$. These regions are therefore predominantly rotated rather than paraelectric. The selected vector potential already admits an intrinsic rotational instability of the Ising wall; the calculation illustrates its interaction with charge-mediated front deformation. It does not attribute the existence of rotation to screening alone.

Both calculations retain a connected reversal front over the recorded interval. The scalar and vector endpoints have root-mean-square front displacements of approximately $4.34\xi$ and $3.87\xi$, respectively. The longer record, larger domain, and weak additional long-wavelength seed are varied together, so this comparison is not a separate finite-size scaling test. It establishes nonlinear deformation and coupled rotation for the selected top-electrode case. Numerical specifications, charge-balance checks, and the limits of the available refinement tests are given in Section S13.

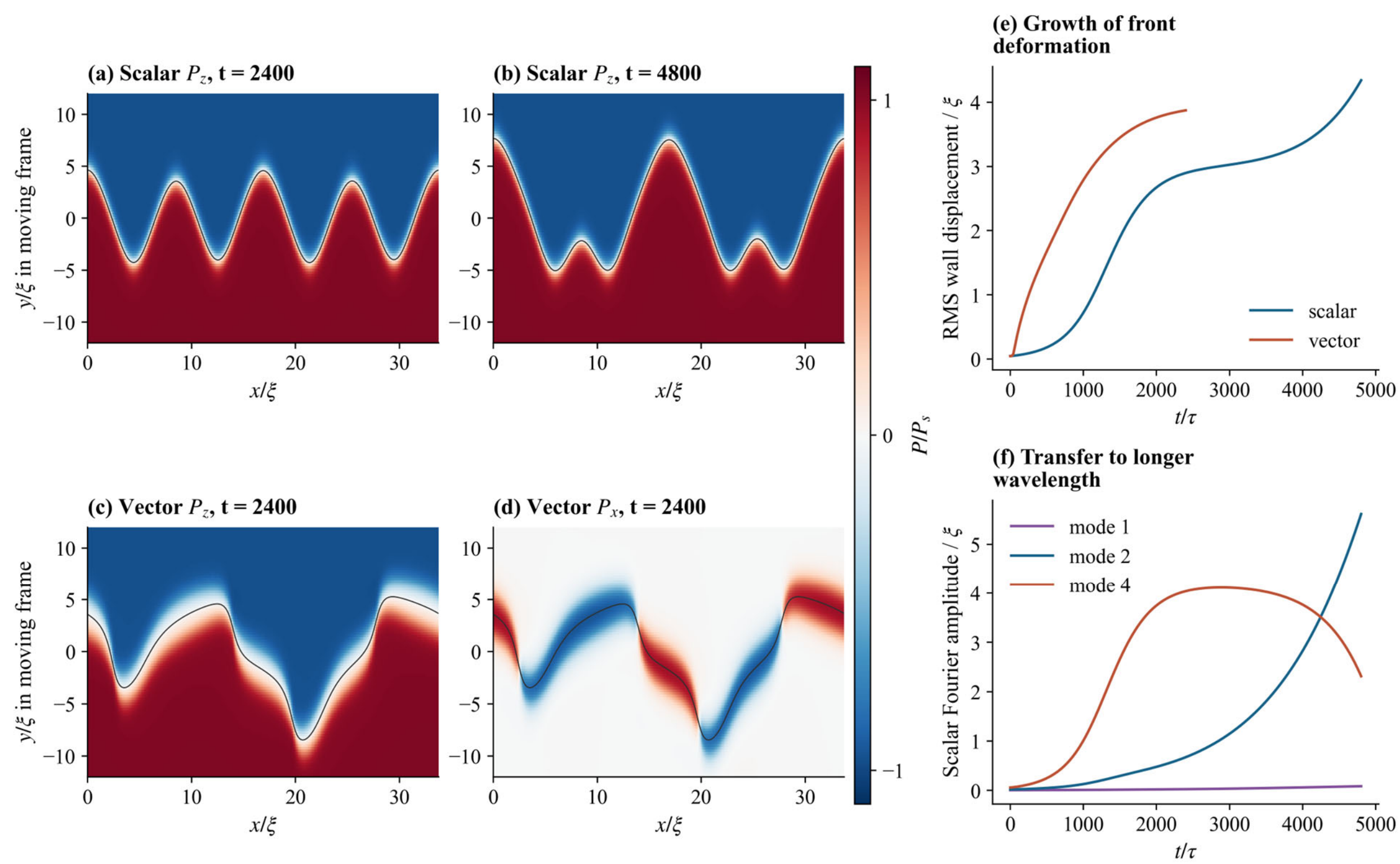

**Figure 8**. Nonlinear evolution in the larger $xy$ field for the top-electrode geometry with exchange and lateral transport. (a,b) Scalar normal polarization at $t/\tau = 2400$ and 4800. (c,d) Normal and tangential polarization in the vector model at $t/\tau = 2400$. Black contours trace the zero crossing of normal polarization. (e) Root-mean-square displacement of the surface front. (f) Scalar front Fourier amplitudes; modes 1, 2, and 4 have wavelengths 33.89, 16.94, and 8.47, respectively. The displayed window follows the wall and shows the central part of the full domain. These are second-order reference models with identical electrical geometry and driving; the vector calculation additionally permits polarization rotation.

### 8.3. Temporal instability and oscillatory behavior

A Hopf bifurcation occurs when a complex-conjugate pair of growth rates crosses the imaginary axis at nonzero frequency while the other nonneutral modes remain stable. Translation is removed by fixing the wall phase. A complex susceptibility or damped oscillation is not sufficient. A stable periodic branch additionally requires the nonlinear amplitude coefficient and must not be inferred from the linear crossing alone.

Several limits cannot undergo an oscillatory instability within the stated models. At passive equilibrium, Eq. (3) gives dissipative relaxation and excludes a Hopf onset. The single-step wall with local exchange and lateral charge transport has no Hopf crossing at any transverse wave number in its displacement dynamics. For a separated pair with local exchange and positive restoring stiffness against changes in separation, the coupled delay equations have no growing mode of the center position or separation. More generally, for the fixed-profile surface response, $\eta > \zeta_{\mathrm{ch}}/8$ is sufficient to exclude both a spatially uniform Hopf crossing and a fold in the force–velocity relation. When this inequality is not satisfied, neither instability follows automatically; the characteristic equation must be solved.

A passive driven finite-mode model demonstrates an allowed oscillatory mechanism without assigning it to every geometry. Let a structural amplitude $A$ couple to a complex screening variable $h$, with $\dot{h} = r(cA - h) + i\omega h$ and $\eta_A \dot{A} = -\eta_A r\mu A - \Delta H c(cA - \mathrm{Re}h)$. Put $U = \omega/r$, $G = \Delta H c^2/(\eta_A r)$ and $p = \mu + G$. The characteristic polynomial is

$$D(z) = (z + p)[(z + 1)^2 + U^2] - G(z + 1) \quad (37a)$$

$$z = s/r. \quad (37b)$$

The Routh–Hurwitz conditions and the Hopf boundary are

$$p + 2 > 0 \quad (38a)$$

$$1 + U^2 + 2p - G > 0 \quad (38b)$$

$$p(1 + U^2) - G > 0 \quad (38c)$$

$$2[(p + 1)^2 + U^2] - G(p + 1) > 0, \quad (38d)$$

$$G_H = \frac{2[(p+1)^2+U^2]}{p+1} \quad (39a)$$

$$(\omega_H/r)^2 = \frac{p-1}{p+1}U^2 - 1. \quad (39b)$$

Only the part of this boundary satisfying the other inequalities is a Hopf onset. If $\mu \geq 0$, the constrained passive linear-source model has no growing mode. For $\mu < 0$, translation may stabilize a mode that is unstable at rest and its subsequent loss of stability can be stationary or oscillatory. At $G = 12, p = 2, U = 3$, $D(z) = (z + 4)(z^2 + 2)$, an explicit nonzero-frequency crossing. A purely charge-coupled Bloch mode has $c = 0$ at $k = 0$; its electrical oscillatory mechanism is then spatially modulated. A first-order width mode has a different nonlinear source and its separate force-controlled construction is given in S9.

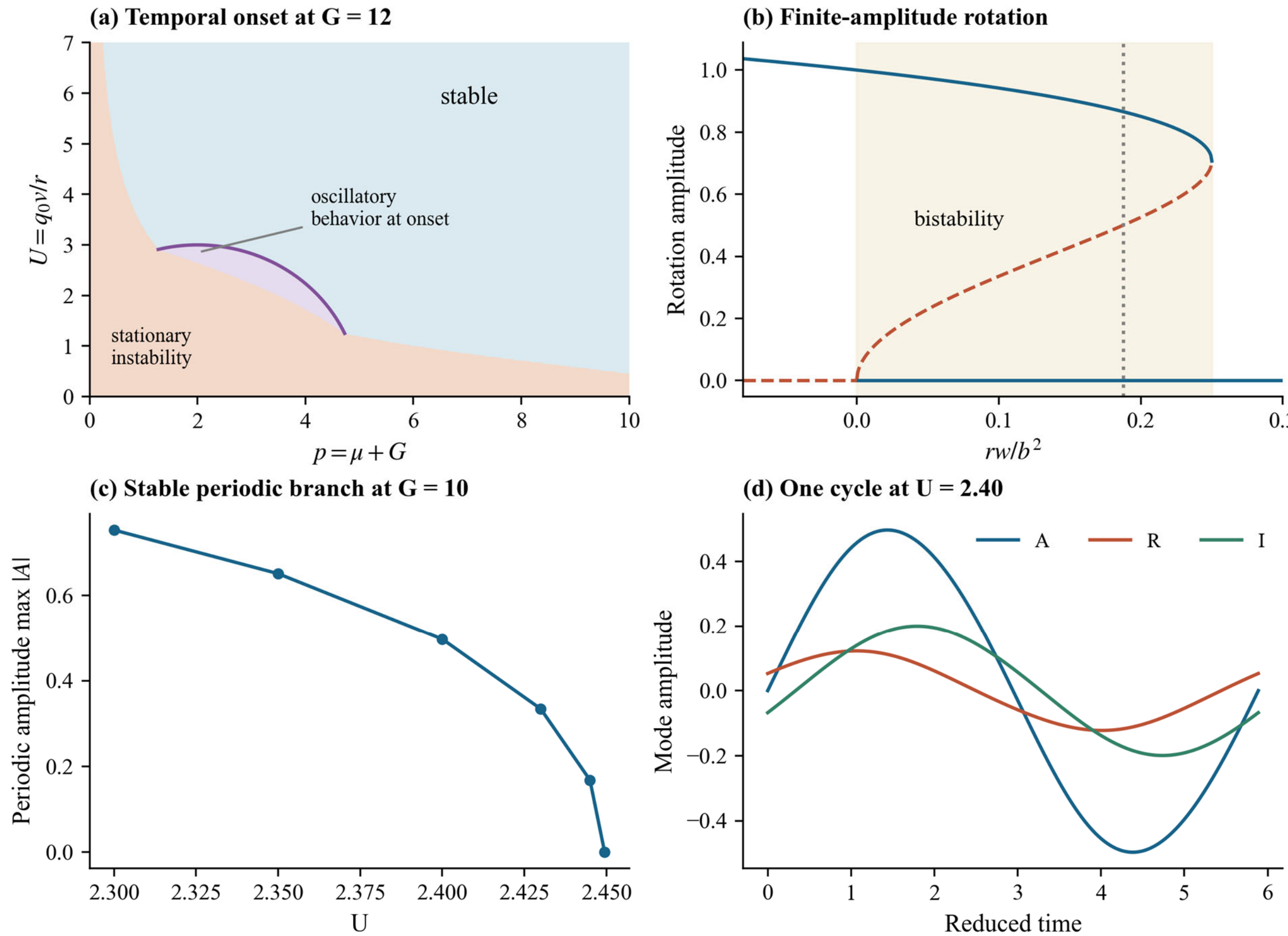

**Figure 9.** Temporal and finite-amplitude structural behavior. (a) Linear stability of the driven structural–screening reduction at G = 12. (b) Equilibrium rotational bistability for a negative quartic amplitude coefficient; solid and dashed branches are locally stable and unstable, respectively, and the dotted line marks equal energies. (c) Stable periodic amplitude at G = 10 and p = 2, ending at the supercritical Hopf point. (d) One computed cycle at U = 2.40. The parameter slices in (a) and (c,d) differ. The periodic branch is a result of the stated reduction, not a material-specific oscillation prediction.

A nonlinear extension of the driven structural–screening reduction establishes that stable periodic behavior can emerge after the linear instability. Adding a positive cubic restoring term gives $A' = -pA + GR - A^3$, $R' = A - R - UI$, and $I' = UR - I$, where time is measured in charge-relaxation units. At $G = 10$ and $p = 2$, the Hopf point is $U_H = \sqrt{6}$. Nonlinear continuation gives a supercritical branch of stable periodic solutions (Fig. 9c). The bifurcation coefficient, crossing rate, and continuation method are given in S12.

At $U = 2.40$, the computed orbit has structural amplitude approximately 0.497 and period 5.89 (Fig. 9d). The translating base supplies energy without a periodically varied external parameter. This example demonstrates stable oscillatory behavior in the specified reduction. It does not assign that behavior to the isolated blocking-MIEC junction or to either material example. The full-field spatial evolution and this temporal reduction answer complementary questions; their different assumptions are retained explicitly.

### 8.4. Characteristic scales and dimensionless parameters

To compare materials and screening conditions, we measure length, time and driving field relative to the properties of a stable polar domain. Its free-energy curvature is $K_b = f_L''(P_s) > 0$.

The polarization length $\xi_P = (g_y/K_b)^{1/2}$ compares gradient energy with this bulk restoring stiffness, and $\tau_P = \Gamma/K_b$ is the corresponding polarization-relaxation time. The reduced field $e = E_d/(K_b P_s)$ measures the driving relative to the bulk polarization scale. These scales differ from the tanh-width and quadratic-coefficient units used for the numerical examples in S12, and conversion must be made before comparing their numerical values. The parameter $\nu = c_6 P_s^4/K_b$ measures the sixth-order contribution at fixed spontaneous polarization and curvature. With polarization measured in units of $P_s$ and energy density in units of $K_b P_s^2$, the local potential becomes

$$\bar{f}(p) = \frac{\nu - 1/2}{2} p^2 + \frac{1/2 - 2\nu}{4} p^4 + \frac{\nu}{6} p^6. \tag{40}$$

The quartic reference has $\nu = 0$; a stable central minimum appears for $\nu > 1/2$ and zero-field coexistence is at $\nu = 3/4$. Negative quartic coefficient begins at $\nu > 1/4$, which alone does not define the tricritical point. This normalization fails when $P_s$ or $K_b$ vanishes.

Surface diagrams additionally require screening strength $G_0 P_s^2/(h K_b P_s^2) = G_0/(h K_b)$, chemical ratio $\chi/G_0$, $r_0 \tau_P$, $D\tau_P/\xi_P^2$, $\alpha h/\xi_P$, $d/\xi_P$ where applicable, and wall-gradient anisotropies. The MIEC requires $\kappa \xi_P$, $D_i/D_e$, carrier susceptibility fractions and the actual outer supply geometry. A transient diagram also uses observation time and preparation. These coordinates cannot all be collapsed into a single relaxation time.

## 9. Regimes of wall motion and structure

The preceding calculations identify several distinct responses to the same imposed driving: a wall can translate with a delayed screening wake, become corrugated, change its internal polarization path, or undergo transient displacement as conserved charge redistributes. We now bring these results together in reduced coordinates and relate the relevant thermodynamic scales to $BaTiO_3$ and $BiFeO_3$ in Section 10. Figure 10 summarizes the physical alternatives. Its panels retain the electrical geometry, charge law, and structural reduction needed for each result; the supplementary materials give the evaluations and the additional parameter slices.

### 9.1. Driving and screening coordinates

The polarization scales in Eq. (40) separate the material free-energy landscape from the electrical environment. Reduced driving $e$ measures the domain free-energy difference, while $\nu$ controls the scalar transition character at matched spontaneous polarization and far-domain curvature. Screening adds its strength, exchange rate, diffusion length, and available charge inventory. An electrode also introduces the gap-to-film capacitance ratio. In an adjoining MIEC, the Debye length, the relative ionic and electronic diffusivities, and the distance over which charge must redistribute remain independent parameters. These distinctions are essential: a metal electrode can screen a field electronically while the interfacial ionic population remains strictly conserved.

For mixed surface transport with $r_0 > 0$ and $D > 0$, the length $\ell_s = \sqrt{D/r_0}$ and speed $v_s = 2\sqrt{D r_0}$ define the displacement variables $u = v/v_s$, $f = F/B$, and $\theta = T\sqrt{r_0/D}/B$. The local unresolved-front law becomes $f = \epsilon u + u/\sqrt{1 + u^2}$, where $\epsilon = \eta v_s/B$ measures intrinsic friction relative to screening drag. Figure 10a fixes $\epsilon = 0.1$ and maps the corrugation criterion of Eq. (35a) from velocity to driving. Increasing driving first destabilizes and can subsequently restabilize the displacement sector at fixed $\theta$. The intermediate instability occurs while the force–velocity curve remains monotone. Changes in wall stiffness, screening strength, or the diffusion-to-exchange ratio move the system vertically in this slice.

### 9.2. Steady motion and conserved transients

With local exchange, the screening charge on each side of a moving wall can relax toward the equilibrium value set by the reservoir and the local polarization. Lateral transport spreads this charge over an additional distance and changes the stability of the wall shape. If a first-order wall separates into two interfaces that interact through their screening wakes, the required force can also decrease over part of the velocity range (Fig. 4d). An exposed surface and an interface beneath a top electrode generate different electric fields for the same charge mismatch, so each calculation must use the electrostatic response of its own geometry. They approach the same reduced diagram only where both responses can be approximated by a local capacitance.

Strict conservation gives a different class of motion behaviors. For an isolated front connecting unequal compensated populations with zero far-field particle currents, no steady screening profile can accompany nonzero constant velocity. Figure 10b therefore plots displacement against driving and elapsed time using the charge-dominated similarity solution, rather than assigning a steady velocity. The square-root spreading regime can give way to finite-inventory arrest in a closed specimen, as demonstrated in Fig. 3. Observation time, preparation, and specimen size are then part of the physical specification. The equal-background traveling-pulse family beneath an electrode is a separate preparation whose excess charge must be balanced elsewhere; it does not remove this conservation constraint.

In a blocking MIEC, compensating charge moves through the adjoining volume. As an isolated straight wall–surface junction advances, the charge needed to screen the growing domain must be redistributed over increasing distances. Figure 10f shows that the required potential drop increases logarithmically with this distance: $\Delta\phi/\Phi_* = j_* \ln(R/a_{\text{in}})$, where $j_* = \Delta P v/(\pi\sigma_m \Phi_*)$. Here $R$ is the outer transport distance, $a_{\text{in}}$ is the inner junction scale, and $\Phi_*$ is the reference potential. A contact at finite distance limits the potential drop, but its position and electrical boundary condition then enter the velocity and stability calculation. For the isolated half-space, a bounded screening distribution of fixed extent cannot accompany indefinite steady translation.

### 9.3. Wall structure and competing instabilities

A low value of the normal polarization inside a wall has several possible meanings. The quartic scalar reference crosses zero within an ordinary Ising wall, but has no equilibrium paraelectric minimum. Frozen screening can temporarily suppress polarization; after charge relaxation, the central region is unstable if the equilibrated curvature remains negative. A persistent paraelectric interpretation therefore requires more than a broad region of small normal polarization. The vector simulation in Fig. 8 gives a direct example: the normal component passes through zero while a large tangential component maintains the polarization magnitude.

For the normalized sixth-order potential, a central scalar minimum first appears at $\nu > 1/2$, and zero-field coexistence occurs at $\nu = 3/4$. Figure 10c evaluates the bias range in which that minimum survives. It is a necessary local thermodynamic condition for a weakly polarized core. Formation of an extended layer additionally depends on the energy of the two bounding interfaces, their interaction, mechanical compatibility, and stability against transverse polarization. The first-order separation coordinate is particularly important near coexistence because its restoring force becomes small while the translational friction remains finite.

Delayed screening can change that separation nonlinearly. Figure 10e evaluates the fixed-speed two-front reduction for which the rear interface experiences more screening drag than the front interface. Over a finite interval of central-phase cost and screening pressure, the separation equation has two stable widths separated by an unstable width. At the marked point these widths are approximately 6.71 and 32.18, with an unstable intermediate separation of 11.95, in units of the elementary-interface width. This is bistability of the stated separation equation; the center-force equation and coupled charge dynamics are still needed to establish bistability of complete traveling walls at the same applied driving.

Multiaxial polarization adds an independent structural route. Figure 10d separates uniform and finite-wave-number rotational instabilities of the projected amplitude operator $r_A + c_A k^2 + d_A k^4$, with $d_A > 0$. For $c_A \geq 0$, the uniform mode becomes unstable at $r_A = 0$. For $c_A < 0$, a modulated mode becomes unstable first at $r_A = c_A^2/(4d_A) > 0$. These coefficients must come from the same electrostatic, mechanical, and polarization model as the base wall. The coupled displacement–rotation example in Fig. 7 shows why positive uniform rotational stiffness alone does not establish stability. A negative quartic amplitude coefficient also permits finite-amplitude bistability before linear instability, as illustrated in Fig. 9b.

Temporal and spatiotemporal behavior require the frequency-dependent coupled spectrum. The local unresolved displacement model excludes a Hopf crossing, whereas the driven structural–screening reduction in Fig. 9 admits a stable periodic branch. A crossing at nonzero wave number would instead describe oscillatory behavior with spatial modulation. Neither a force fold nor a structural soft mode alone fixes which of these outcomes occurs. The maps thus identify the relevant mechanisms and the conditions separating them, while the nonlinear field calculations show how corrugation and rotation can develop together in a specified geometry.

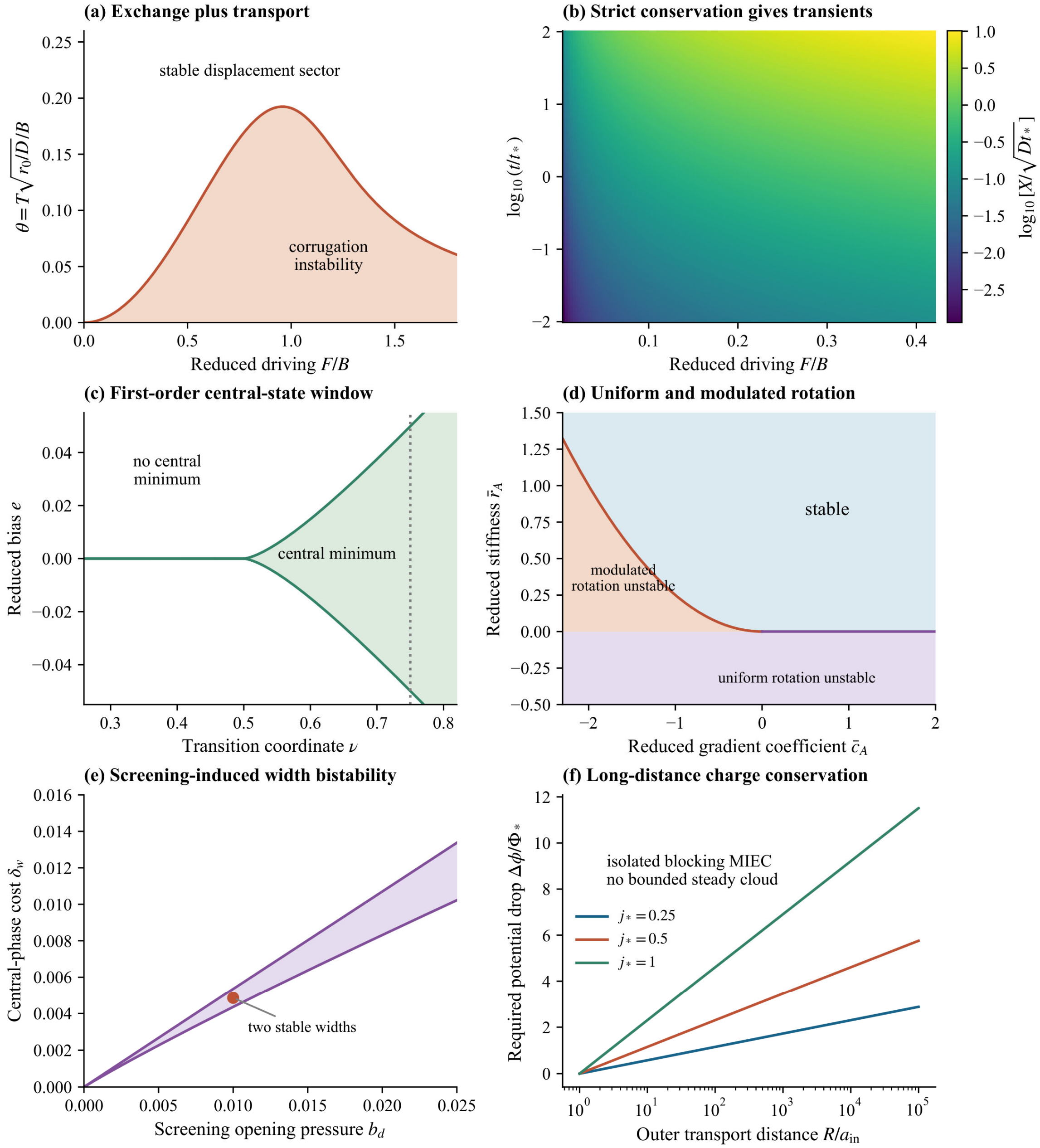

**Figure 10.** Reduced-variable overview of wall motion and structure. (a) Corrugation boundary of the local mixed-transport front, mapped to driving at $\epsilon = 0.1$. Stability refers to the displacement sector. (b) Conserved-front similarity displacement before finite-size arrest; $t_*$ is an arbitrary reference time. (c) Bias range supporting the central minimum of Eq. (40); the dotted line marks zero-field coexistence. (d) Rotational onset for the projected fourth-gradient operator, with $\bar{r}_A = r_A/(d_A k_*^4)$ and $\bar{c}_A = c_A/(d_A k_*^2)$ for a fixed reference wave number $k_*$. (e) Two stable widths in the fixed-speed separation model at wake lengths 50 and 10 elementary-interface widths. (f) Long-distance potential drop required by charge conservation at a blocking MIEC junction. These are evaluated slices of the stated reductions, not a material-independent nonlinear phase diagram. Section S14 gives the definitions and calculations.

## 10. Material examples: $BaTiO_3$ and $BiFeO_3$

Here we analyze these behaviors for $BaTiO_3$ and $BiFeO_3$. Their actual polarization and structural potentials are treated separately from the quartic comparator. To compare transition-order effects alone, match $P_s$, $K_b$, $g_y$, $\Gamma$, electrical geometry and the entire charge law. At that matching the first-order coexistence limit has $I_6/I_4 \to 3/8$, while comparisons at the same temperature with unrelated coefficients do not have a universal mobility ordering. We use the $BaTiO_3$ parameterization of Marton, Rychetsky and Hlinka [11] consistently for the principal material example. Alternative fitted potentials [78] are parameter-systematic comparisons and are not spliced into the same wall calculation. For $BiFeO_3$ we use the polar coefficients and structural convention of Karpinsky and colleagues [79]. Screening parameters are independent inputs. Bulk Landau coefficients do not determine adsorption rates, surface diffusivity, MIEC carrier density or the source impedance.

### 10.1. $BaTiO_3$ and mechanical compatibility

For a tetragonal $c$ wall with polarization [001] and normal [010], use $f = \alpha_I p^2 + \alpha_{II} p^4 + \alpha_{III} p^6$ and $g_y = 2 \times 10^{-11}$ J m³ C⁻². The temperature-dependent coefficients and elastic constants are listed in S10. The relaxed polar branch and central-state cost are

$$S = \frac{-\alpha_{II} + \sqrt{\alpha_{II}^2 - 3\alpha_{III}\alpha_I}}{3\alpha_{III}} \tag{41a}$$

$$K_b = 8S(\alpha_{II} + 3\alpha_{III} S) \tag{41b}$$

$$\Delta f = \alpha_{II} S^2 + 2\alpha_{III} S^3. \tag{41c}$$

The coherent planar correction has the form $A_{\mathrm{el}}(p^2 - S)^2$. The same far polarization is retained, but $a_{\mathrm{coh}} = 2\alpha_I - 4A_{\mathrm{el}}S$ and $b_{\mathrm{coh}} = 4\alpha_{II} + 4A_{\mathrm{el}}$. For the listed elastic and electrostrictive coefficients, $A_{\mathrm{el}} = 8.1662 \times 10^8$ J m⁵ C⁻⁴. A stable central scalar state requires $2\alpha_I - 4A_{\mathrm{el}}S > 0$. This condition is not met by the coherent reference over 300–390 K, even though the relaxed bulk model has a first-order transition. The material therefore illustrates a restriction on scalar core formation, not an inevitable paraelectric slab.

The fully relaxed and coherent results are limiting mechanical assumptions. A junction near a free surface can relax in depth and does not generally coincide with either local limit. The multiaxial Hessian must be derived under the same constraint as the longitudinal profile. Combining a coherent scalar wall with a fully relaxed transverse stiffness would compare different models.

### 10.2. $BiFeO_3$ and coupled structural order

For the [111] polar path in a fixed structural environment, write $f_R = (a_I^P + h_R)P^2 + b_R P^4 + c_R P^6$, where

$$b_R = (a_{II}^P + a_{12}^P)/3 \tag{42a}$$

$$c_R = (3a_{III}^P + 6a_{112}^P + a_{123}^P)/27. \tag{42b}$$

The structural shift $h_R$ contains the specified octahedral, magnetic and mechanical contributions. At fixed $h_R$, the central-state, coexistence and polar-spinodal landmarks are

$$h_0 = -a_I^P(T) \tag{43a}$$

$$h_t = \frac{b_R^2}{4c_R} - a_I^P(T) \tag{43b}$$

$$h_{\mathrm{sp}} = \frac{b_R^2}{3c_R} - a_I^P(T). \tag{43c}$$

These produce a coefficient-based conditional diagram without inventing an octahedral amplitude. The nonpolar high-temperature structure has additional structural order; setting $P = 0$ along a fixed polar path does not by itself create that phase. For an orthorhombic core with $\mathbf{\Phi} = (\varphi, \varphi, 0)$ and $\mathbf{\Psi} = (0, 0, \psi)$ in the source convention, all three polar Hessian eigenvalues and the structural Hessian must be positive. The conditions are given in S10.

When structural fields relax, the effective polar curvature is a Schur complement, $H_{PP} - H_{PS}H_{SS}^{-1}H_{SP}$. Near structural softening those fields cannot be eliminated. $BiFeO_3$ therefore requires a joint polarization–structure wall for a quantitative core-splitting or rotation prediction. The scalar curves in Fig. 11 identify necessary constrained-state conditions, not a completed dynamical phase diagram.

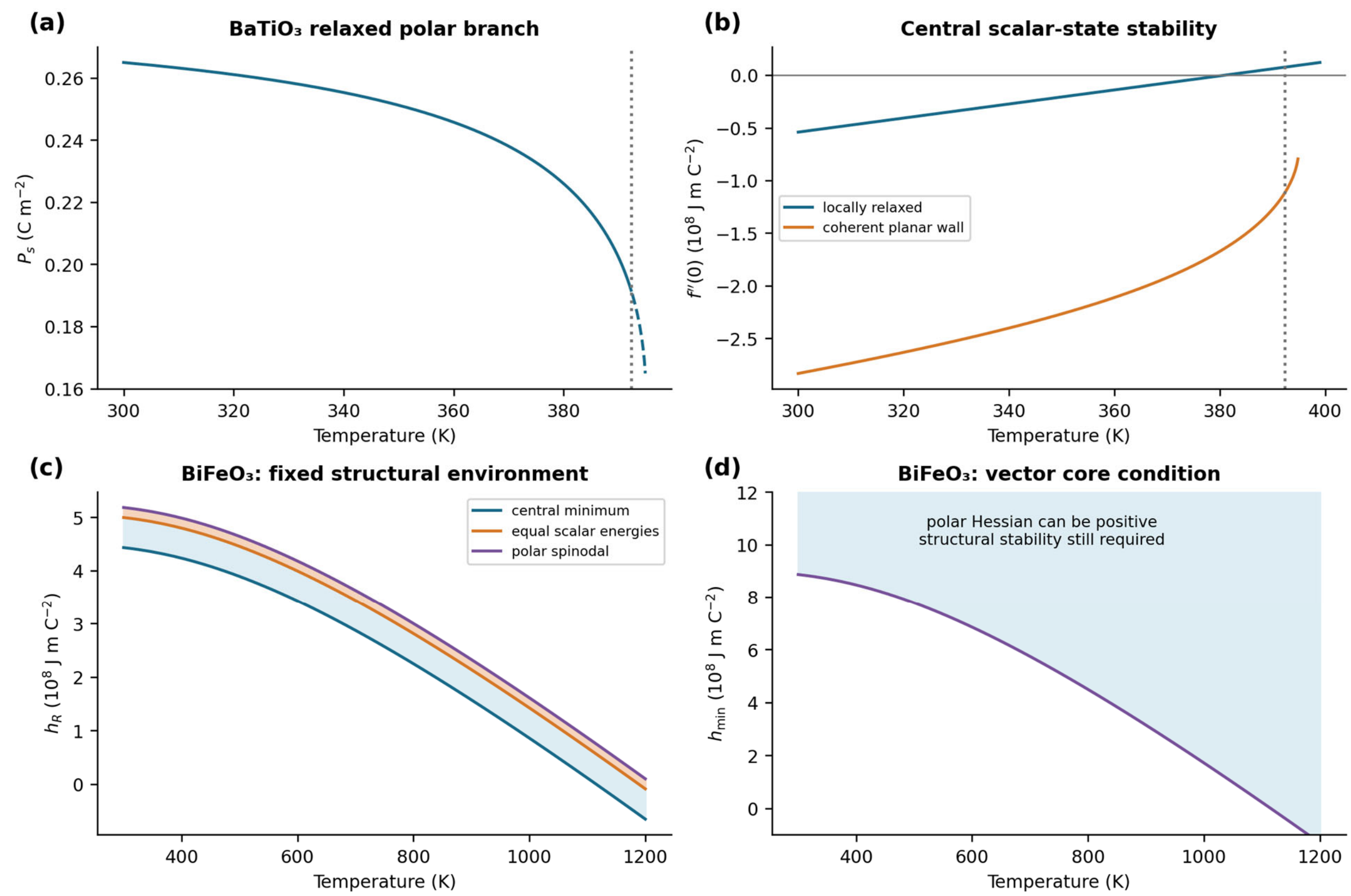

**Figure 11.** Material examples with assumptions retained. (a,b) $BaTiO_3$ polar amplitude and central scalar curvature for the specified relaxed and coherent paths. Dashed extensions, where present, are metastable polar branches. (c) $BiFeO_3$ constrained [111] polar landmarks as a function of the structural quadratic shift, Eqs. (43a–c). (d) The minimum additional polar-Hessian shift required to stabilize all polar directions at $P = 0$, $h_{\min} > -2a_1^P(T)$. Structural stability is an additional requirement. These are evaluated thermodynamic slices; no wall-stability simulation is represented.

The material coefficients determine polarization branches, local spinodals, reference wall profiles, and energy scales. Absolute wall velocities and relaxation times additionally require independently specified polarization kinetics, surface diffusion, exchange rates, and electrical boundary conditions. The conditional $BaTiO_3$ and $BiFeO_3$ calculations in Fig. S6 illustrate how these inputs convert the reduced framework into dimensional trajectories and tested stability points. All sampled scalar and separately tested uniform rotational modes decay in those parameter slices. They do not define an uncomputed material instability boundary. The parameter provenance, dimensional values, and sensitivity to the assumed kinetics are retained in S10, S12, and S15.

## 11. Frequency-dependent response of domain-wall displacement

The coupled wall and screening-charge dynamics determine several responses to a weak alternating perturbation. An applied electric field changes the polarization within the domains and can displace an existing wall; a chemical perturbation changes the screening population; and an electrode voltage produces a current that includes geometric charging and the coupled polarization and charge response. Local measurements can instead detect wall position or surface potential. For a wall with internal structure, the separation of two interphase boundaries and the transverse polarization provide additional response coordinates. Their spectra depend on the excitation and detection mechanisms.

Here we retain the collective displacement of the domain wall and its associated force-to-velocity impedance. This coordinate describes translation of the polarization profile. In the adjoining-conductor geometry, we use a localized wall–surface-junction displacement with finite projected friction. The bulk polarization, chemical, surface-potential and terminal electrical responses, together with the structural modes and complete derivations, are given in Section S16. Figure 12 and Table 2 summarize the displacement response under the different screening conditions.

### 11.1. Definition of the wall response

For a stable reference wall, a small force produces a harmonic displacement with complex amplitude $\delta X$. With the convention $e^{i\omega t}$, define

$$\mathcal{X}_{XF}(\omega) = \frac{\delta X(\omega)}{\delta F(\omega)}. \tag{44a}$$

$$Z_X(\omega) = \frac{\delta F(\omega)}{\delta \dot{X}(\omega)} = \frac{1}{i\omega \mathcal{X}_{XF}(\omega)}. \tag{44b}$$

The displacement amplitude is $|\mathcal{X}_{XF}|\,|\delta F|$, and the phase of $\mathcal{X}_{XF}$ gives its lag relative to the force. Nyquist plots show $-\mathrm{Im}Z_X$ against $\mathrm{Re}Z_X$. The quantity $Z_X$ is a wall impedance, with units fixed by the force and displacement normalization. A measured electrode impedance additionally requires the voltage-to-force coupling and the electrode-charge response. Under a prescribed uniform thermodynamic field, a rigid wall contributes to the average polarization through the change in domain fractions; an applied electrode voltage can also act directly on screening charge. Dielectric and terminal-impedance analyses of ferroelectrics explicitly include these excitation and readout factors [12, 13, 14]. Their electrical equivalent-circuit elements are distinct from the force-to-velocity quantity defined here.

Eliminating the surface charge perturbation gives the displacement response in the form

$$\mathcal{X}_{XF}^{-1}(s,k) = K_{\mathrm{eq}}(k) + s[\eta + \mathcal{M}(s,k)]. \tag{45}$$

Here $s = i\omega + 0^+$, $k$ is the wave number along the wall, $K_{\mathrm{eq}}$ is the equilibrium restoring stiffness, $\eta$ is the intrinsic wall friction, and $\mathcal{M}$ describes the delayed screening contribution. For an unconfined, uniformly displaced wall with equilibrating charge, the static translational stiffness vanishes. Conservation and finite boundaries must be retained when evaluating $\mathcal{M}$; a conserved zero mode can itself generate a restoring term. The exposed and electrode-covered surfaces enter through their respective electrostatic kernels.

### 11.2. Charge exchange and conserved redistribution

With local exchange, screening relaxes as the wall moves. The low-frequency motion has finite mobility, so the displacement amplitude varies as the inverse frequency and its phase approaches $-90°$. In the local-capacitance limit, a single screening rate produces a semicircular wall-impedance arc between the intrinsic friction and the total low-frequency friction. The full electrostatic kernel broadens the distribution of rates. Adding lateral transport preserves the low-frequency limit when exchange remains possible, but changes the intermediate dispersion (Fig. 12a–c).

Strict conservation changes the low-frequency response of an isolated wall. On an infinite surface, charge redistribution extends over a distance proportional to the square root of the oscillation period. In the long-wavelength, diffusion-controlled regime, the displacement amplitude therefore varies as $\omega^{-1/2}$ and lags the force by $45°$. The corresponding impedance has a diffusion branch with equal real and negative imaginary parts. This harmonic response about a stationary wall is consistent with the exclusion of indefinite steady translation under the same conservation law.

In a closed specimen, the finite charge inventory limits the low-frequency displacement. For a sharp wall at the center of an interval of length $L$ with reflecting charge boundaries,

$$Z_X^{\mathrm{closed}}(s) = \eta + \frac{B}{2\sqrt{Ds}} \coth\left(\frac{L}{2}\sqrt{\frac{s}{D}}\right). \quad (46)$$

The positive coefficient $B$ measures electrostatic coupling to displacement, and $D$ is the effective surface charge diffusivity in this reduction. The static displacement susceptibility is $L/B$, with phase approaching zero. A diffusion-like intermediate regime appears when the redistribution distance is smaller than the specimen but larger than the wall width (Fig. 12d). Frozen screening prepared around the initial wall also produces a finite static compliance, because displacement separates the wall from its compensation pattern. The crossover and size dependence distinguish this response from that of a mobile conserved population. At sufficiently high frequency, the finite-width overdamped wall approaches the intrinsic viscous response, with inverse-frequency amplitude and a 90° lag; a sharp-wall diffusion asymptote must not be extrapolated beyond its spatial resolution.

### 11.3. Screening in an adjoining conductor

In a blocking mixed ionic–electronic conductor, charge redistributes through the adjoining volume. Figure 12e shows wall-junction impedance arcs for equal carrier diffusivities and several outer transport distances. Their size increases with the distance over which charge must redistribute. Figure 12f compares equal and unequal ionic and electronic diffusivities at fixed outer distance. Slower ionic transport increases the low-frequency resistance and introduces additional dispersion. Normalizing the latter curves by their zero-frequency screening resistance separates the change in spectral shape from the change in magnitude; the unnormalized spectra are given in Section S16.

These examples retain zero normal carrier flux at the ferroelectric interface and use an explicit long-distance cutoff. They describe sensitivity to the remote geometry rather than a solution for a particular electrode shape. For an isolated straight junction in an infinite half-space, uniform displacement changes the integrated bound charge and the projected impedance retains a logarithmic long-distance divergence, including at nonzero frequency. A finite return geometry, a charge-balanced pattern or a nonzero wave number along the wall is required for a finite physical response. Likewise, finite friction requires a specified wall depth profile; rigid translation of an infinitely deep wall is not a finite-friction coordinate.

Where a stable traveling state exists, its low-frequency differential velocity response is the inverse slope of the force–velocity curve. Approaching a spatial instability enhances the displacement response at the corresponding wave number. Structural softening and oscillatory behavior involve additional coordinates and are treated in Section S16. The spectra therefore connect charge conservation to the mobility and stability of the wall.

**Table 2.** Small-signal wall displacement under different screening conditions. The low-frequency limits refer to a stable reference and the stated geometry. Phase denotes displacement relative to force. The unconfined cases exclude additional pinning. Within the finite-width overdamped model, the high-frequency displacement approaches $1/(i\omega\eta)$ when the projected friction and geometry are finite.

| Screening and geometry | Low-frequency displacement | Wall-impedance signature |
|---|---|---|
| I. Local exchange | Amplitude proportional to $\omega^{-1}$; phase $-90°$. | Finite friction; one semicircle only in the single-rate local-capacitance limit. |
| II. Exchange and lateral transport | Amplitude proportional to $\omega^{-1}$; phase $-90°$. | Finite friction at low frequency; diffusion can broaden the intermediate dispersion. |
| III. Conserved charge on an infinite surface | Amplitude proportional to $\omega^{-1/2}$; phase $-45°$ in the diffusion limit. | A $45°$ diffusion branch; no finite zero-frequency friction. |
| III. Conserved charge in a closed interval | Static susceptibility $L/B$; phase tends to zero. | Restoring term $B/(i\omega L)$; crossover controlled by $D/L^2$. |
| Frozen equilibrium screening pattern | Static susceptibility $1/K_{\mathrm{fr}}$; phase tends to zero. | Vertical branch for $Z_X = \eta + K_{\mathrm{fr}}/(i\omega)$. |
| Blocking MIEC with equal diffusivities and finite outer cutoff | Unconfined displacement proportional to $\omega^{-1}$; phase tends to $-90°$. | Relaxation arcs with resistance dependent on the outer distance. |
| Blocking MIEC with unequal diffusivities at fixed outer cutoff | Same unconfined exponent; the mobility depends on both carrier diffusivities. | Additional dispersion; the normalized arc can differ from a single semicircle. |

For either MIEC row, removing the outer cutoff does not give a finite uniform isolated-junction impedance. A fixed finite geometry is essential before taking the zero-frequency limit. Surface results apply to both exposed and electrode-covered geometries with their own electrostatic kernels. The finite-interval formula assumes fixed far-domain polarization; domain compressibility changes its coefficients.

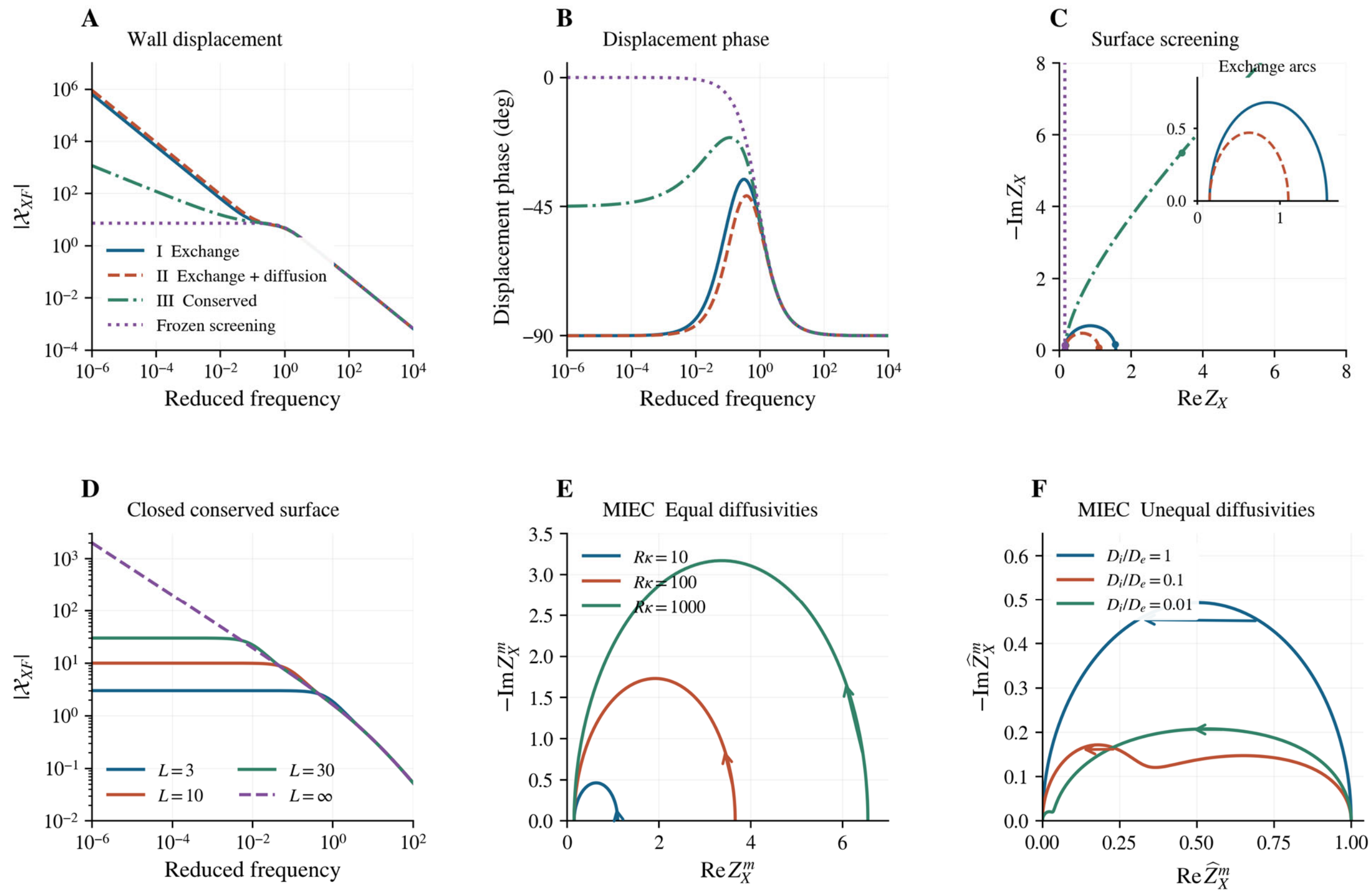

**Figure 12.** Wall displacement and impedance. (a–c) Amplitude, phase and Nyquist curves beneath a dielectric and top electrode. Colors and line styles follow (a); the inset enlarges the exchange arcs. Points mark reduced frequencies $10^{-2}$, 1 and $10^{2}$; unbounded branches extend beyond the displayed window. (d) Finite-size crossover on a closed conserved surface. (e) Blocking-MIEC junction spectra with equal diffusivities and increasing outer distance. (f) Unequal diffusivities at $R\kappa = 100$, normalized as $\hat{Z}_X^m = [Z_X^m - \eta]/[Z_X^m(0) - \eta]$. Arrows indicate increasing frequency. Panels (e,f) use $q_{\min} = 1/R$ and finite junction friction. All quantities use reduced units; Section S16.13 gives parameters and the unnormalized MIEC spectra.

## 12. Discussion and conclusions

The analysis connects wall translation, shape instability and changes in internal polarization structure through the dynamics of screening charge. Charge conservation determines whether a steadily translating compensated wall is possible, while the electrical geometry and polarization free energy determine the forces and structural modes that remain accessible. This distinction provides the basis for comparing the universal cases with material-specific behavior and experimental observations.

### 12.1. Physical consequences of coupled wall and charge dynamics

The electrical boundary condition and the charge dynamics determine which wall motions are possible. Local exchange can replenish compensation as a domain grows, whereas a conserved population must redistribute the charge already present. For an isolated wall connecting differently compensated domains, this distinction separates steady screened translation from transient motion and finite-inventory arrest. A top electrode changes the electrostatic cost of charge mismatch but does not relax ionic conservation at a sealed interface. An adjoining blocking conductor replaces surface redistribution by volume transport and introduces a dependence on the remote electrical geometry. These are different physical cases, even when a local velocity law can be written in a similar form.

Screening affects more than the wall velocity. A spatially delayed charge pattern changes the field at the junction and can make a straight front unstable while its force–velocity relation remains monotone. The finite-depth calculations establish this possibility, and the larger-field evolution shows the subsequent growth of corrugation and transfer toward longer spatial scales. The wavelength selected during initial growth therefore need not be the wavelength observed after a long evolution. Earlier screening-retardation and shape-instability theories provide the starting points for this behavior [3, 8, 46]; the present analysis connects them to conserved transients, deformable walls, and internal polarization modes within consistently specified electrical geometries.

The polarization free energy determines how the wall can accommodate this changing electrostatic environment. A second-order scalar wall can broaden without creating a stable paraelectric phase. A first-order potential can support two interphase fronts with a separation that evolves independently of their common position. A multiaxial material can instead reverse through polarization rotation, and a favorable intermediate variant can develop into a finite domain. These alternatives connect classical first-order wall structure and wetting [66, 67, 69] with established multiaxial wall transformations [71, 72, 74]. The vector simulation makes the distinction directly visible: a region of small normal polarization can remain strongly polar because the polarization has rotated into the wall plane.

The reduced diagrams organize these responses by driving, screening, wall stiffness, and the relative energies of the competing polarization states. They show where the displacement, rotational, or separation mode must be retained and where a steady-state description gives way to a transient one. The $BaTiO_3$ and $BiFeO_3$ examples then show what is needed to connect this framework to a material: mechanical compatibility and additional structural order can suppress or redirect a transformation allowed by a scalar bulk potential. Screening kinetics remain separate material and interface inputs. This separation makes it possible to compare experiments through the physical mechanism being tested, rather than through a fitted wall mobility alone.

### 12.2. Conclusions and experimental signatures

We have developed a common description of the principal electrical and charge-conservation cases for an isolated ferroelectric wall and used it to examine both motion and internal structure. The analytical results identify conservation constraints, screening forces, exact scalar profiles, interphase separation, and selected spatial and temporal stability boundaries. Numerical calculations add deformable traveling walls, conserved displacement and arrest, nonlinear corrugation over larger lateral fields, coupled wall rotation, and stable periodic behavior in an explicit driven reduction. Together they establish a set of physical mechanisms whose realization depends on the electrical geometry, the charge dynamics, and the polarization energy. The supplementary materials retain the derivations, numerical specifications, extended maps, and the limits of the material calculations.

The experimental challenge is to identify the signatures of these mechanisms in published and future measurements. Wall displacement correlated with a delayed potential wake tests the screening contribution to motion. A ripple whose wavelength grows near onset tests a spatial instability; a broad potential feature alone does not establish a broad polarization wall. Two polarization-gradient peaks identify an interphase pair, while vector imaging distinguishes localized rotation from a finite intermediate domain. A repeatable temporal signal must be compared with the electrical circuit, stimulus and charge relaxation before assigning an internal oscillatory instability. These comparisons connect the theoretical cases to observables while preserving the charge balance and boundary conditions of each experiment.

**Author contributions and use of AI**

Sergei V. Kalinin conceived and directed the scientific problem, selected and organized the physical cases, established the electrical boundary conditions and charge-conservation requirements, and defined the questions of wall motion, stability, and structural change. He guided the literature analysis and theoretical formulation, identified and corrected inconsistencies in the calculations and their physical interpretation, directed the numerical tests, and revised the manuscript. His contribution includes the central physical formulation, interpretation, and critical scientific judgment underlying this work. ChatGPT Astra 6 assisted under Kalinin's direction with literature retrieval, algebraic derivations and consistency checks, development and execution of numerical scripts, figure preparation, drafting, editing, and document assembly. Fable 5.1 provided critical comments on preceding analyses; the present large-field calculations and manuscript integration are not represented as newly reviewed by Fable. The analytical assumptions, numerical inputs, and verification checks are documented in the supplementary materials. AI systems were used as research and writing tools. The human author retains responsibility for the final scientific assessment and approval of the manuscript.

**Acknowledgments**

This work (S.V.K.) was supported by the DOE BES project DE-SC0026253, "Deciphering Electrochemical Transformation Pathways on the Nanometer Scale: Advancing Fundamental Discovery for Material Innovation."

## SUPPLEMENTARY MATERIALS

These materials derive the analytical results and document the numerical calculations used in the main paper. The notation and reference numbering are shared with the main text. Sections S1–S11 retain the analytical foundations and comparison with prior work; S12 documents the earlier numerical calculations, S13 the larger-field scalar and vector evolution, S14 the reduced-variable summary, S15 the additional figures and dimensional scenarios, and S16 the small-signal response functions and impedance spectra.

### S1. Electrostatic kernels, source work and charge response

#### S1.1. Exposed film and finite electrode stack

For a depth-independent scalar polarization, the induced potential in the ferroelectric has no volume source away from a bent wall. Its Fourier solution satisfying the grounded bottom condition is

$$\phi_{f,Q}(z) = \phi_{s,Q} \frac{\sinh[\alpha Q(z+h)]}{\sinh(\alpha Q h)}. \tag{S1}$$

For the exposed exterior, $\phi_{e,Q} = \phi_{s,Q} e^{-Qz}$. For the finite dielectric at nonzero $Q$, $\phi_{d,Q} = \phi_{s,Q} \sinh[Q(d-z)]/\sinh(Qd)$. Evaluating the displacement jump $D_{\mathrm{out},z} - D_{f,z} = -c$ gives $Y_Q \phi_{s,Q} = p_Q - c_Q$, with $Y_Q = G_Q^{-1}$ as in Eqs. (4) and (14). The $Q = 0$ limits follow from $Q\coth(Ql) \to 1/l$. In the top-electrode geometry the background linear solution adds $-b_V V$ to the surface potential.

The local-capacitance approximation requires $\alpha Q h \ll 1$ and, for the top stack, $Qd \ll 1$. For the exposed film the exterior term additionally requires $\epsilon_e Q h/\epsilon_z \ll 1$. Across the entire wall spectrum this is a restrictive condition involving the microscopic wall width. It can still be valid for long-wave outer transport even when it is invalid inside the wall. The top-stack expansion is

$$G_T(Q) = G_0 - G_0^2 \frac{\epsilon_0}{3}(\epsilon_\parallel h + \epsilon_d d) Q^2 + O(Q^4). \tag{S2}$$

The exposed admittance contains a term linear in $Q$, so the same analytic expansion cannot be used there. Sending $d \to \infty$ reproduces the exposed nonzero-mode kernel if the exterior permittivity is matched. The uniform applied-voltage mode must still be re-specified; a fixed voltage at an electrode sent to infinity does not preserve a finite remote field.

#### S1.2. Uniform ensembles and passive evolution

For the exposed local reference, the areal energy is $g = h f_L(p) - h E_b p + G_0 (p-c)^2/2 + \chi c^2/2 - \Psi c$. Its $c$ derivative is the chemical potential in Eq. (2). Minimizing over $c$ gives $c = (G_0 p + \Psi)/(G_0 + \chi)$ and Eq. (6). In the top stack, replace the background work by $-b_V V(p-c)$; minimization gives $c = (G_0 p - b_V V + \Psi)/(G_0 + \chi)$ and Eq. (15). A conserved uniform $c$ is not a minimization variable; the appropriate condition is its fixed inventory.

The nonuniform charge equation is $\partial_t c_Q = -r_Q (c_Q - a_Q p_Q)$. Laplace transformation retains the initial term and gives Eq. (5). The induced potential response is therefore

$$Z(Q,s) = K_{\mathrm{eq}}(Q) + H_Q \frac{s}{s + r_Q} \tag{S3a}$$

$$K_{\mathrm{eq}} = \frac{\chi G_Q}{G_Q + \chi} \tag{S3b}$$

$$H_Q = \frac{G_Q^2}{G_Q + \chi}. \tag{S3c}$$

At $Q = 0$ and $\Lambda = 0$, the time-domain conservation equation is used directly. Assigning the formal $s = 0$ value of Eq. (S3) to that mode would erase the inventory constraint. The physical populations formulation similarly retains every independently conserved uniform population, including electrically neutral combinations.

### S1.3. Energy balance and equilibrium spectrum

Vary the electrostatic energy only after enforcing Poisson's equation and the electrical boundary conditions, or use the electric enthalpy with the potential eliminated by stationarity. The resulting variations are $\delta\mathcal{G}/\delta p = f_L{}'(p) - \sum_j g_j\,\partial_j^2 p - E_z$ and $\delta\mathcal{G}/\delta c = \mu$. Multiplying their evolution equations by these variations, integrating by parts and using no-flux species conditions gives Eq. (3). For fixed-voltage electrodes the electrical source work is already included in the enthalpy.

At equilibrium write $\dot{u} = -\mathsf{M}\mathsf{H}u$, where $\mathsf{H}$ is the constrained symmetric Hessian and $\mathsf{M}$ the positive mobility operator. On the nonconserved, nonneutral subspace this is similar to $-\mathsf{M}^{1/2}\mathsf{H}\mathsf{M}^{1/2}$, which is self-adjoint. The spectrum is real. Conserved uniform modes and translation must be removed or treated as constraints. The proof applies to equilibrium, not automatically to a translating state with comoving advection and sustained energy release.

## S2. Common scalar walls, exact profiles and interphase dynamics

### S2.1. Quartic wall and biased traveling front

At zero field, multiplying $g_y p'' - f_L{}'(p) = 0$ by $p'$ gives $g_y p'^2/2 = f_L(p) - f_L(P_s)$. For $c_6 = 0, b > 0$, integration gives the tanh profile used in Section 4, $I_4 = 4P_s^2/(3w)$ and wall energy $\gamma_4 = g_y I_4$.

For a uniform effective field, let $p_1 < p_2 < p_3$ be the three roots of $ap + bp^3 - E = 0$. The descending front is

$$p(\xi) = p_1 + \frac{p_3 - p_1}{1 + \exp\left[(p_3 - p_1)\sqrt{b/(2g_y)}\,\xi\right]} \tag{S4a}$$

$$v = \frac{\sqrt{b g_y/2}}{\Gamma}(p_1 + p_3 - 2p_2). \tag{S4b}$$

Substitution uses $p' = -\sqrt{b/(2g_y)}(p - p_1)(p_3 - p)$. Matching the remaining coefficient gives the velocity. The outer roots are locally stable for $|E| < 2|a|^{3/2}/(3\sqrt{3b})$. At weak field, $v/E = 2P_s/(\Gamma I_4) = 3w/(2\Gamma P_s)$. This exact local result presumes that screening and other eliminated fields do not introduce an additional spatial force.

The longitudinal equilibrium Hessian is $(g_y/w^2)[-\partial_u^2 + 4 - 6\,\mathrm{sech}^2 u]$, $u = \xi/w$. It has a translation zero mode, a localized shape eigenvalue $3g_y/w^2$ and continuum threshold $4g_y/w^2$. The ordinary quartic shape mode is not the first-order interphase-separation mode.

### S2.2. First-order thermodynamics and factorization

For $b < 0, c_6 > 0$, zero-field coexistence occurs at $a_t = 3b^2/(16c_6)$ and $S_t = -3b/(4c_6)$. The paraelectric spinodal is $a = 0$ and the polar spinodal is $a = b^2/(4c_6)$. At finite field, local spinodals satisfy

$$p_{\mathrm{sp}}^2 = \frac{-3b \pm \sqrt{9b^2 - 20ac_6}}{10c_6} \tag{S5a}$$

$$E_{\mathrm{sp}} = -2p_{\mathrm{sp}}^3(b + 2c_6 p_{\mathrm{sp}}^2). \tag{S5b}$$

Only nonnegative squared roots on the followed branch are retained. These are homogeneous branch limits, not wall depinning or splitting thresholds.

Using $a = -bS - c_6S^2$ and $d_6 = 2S + 3b/(2c_6)$ expands Eq. (28) exactly. The first integral then becomes $p' = -\sqrt{c_6/(3g_y)}(S - p^2)\sqrt{p^2 + d_6}$. Separating variables gives Eq. (29). A static scalar heteroclinic crossing $p = 0$ requires $d_6 > 0$; metastability of the outer endpoints alone is insufficient when the central phase has lower energy.

Changing the integration variable from $\xi$ to $p$ gives

$$I_6 = 2\sqrt{\frac{c_6}{3g_y}}\int_0^{\sqrt{S}}(S - p^2)\sqrt{p^2 + d_6}\,dp. \tag{S6}$$

The primitives of $\sqrt{p^2 + d_6}$ and $p^2\sqrt{p^2 + d_6}$ reduce Eq. (S6) to Eq. (30). It remains finite at coexistence. The longitudinal operator factorizes as $\mathsf{H}_\parallel = g_y\mathcal{A}^\dagger\mathcal{A}$, with $\mathcal{A} = \partial_\xi - p_6''/p_6'$, since differentiation of the stationary equation gives $\mathsf{H}_\parallel p_6' = 0$. The nodeless derivative is the translation ground state; the local scalar wall has no negative longitudinal eigenvalue for $d_6 > 0$.

### S2.3. Width conventions and matched comparisons

Solving $p_6^2/S = f^2$ gives the full width between $p = \pm fP_s$:

$$w_f = \frac{2}{\kappa_w}\operatorname{asinh}\sqrt{\frac{f^2(1+\rho)}{\rho(1-f^2)}}. \tag{S7}$$

Thus $f = 1/\sqrt{2}$ defines $\ell$, whereas $f = 1/2$ defines $w_{1/2}$. Confusing these definitions changes the constant inside the coexistence logarithm and the interaction prefactor. The definition used in Eq. (31) is $\ell = w_{1/\sqrt{2}}$. In the large-separation limit, elementary interface positions are specified by these half-squared-amplitude locations.

The polar curvature is $K_b = 4g_y\kappa_w^2$. Matching $P_s, K_b, g_y$ therefore fixes $\kappa_w$ and the quartic width $w = 1/\kappa_w$. Equation (30) gives

$$\frac{I_6}{I_4} = \frac{3}{16\sqrt{1+\rho}}\left[(2-\rho)\sqrt{1+\rho} + \rho(4+\rho)\operatorname{asinh}\left(\rho^{-1/2}\right)\right] \to \frac{3}{8}. \tag{S8}$$

The limit is taken with the weak-field mobility evaluated before complete unbinding. It is not a statement about a finite rigid wall at exact coexistence.

### S2.4. Elementary interfaces and separation energy

At coexistence, an elementary front from $+P_t$ to zero is $p_{F0} = P_t\left[1 + e^{2\kappa_t\xi}\right]^{-1/2}$. Direct integration gives $I_{F0} = S_t\kappa_t/4$ and $\gamma_{F0} = g_yI_{F0}$. The two-front limit has $I_6 = 2I_{F0}$. Matching the exact $\ell \sim \kappa_t^{-1}\ln(4/\rho)$ to $U' = 0$ gives $A_w = 8\gamma_{F0}$ in Eq. (31).

With elementary interface frictions $\eta_h = h\Gamma I_{F0}$, the center and separation frictions are $\eta_X = 2\eta_h$ and $\eta_\ell = \eta_h/2$. Consequently $s_\ell = -U''/\eta_\ell \simeq -h\kappa_t\Delta f/\eta_\ell$. The mode softens while $K_b$ remains finite. At coexistence and without charge drag,

$$e^{\kappa_t\ell(t)} = e^{\kappa_t\ell(0)} + \frac{h\kappa_t^2A_w}{\eta_\ell}t. \tag{S9}$$

For nonzero $\Delta f$, set $Y = e^{\kappa_t\ell}$; then $\dot{Y} = (h\kappa_t/\eta_\ell)(\kappa_tA_w - \Delta fY)$. This solves binding, coexistence growth and expansion explicitly as long as the interfaces remain separated and the local phase description is valid.

An exact zero-field isolated interphase-front result away from coexistence is obtained with the two positive roots $S > R$ of $a + bz + c_6z^2 = 0$. With $\kappa_S = S\sqrt{c_6/(3g_y)}$,

$$p(\xi) = \sqrt{\frac{S}{1+e^{2\kappa_S\xi}}} \tag{S10a}$$

$$v_{F0} = \frac{\sqrt{3g_y c_6}}{\Gamma}\left(\frac{S}{3} - R\right). \tag{S10b}$$

Substitution into $g_y p'' + \Gamma v p' - f_L{}'(p) = 0$ verifies this expression. It applies when both endpoints are locally stable. It gives the direction of thermal interphase advance without delayed screening; charge drag must be added for the coupled junction.

## S3. Whole-wall force projection, local exchange and mixed transport

### S3.1. Projection and charge drag

For a straight reference $p_0(y - X)$, $\dot{p} = -\dot{X}p_0{}'$. Multiply the polarization equation by $-p_0{}'$ and integrate over $y$ and film thickness. The intrinsic term gives $h\Gamma I\dot{X}$. The background drive gives the domain conversion force; the induced field integrates through depth to the surface-potential term. With $c_q = a_q r_q p_q/(r_q - iqv)$ and $iqp_q = W(q)$, the retarding force is Eq. (7), or generally

$$F_{\mathrm{ch}}(v) = \int_{-\infty}^{\infty} \frac{|W(q)|^2}{2\pi} H_q \frac{v r_q}{r_q^2 + q^2 v^2}\, dq. \tag{S11}$$

The factor $|W|^2$ comes from the polarization source and the force projection. Sampling the field only at the junction gives a different weight. Expanding Eq. (S11) at small $v$ gives Eq. (8). At $\chi = 0, M = 0$, Parseval's identity yields $\zeta_{\mathrm{ch}} = I/\Lambda$. The small-field velocity is $v = 2hP_S E_d/[I(h\Gamma + \Lambda^{-1})]$ for either scalar profile under complete compensation.

### S3.2. Monotonicity and stability for local exchange

In local capacitance, Fourier inversion of $v r_0/(r_0^2 + q^2 v^2)$ gives the positive exponential kernel in Eq. (16). A monotone profile has $p_0{}'(\xi)p_0{}'(\eta) \geq 0$, proving $F'_{\mathrm{ch}} \geq 0$. Define $C_p(a) = \int p_0{}'(y)p_0{}'(y + a)dy \geq 0$. The uniform perturbation equation is

$$\eta s + H_0 r_0 \int_0^{\infty} e^{-r_0 t}\, C_p(vt)(1 - e^{-st})dt = 0. \tag{S12}$$

For $\mathrm{Re}s \geq 0$, the real part of each integrand is nonnegative. For a nonzero frequency or positive growth rate the equality cannot hold. Translation is the only neutral mode. Adding a positive $Tk^2$ retains this exclusion for a fixed-profile graph in this strictly local model. The argument does not freeze out a physical internal mode of a deformable wall.

For the exact exposed-film kernel with the tanh form factor, put $d(q) = 1/G_O(q)$ and $t = q/r_q = qd/[\Lambda(1 + \chi d)]$. The variable $t$ increases with $q$. Equation (S11) becomes

$$F_{\mathrm{ch}}(v) = \frac{\Delta P^2}{\pi} \int_0^{\infty} \frac{\mathcal{W}[q(u/v)]}{1 + u^2}\, du \tag{S13a}$$

$$\mathcal{W}(q) = \frac{f_w(q)^2}{d + \chi d^2 + qd'}. \tag{S13b}$$

The numerator decreases. The denominator increases because $d, d' > 0$ and the derivative of $q\coth(aq) + q\left[d\left(q\coth(aq)\right)/dq\right]$ is positive. With $x = aq$, its nontrivial sign reduces to $\cosh x(\sinh^2 x + x^2) - 2x\sinh x > 0$, following from $\cosh x \geq 1$ and $\sinh^2 x + x^2 \geq 2x\sinh x$. Thus $\mathcal{W}$ decreases and Eq. (S13) increases with $v$. A split-wall oscillatory Fourier factor does not satisfy the same numerator hypothesis.

### S3.3. Mixed transport, asymptotics and folds

For $r_q = r_0 + Dq^2$, separate the small-$q$ contribution before expanding Eq. (S11) at large $v$. The tanh integrals are $\int_0^{\infty} f_w^2\, dq = \pi/(3w)$ and $\int_0^{\infty} (f_w^2 - 1)\, q^{-2} dq = -\pi w/2$. These give Eq. (9). Expanding the entire integrand first would miss the small-wave-number plateau. For every finite width the plateau is $B/2$; the zero-width-first result has plateau $B$. The limits do not commute.

Each spectral component has normalized differential slope $(1-u^2)/(1+u^2)^2$, whose minimum is $-1/8$ at $u^2=3$. Summing positive weights proves $F'_{\rm ch} \geq -\zeta_{\rm ch}/8$. This bound rules out folds when intrinsic friction exceeds $\zeta_{\rm ch}/8$ but does not prove a fold otherwise.

For weak exchange the relevant $q \sim \sqrt{r_0/D}$ are small. Replacing the smooth form factor and kernel by their zero-mode limits gives $\zeta_{\rm ch} \sim B/\left(2\sqrt{Dr_0}\right)$. If $W=-\Delta P$ over the entire contributing spectrum, residue integration gives $F_{\rm ch} = Bv/\sqrt{v^2+4Dr_0}$. This approximation requires $wa_+ \ll 1$, where $a_+ = \left(\sqrt{v^2+4Dr_0}+v\right)/(2D)$. Its formal high-speed limit is not valid at fixed nonzero $w$.

**S4. Re-derivation of the historical capacitor screening law**

**S4.1. Traveling charge and field**

The historical relaxation target is the translating polarization, $\tau\dot{c} = p_0(y-vt) - c$ [3]. At $v>0$, boundedness ahead gives Eq. (17). This differs from relaxation toward a potential-dependent electrochemical equilibrium, and $\tau$ is an independent parameter of that reduced law. With $W(q) = iqp_q$,

$$\phi_{s,q} = -\frac{v\tau G_T(q)W(q)}{1-iqv\tau} \tag{S14a}$$

$$E_{z,q}(z) = -L(q,z)\phi_{s,q} \tag{S14b}$$

$$L(q,z) = \frac{\alpha|q|\cosh[\alpha|q|(z+h)]}{\sinh(\alpha|q|h)}. \tag{S14c}$$

For an even derivative $W(q)$ and junction position $\xi=0$,

$$E_{\rm ret}(0,z) = \frac{v\tau}{\pi}\int_0^\infty \frac{L(q,z)G_T(q)W(q)}{1+(qv\tau)^2}\,dq. \tag{S15}$$

For a descending wall $W<0$, this field opposes positive advance. At the top surface $L = \alpha q\coth(\alpha qh)$; after averaging through depth, $L$ is replaced by $1/h$. These two explicit integrals reproduce the distinct field samplings used in the historical velocity closures. The material width enters through the single factor $W$ in the sampled field.

**S4.2. Local and whole-wall velocity closures**

A local-junction prescription is $v = \mathcal{M}\left[E_{\rm app} + E_{\rm ret}(0,0)\right]$, with the assigned junction mobility $\mathcal{M}$. Its algebraic turning condition is $1/\mathcal{M} - dE_{\rm ret}/dv = 0$. It treats the surface junction as the rate-controlling coordinate and assumes that the rest of the wall can follow the resulting deformation. It does not calculate that deformation by itself.

The whole-wall projection instead gives

$$F = h\Gamma Iv + v\tau\int_{-\infty}^{\infty} \frac{G_T(q)|W(q)|^2}{1+(qv\tau)^2}\frac{dq}{2\pi}. \tag{S16}$$

This has two profile factors and the mechanical friction of the entire finite-depth wall. A calculation using a sampled depth-averaged field remains a further approximation unless the profile projection is performed. The published distinction between local and averaged driving is therefore retained without identifying either with every possible deformable-wall closure. Equation (S16), the electrochemical Eq. (S11), and the junction equation must be compared using their own relaxation laws and force normalizations.

For an exposed surface, $G_T$ can be replaced by the exposed nonuniform kernel only after the exterior drive and uniform charge condition have been specified. This yields a new boundary-value problem with the same general screening-delay mechanism, not disappearance of the mechanism. A full calculation must then solve the depth-bending equation in S6 rather than infer bending solely from a negative differential junction law.

## S5. Conserved charge, transient motion and two-interface dynamics

### S5.1. The traveling jump condition and its scope

For $\partial_t c + \partial_y J = 0$, substitution of $c(y - vt)$ gives $-vc' + J' = 0$. Integration yields Eq. (11). The same identity holds for every conserved physical species. On an infinite surface with different nonzero far charges, the absolute total charge integral diverges; conservation is expressed through the flux identity and finite changes relative to the initial state, or through a finite closed interval followed by a large-size limit. A reservoir chemical offset is a constant on a fixed-inventory manifold and supplies no additional dynamical force.

### S5.2. Impedance, similarity solution and finite-size arrest

Linearize the conserved response about a compensated wall at rest and retain the initial-charge term. The displacement impedance is

$$Z(s) = \eta + \int_{-\infty}^{\infty} \frac{|W(q)|^2 G_q a_q}{s + Mq^2(G_q + \chi)} \frac{dq}{2\pi}. \tag{S17}$$

At small $s$ for a finite film, $\int dq/(2\pi)(s + Dq^2)^{-1} = 1/\left(2\sqrt{Ds}\right)$, giving Eq. (12). For a step force, $F(s) = F/s = sZ(s)X(s)$ and inverse Laplace transformation gives $X = 4F\sqrt{Dt}/\left(B\sqrt{\pi}\right)$ in the charge-dominated limit.

For finite reduced force in the sharp-interface outer model, let $m = c - a_0 p$. Away from the moving interface $m_t = Dm_{yy}$. Conservation and the wall force impose

$$D\left(m_y^+ - m_y^-\right) = a_0 \varDelta P \dot{X} \tag{S18a}$$

$$\eta \dot{X} = F + G_0 \varDelta P m_*. \tag{S18b}$$

At long times $m_* = -F/(G_0 \varDelta P)$. With $z = y/\left(2\sqrt{Dt}\right)$ and $X = 2\lambda\sqrt{Dt}$, the solutions are $m_-(z) = m_* \mathrm{erfc}(-z)/\mathrm{erfc}(-\lambda)$ and $m_+(z) = m_* \mathrm{erfc}(z)/\mathrm{erfc}(\lambda)$. Their derivatives in Eq. (S18) give Eq. (13). For small force, $\lambda = 2(F/B)/\sqrt{\pi}$, agreeing with the impedance result. The formal sharp fixed-domain branch has $0 < F/B < 1$; near a domain spinodal the fixed-domain approximation fails first.

In a closed interval of length $L_y$, a small displacement produces a uniform mismatch $m = -a_0 \varDelta P X/L_y$ after full redistribution. The resulting force is $-BX/L_y$ and the interior arrest estimate is $X_\infty = FL_y/B$. It requires both domains to remain present and stable. The crossover scale is $L_y^2/D$, but the full crossover function is not obtained by joining the early and late asymptotes.

Domain susceptibility modifies the outer diffusion. With $f_L{}'(p) + G_0(p - c)/h = E_b$ on each stable branch,

$$\frac{dp}{dc} = \frac{G_0}{hf_L{}''(p) + G_0} \tag{S19a}$$

$$\frac{d\mu}{dc} = \chi + \frac{G_0 h f_L{}''(p)}{hf_L{}''(p) + G_0} \tag{S19b}$$

$$D_{\mathrm{eff}} = M\frac{d\mu}{dc}. \tag{S19c}$$

Near loss of a branch the coupled polarization problem must be retained. For the nonlinear outer problem, $c = C(z)$ gives $-2zC' = H_{\mathrm{ref}}^{-1}[\mu(C)]''$ with $D_{\mathrm{ref}} = MH_{\mathrm{ref}}$. Chemical-potential continuity, the conserved jump and the interface conversion force close the problem. Solving it is a specific deferred calculation, not an analytic error-function solution for arbitrary susceptibility.

**S5.3. Translation and width friction under conservation**

For collective coordinates $q_i$, slow equilibrium elimination of the charge gives the positive metric

$$\zeta_{ij}^{\mathrm{ch}} = \int \frac{dq}{2\pi} \frac{a_q^2\left(\partial_{q_i} p_q\right)^*\left(\partial_{q_j} p_q\right)}{\Lambda + M q^2}. \tag{S20}$$

For a separated pair, $p' = -P_s[\delta(y+\ell/2) + \delta(y-\ell/2)]$. In local capacitance, parity separates center and width modes and gives

$$\zeta_X^{\mathrm{ch}} = \frac{a_0^2 P_s^2}{\sqrt{\Lambda M}}\left(1 + e^{-\ell\sqrt{\Lambda/M}}\right) \tag{S21a}$$

$$\zeta_\ell^{\mathrm{ch}} = \frac{a_0^2 P_s^2}{4\sqrt{\Lambda M}}\left(1 - e^{-\ell\sqrt{\Lambda/M}}\right). \tag{S21b}$$

As $\Lambda \to 0^+$ at fixed $\ell$, $\zeta_X$ diverges but $\zeta_\ell \to a_0^2 P_s^2 \ell/(4M)$. The center source has a nonzero value at $q = 0$; the width source vanishes there. Combining this finite width friction with Eq. (31) gives a real relaxation rate $s_\ell = -h\kappa_t \Delta f/[\eta_\ell + a_0^2 P_s^2 \ell/(4M)]$. At exact coexistence, the leading conserved-width growth has $\kappa_t \ell \sim \ln t - \ln\ln t$ up to matching constants. This is slow internal wetting dynamics, not common translation or a periodic oscillation.

**S5.4. Charge wakes of two resolved interfaces**

The steady local reaction–transport operator has inverse lengths $a_\pm$ and Green function $G_v(y) = e^{-vy/(2D) - S_v|y|/(2D)}/S_v$. Superposing the responses of two jumps and sampling each sharp interface with the symmetric self-value gives Eq. (21). The center is driven by the sum of their forces and the width by half their difference. The local-exchange limit instead gives the downstream delay $\ell/v$ and the exponential $e^{-r_0\ell/v}$.

At fixed $\ell$, $r_0\ell^2/D \ll 1$ and $x = v\ell/D = O(1)$ yield $F_{\mathrm{ch}}^{(2)}/B = (3 + e^{-x})/4 + O(r_0\ell^2/D)$. Hence $\partial_v F = \eta_X - B\ell e^{-x}/(4D) + \cdots$. Along a stationary freely relaxing pair,

$$\frac{d\ell}{dv} = \frac{\partial_v F_\ell^{\mathrm{ch}}}{U'' - \partial_\ell F_\ell^{\mathrm{ch}}} \tag{S22a}$$

$$\frac{dF}{dv} = \eta_X + \partial_v F_{\mathrm{ch}}^{(2)} + \partial_\ell F_{\mathrm{ch}}^{(2)} \frac{d\ell}{dv}. \tag{S22b}$$

If the retained intrinsic friction depends on $\ell$, its derivative adds $v(d\eta_X/d\ell)(d\ell/dv)$. Vanishing denominator in Eq. (S22) is a width-branch condition and invalidates a one-coordinate elimination.

For the conserved equal-background preparation, integrating the steady charge equation gives $Dc' + v(c - c_b) = Da_0 p'$. Two negative polarization jumps produce two negative compensation pulses ahead of their positions. Their total integral is $-2a_0 P_s D/v$, independent of separation. The center and width forces are those in Section 5.5 and use the frozen-background phase energy. The pulse family cannot be scanned in velocity while silently keeping its inventory fixed.

**S6. Deformable walls and the full linearized surface problem**

For $p = p_0[y - X(x,z,t)]$, $\rho_b = -\partial_z p = p_0{}' \partial_z X$. The small-slope projection is

$$\Gamma I\, \partial_t X = g_x I\, \partial_x^2 X + g_z I\, \partial_z^2 X + f_L(p_R) - f_L(p_L) - \int E_z p_0{}'\, dy. \tag{S23}$$

At $z = 0, -h$, the natural polarization condition gives $\partial_z X = 0$ within the graph approximation. The electrostatic field must include the volume charge proportional to $\partial_z X$. Replacing the polarization by its surface value in the straight-wall kernel would omit this source and is not a bending calculation.

For a solved traveling base $p_0(\xi,z), c_0(\xi)$, write perturbations $u(\xi,z)e^{st+ikx}$, $\delta c(\xi)e^{st+ikx}$ and $\varphi(\xi,z)e^{st+ikx}$. The scalar equations are

$$\Gamma s u = \left[g_y \partial_\xi^2 + g_z \partial_z^2 - g_x k^2 + \Gamma v \partial_\xi - f_L''(p_0)\right] u - \partial_z \varphi, \tag{S24}$$

$$\epsilon_0 \left[\epsilon_\parallel \left(\partial_\xi^2 - k^2\right) + \epsilon_z \partial_z^2\right] \varphi = \partial_z u \tag{S25a}$$

$$\left(s - v \partial_\xi\right) \delta c = -\left[\Lambda - M\left(\partial_\xi^2 - k^2\right)\right] (\chi \delta c - \varphi_s). \tag{S25b}$$

The electrical perturbations satisfy the homogeneous versions of the original boundary conditions, including the correct exterior or top-electrode solution. The source jump is $u_s - \delta c$. For a nonlinear chemical law, replace $\chi$ by the local chemical Hessian and retain derivatives of reaction and mobility coefficients where the base is nonuniform.

At equilibrium a positive constrained Hessian establishes stability. A small-drive continuation can be obtained after fixing translation when the remaining operator is invertible on the chosen function space and the far states remain stable. In an infinite lateral domain, diffusive and long-wave modes require separate treatment; one must not assert a uniform spectral gap by removing translation alone. Equations (S24)–(S25) specify the actual continuation and eigenvalue calculation in N1 and N3.

For a conserved transient, the perturbation operator depends on time: $\mathsf{M}\, \partial_t u = \mathsf{L}_k(t) u$. The propagator measures finite-time amplification. In the similarity outer problem, a uniform transverse mode of the form $\delta m = t^\beta \widehat{m}\left[y/\left(2\sqrt{Dt}\right)\right]$ obeys $\widehat{m}'' + 2z\widehat{m}' - 4\beta\widehat{m} = 0$, with linearized interface conditions. A fixed nonzero transverse wave number adds $-4Dk^2 t\widehat{m}$; the problem remains nonautonomous. Finite amplification or overshoot is not an autonomous Hopf bifurcation.

## S7. Spatial and temporal stability of the projected surface front

### S7.1. Translational invariance and the characteristic equation

Vary both the source $p_0(y - X)$ and the factor $p_0{}'$ in the projected electrical force. The source variation produces $Z(Q, s - iqv)$; the force-weight variation subtracts its base value $Z(|q|, -iqv)$. This gives Eq. (34). Omitting the second term gives a spurious restoring force for uniform translation. The stable charge-response poles and any uncoupled charge modes must also be retained in a full spectral interpretation.

At $k = 0$, $\mathcal{C}(0,0) = 0$ and $\partial_s \mathcal{C}(0,0) = \eta + F'_{\mathrm{ch}}$. If this derivative is negative, $\mathcal{C}(s,0)/s$ is negative near $s = 0^+$ and tends to positive $\eta$ at large positive $s$, proving a positive real root. A fold is therefore associated with real instability in this reduction. At a regular long-wave branch,

$$s(k) = -\frac{\partial_{k^2} \mathcal{C}(0,0)}{\partial_s \mathcal{C}(0,0)} k^2 + O(k^4). \tag{S26}$$

The expansion requires convergent derivatives. The exposed-surface nonanalytic case must use the full integral.

### S7.2. Exact local unresolved-interface dispersion

With $W = -\Delta P$, $G = G_0$ and $r = r_0 + DQ^2$, integration over $q$ gives Eq. (19). Its derivative in $k^2$ at zero gives $T_{\mathrm{eff}} = T - BDv^2/S_v^3$. Setting $s = 0$ and factoring the neutral $k = 0$ root gives the band edge stated in Section 8.2. With $\beta_f = \eta\sqrt{Dr_0}/B$, the dimensionless expansion has

$$c_\lambda = \beta_f + \frac{1}{2(1+u^2)^{3/2}} \tag{S27a}$$

$$a_2 = \theta_c - \theta \tag{S27b}$$

$$a_4 = \frac{3u^2}{8(1+u^2)^{5/2}}. \tag{S27c}$$

The near-onset formulas require $a_2 > 0$ small enough that neglected $sQ_k^2$ and higher powers are subleading. In a finite width $L_x$ the smallest periodic wave number is $2\pi/L_x$; the mathematical long-wave onset can therefore be inaccessible experimentally.

To exclude Hopf in this reduction, write the square root at $s = i\omega$ as $u_s + iw_s$, with $u_s > 0$, $u_s w_s = 2D\omega$ and $u_s^2 - w_s^2 = v^2 + 4D(r_0 + Dk^2)$. Then

$$\mathrm{Im}\frac{i\omega + v^2/(2D)}{u_s + iw_s} = \frac{\omega(u_s^2 - v^2)}{u_s(u_s^2 + w_s^2)}. \quad \text{(S28)}$$

It has the sign of $\omega$, as does $\eta\omega$. The imaginary part cannot vanish at nonzero frequency. The conclusion does not apply to an independently evolving separation or rotation mode.

### S7.3. A resolved pair and its local-exchange exclusion

Let the two base positions be $a_1 = -\ell/2, a_2 = \ell/2$ and $J = U''(\ell)$. Put $S = \sqrt{v^2 + 4D(r_0 + Dk^2 + s)}$, $G_v(d; s, k) = e^{-vd/(2D) - S|d|/(2D)}/S$ and $R(d; s, k) = H_0[sG_v - v\,\partial_d G_v]$. Use the symmetric derivative at $d = 0$. The two-interface matrix is

$$C_{ij} = (\eta_h s + T_h k^2)\delta_{ij} + J\begin{pmatrix} 1 & -1 \\ -1 & 1 \end{pmatrix}_{ij} + P_s^2\left[R(a_i - a_j; s, k) - \delta_{ij}\textstyle\sum_l R\,(a_i - a_l; 0,0)\right]. \quad \text{(S29)}$$

Here $\eta_h = h\Gamma I_{F0}$ and $T_h = hg_x I_{F0}$. The base must satisfy both center and width force balances. The row subtraction enforces global translation. A stationary or Hopf candidate follows from $\det C = 0$ with $s = 0$ or $s = i\omega$, respectively; neither is asserted merely by writing this determinant.

For local exchange and two narrow interfaces, $a_d = H_0 P_s^2 (r_0/v) e^{-r_0\ell/v} > 0$ and $\tau_d = \ell/v$ give at $k = 0$

$$C = \begin{pmatrix} \eta_h s + J + a_d & -J - a_d e^{-s\tau_d} \\ -J & \eta_h s + J \end{pmatrix}. \quad \text{(S30)}$$

For $J > 0$ and $\mathrm{Re}s \geq 0$, $|(\eta_h s + J + a_d)(\eta_h s + J)| \geq J(J + a_d)$, whereas $|J(J + a_d e^{-s\tau_d})| \leq J(J + a_d)$. The first inequality is strict except at $s = 0$. No nonneutral unstable root exists. Adding positive line stiffness preserves the result. This specific full-wake exclusion and the finite-Fourier-mode Hopf example in S9 have different retained response spectra; the latter cannot overturn the former for the same local pair problem.

### S7.4. General uniform Hopf test

For the surface response define $b_q = qv$, $R_q = r_q^2 + b_q^2$ and $Q_q = \left(r_q^2 + b_q^2 - \omega^2\right)^2 + 4r_q^2\omega^2$. Removing the translational zero mode from Eq. (34) gives necessary conditions

$$0 = \int_0^\infty |W|^2\, H_q\, \frac{r_q^2(\omega^2 + r_q^2 - 3b_q^2)}{R_q Q_q}\, dq \quad \text{(S31a)}$$

$$0 = \eta + \frac{1}{\pi}\int_0^\infty |W|^2\, H_q\, \frac{r_q(r_q^2 + \omega^2 - b_q^2)}{Q_q}\, dq. \quad \text{(S31b)}$$

The first condition has an irrelevant positive normalization removed. The bound $r(r^2 + \omega^2 - b^2)/[(r^2 + b^2 - \omega^2)^2 + 4r^2\omega^2] \geq -1/(8r)$ gives the sufficient exclusion $\eta > \zeta_{\mathrm{ch}}/8$. Outside it, both equations, simplicity, transversality and the stability of other modes must be checked. The sign of the nonlinear Hopf coefficient remains a separate calculation.

## S8. Blocking-MIEC transport and charge conservation

### S8.1. Two-species linear solution

Let $\rho_j = q_j \delta n_j$ about a homogeneous neutral reference. Linearizing Eq. (22) gives

$$\partial_t \rho_j = D_j\left[\nabla^2 \rho_j - \kappa_j^2(\rho_i + \rho_e)\right] \tag{S32a}$$

$$-\epsilon_m \nabla^2 \phi = \rho_i + \rho_e. \tag{S32b}$$

Seek modes $e^{-pz}$ and put $t = p^2 - Q^2$. The two charge roots satisfy

$$(t - \kappa_i^2 - s/D_i)(t - \kappa_e^2 - s/D_e) - \kappa_i^2\kappa_e^2 = 0. \tag{S33}$$

There is also a harmonic mode. Write $\phi = A_0 e^{-Qz} + A_+ e^{-p_+ z} + A_- e^{-p_- z}$, choosing $\mathrm{Re}p_\pm > 0$ for $\mathrm{Re}s > 0$. The species charge in mode $l$ is $\rho_{j,l} = -\epsilon_m \kappa_j^2 t_l A_l / \left(t_l - s/D_j\right)$. The two blocking conditions and the specified interface potential are the three equations

$$A_0 + A_+ + A_- = \phi_s \tag{S34a}$$

$$QA_0 - \sum_{l=\pm} \frac{s p_l A_l}{D_j t_l - s} = 0 \; (j = i, e). \tag{S34b}$$

Their solution gives $Y_m = \epsilon_m(QA_0 + p_+ A_+ + p_- A_-)/\phi_s$. Degenerate or static limits are taken by continuity from the original equations. The neutral composition mode has $t_- \sim s/D_a$, where $D_a = D_i D_e \kappa^2/(D_i\kappa_i^2 + D_e\kappa_e^2)$. It must not be discarded when diffusivities differ.

For $D_i = D_e = D_m$, the electrically forced neutral mode decouples. Put $p^2 = Q^2 + \kappa^2 + s/D_m$ and $\phi = Ae^{-Qz} + Be^{-pz}$. Blocking gives $QA = spB/(D_m\kappa^2)$. Dividing the interface displacement $\epsilon_m(QA + pB)$ by $A + B$ gives Eq. (23), including both static and frozen-carrier limits.

## S8.2. Nonlinear planar screening and validity

At equilibrium, $\mu_j$ is spatially constant. Dilute populations give $n_j = n_{j\infty}\exp\left[-q_j\phi/(k_B T)\right]$; finite-site statistics replace this by their bounded occupancy relation. Multiplying Poisson's equation by $\phi'$ and using the neutral far state gives $\epsilon_m \phi'^2/2 = W(\phi) = -\int_0^\phi \rho\,(u)du$. The interface displacement is $S = \mathrm{sgn}(\phi_s)\sqrt{2\epsilon_m W(\phi_s)}$ and the screening energy is $U(S) = \int_0^S \phi_s\,(S')dS'$.

For symmetric dilute monovalent carriers, $S = 2\epsilon_m\kappa(k_BT/e)\sinh[e\phi_s/(2k_BT)]$. Linear response has capacitance $C_m = \epsilon_m\kappa$ and requires $|e\phi_s| \ll k_BT$. Large ferroelectric polarization charge can violate this condition. The uniform-coefficient kernel is then a declared approximation; linearizing about the nonlinear screening profile produces depth-dependent coefficients and a different operator.

## S8.3. Long-distance transport and the infrared sector

In a steadily translating source, $s = -iqv$. The small-$q$ response contains the Ohmic spreading term. Equivalently, a current per unit junction length $I' = \Delta P v$ crossing a half-circle of radius $r$ has current density $I'/(\pi r)$ and electric field $I'/(\pi\sigma_m r)$. Integrating from $a_{\mathrm{in}}$ to $R$ gives Eq. (25). This establishes the logarithm independently of the near-field Debye approximation, provided the far medium has finite homogeneous conductivity.

For unequal positive diffusivities the fixed-finite-velocity Ohmic sector still uses $\sigma_m$. Taking zero frequency before zero wave number gives a different quasistatic transport combination because the composition mode has time to relax. The limits must not be interchanged. A frozen ionic species, zero conductivity or a different remote supply geometry requires a separate limit.

For a translated periodic pattern $P(y) = \sum_m P_m\, e^{iq_m y}$ with fixed fractions, the mean does not change. Its finite-wave-number dissipative force density is $\sum_{m\neq 0} q_m\,|P_m|^2[-\mathrm{Im}H(|q_m|, -iq_m v)]$. Expansion of one domain fraction changes the mean and reintroduces a supply problem. An isolated corrugated front also retains that problem if its average position advances.

### S8.4. Passive memory and conditional folds

For each nonzero spatial mode, diagonalizing the passive carrier relaxation operator gives $H(Q,s) = H_\infty(Q) - \int_0^\infty d\,\mu_Q(r)/(s+r)$ with $d\mu_Q(r) \geq 0$. A prescribed wall therefore has a positive drag $D(v) = \int |W(q)|^2 \int v\, d\mu_q(r)/(r^2 + q^2 v^2)\, dq/(2\pi)$ where the infrared sector is physically admissible. Positivity of dissipation does not imply positive differential drag.

Retaining a single spatial mode and relaxation rate gives Eq. (27). Differentiation yields $F'/\eta = 1 + \mathcal{A}(1-u^2)/(1+u^2)^2$. Its minimum is $1 - \mathcal{A}/8$, proving the fold criterion and roots in Section 6.4. The outer supply resistance adds positive mechanical resistance to a finite junction mode and is included in the same force normalization. The one-mode force is not the force on an infinitely deep rigid wall.

## S9. Multiaxial modes, coupled spatial instability and oscillatory reductions

### S9.1. Localized rotation and a true intermediate variant

For Eq. (32), variation gives $p'' = p(p^2 + a_\perp^2 - 1)$ and $a_\perp'' = a_\perp(p^2 + a_\perp^2 - 1 + \kappa_\perp)$. On the Ising wall the transverse operator is $-\partial_y^2 + \kappa_\perp - \mathrm{sech}^2(y/\sqrt{2})$. Applying it to $\mathrm{sech}(y/\sqrt{2})$ gives eigenvalue $\kappa_\perp - 1/2$. Substituting the two fields stated in Section 7.2 verifies the exact rotated branch. The nominal pure in-plane stationary point has negative longitudinal curvature for $\kappa_\perp > 0$; it is not a stable bulk variant.

For the stable-variant potential in Section 7.3, the Ising wall has $P_s^2 = A/B$, $w^2 = 2g_p/A$ and transverse operator

$$\mathsf{H}_\perp = -g_a\,\partial_y^2 + A(C/B - 1) - A(C/B)\mathrm{sech}^2(y/w). \tag{S35}$$

Let $\nu_a(\nu_a + 1) = 2(C/B)(g_p/g_a)$. The nodeless state is $\mathrm{sech}^{\nu_a}(y/w)$ and its stiffness is $\epsilon_\perp = A(C/B - 1) - g_a\nu_a^2/w^2$. This tests localized linear rotation. Domain insertion additionally compares wall/interphase energies and the compatibility of the bounding interfaces. Favorable bulk variant energy alone does not calculate its nucleation barrier.

For a prescribed moving profile, $\Gamma_a s a = g_a a'' + \Gamma_a v a' - m_\perp^2(y)a$. Substitution $a = e^{-\Gamma_a v y/(2g_a)}\psi$ shifts the eigenvalue to $s = -[\epsilon_\perp + \Gamma_a^2 v^2/(4g_a)]/\Gamma_a$. The transformed eigenfunction must still decay at both infinities. This is an exact frozen-profile kinematic result; the real moving profile and charge coupling may change the underlying stiffness.

### S9.2. Nonlinear rotation and a soft separation coordinate

Normalize the transverse mode by $\int \psi^2 dy = 1$. Longitudinal relaxation contributes to the amplitude quartic coefficient as

$$u_{\mathrm{eff}} = \int b_\perp(p_0)\psi^4 dy - 2\langle S, \mathsf{H}_\parallel^{-1} S\rangle \tag{S36a}$$

$$S = 1/2\,\partial_p m_\perp^2(p_0)\psi^2. \tag{S36b}$$

The inverse is restricted away from translation and uses the same constrained coupled Hessian as the base. The negative term follows by minimizing the quadratic longitudinal energy induced at order $A_\perp^2$. Near first-order coexistence its separation contribution can be large. Once that mode is slow, it must be retained explicitly rather than eliminated into a divergent quartic coefficient.

For $u < 0, w > 0$, rescale $x = A_\perp\sqrt{w/|u|}$ and $t = rw/u^2$. The amplitude energy is $tx^2/2 - x^4/4 + x^6/6$, with $x_+^2 = \left[1 + \sqrt{1-4t}\right]/2$. Its branch appearance, equal energy and loss of unrotated stability occur at $t = 1/4, 3/16, 0$. These are equilibrium results. Under driving, the nonlinear amplitude equation must be derived on the actual moving branch.

### S9.3. Coupled displacement and polarization

After choosing physically admissible depth modes and a positive kinetic normalization, a conservative two-field long-wave Hessian can have the form of Eq. (36). Its growth rates are

$$z_{\pm} = -\frac{(b_s+d_s)Q_k^2+\mu}{2} \pm \frac{1}{2}\sqrt{[(b_s-d_s)Q_k^2-\mu]^2+4c_s^2Q_k^2}. \tag{S37}$$

For $\mu > 0$, the determinant gives the instability band $0 < Q_k^2 < (c_s^2/b_s - \mu)/d_s$. Differentiating the upper eigenvalue at $b_s = d_s = D_s$ gives the fastest-mode formulas in Section 8.2. At $\mu < 0$ the uniform transverse mode is already unstable. A finite-$k$ onset requires the full nonlocal kernel or additional gradient terms; it does not follow from a long-wave onset merely because the fastest mode inside its unstable band is finite.

The full displacement–rotation–separation system has a matrix dispersion $\det[s\mathsf{M} + \mathsf{K}(k) + \boldsymbol{\Sigma}(k, s; v)] = 0$. Each entry follows from the charge sources in Eq. (33), the separation source and the geometry-specific response. A single-pole replacement is a declared reduction. For the strict isolated MIEC or a conserved transient, a moving-base diagram is applicable only after the supply and preparation problem has been resolved.

### S9.4. Passive driven linear-source model

Eliminate the complex screening variable in Section 8.3: $h = cA/(z + 1 - iU)$. Its real response produces Eq. (37), with coefficients $a_1 = p + 2$, $a_2 = 1 + U^2 + 2p - G$, $a_3 = p(1 + U^2) - G$. Routh–Hurwitz stability requires $a_1, a_2, a_3 > 0$ and $a_1a_2 > a_3$, giving Eq. (38). At equality, $D = (z + a_1)(z^2 + a_2)$, and eliminating $G$ gives Eq. (39).

The Hopf arc adjacent to a stable region requires $p > 1$ and $U^2 > (p + 1)/(p - 1)$. It implies $G > 8$; equality is a zero-frequency degeneracy. At $G = 12, p = 2, U = 3$, $a_1 = 4, a_2 = 2, a_3 = 8$. The other mode is stable and the imaginary pair has frequency $\sqrt{2}\,r$. At fixed $G = 12$ and $U = 3$, varying $p$ is tangent to the Hopf boundary because $d(a_1a_2 - a_3)/dp = 4(p + 1) - G = 0$; varying $U$ gives $d(a_1a_2 - a_3)/dU = 4U = 12 \neq 0$ and is the appropriate transverse control. This explicitly distinguishes a point on the boundary from a specified crossing path.

At zero velocity the model dissipates the sum of the structural energy and $\Delta H|h - cA|^2/2$. At finite prescribed translation, the phase rotation supplies work. If $\mu \geq 0$, the passive feedback and restoring terms are constrained together and do not produce growing modes. Choosing them independently would violate the model's energy structure.

### S9.5. Width-dependent source at fixed driving force

A separate example retains the interference of two interfaces in one complex Fourier component. Let $B(\ell) = 2\cos(\ell/2)$ and use energy $U(\ell) - FX + \left|c - B(\ell)e^{-iX}\right|^2/2$, with positive gradient-flow mobilities. In units with the retained wave number, charge relaxation rate and electrical stiffness equal to one, put $z = ce^{iX} = R + iI$. The exact reduced equations are

$$\eta_X v = F - BI \tag{S38a}$$

$$\eta_\ell \dot{\ell} = -U' + B'(R - B) \tag{S38b}$$

$$\dot{R} = -R + B - vI \tag{S38c}$$

$$\dot{I} = -I + vR. \tag{S38d}$$

At $\ell_* = 2\pi/3$, $B = 1$, $B' = -\sqrt{3}/2$ and $B'' = -1/4$. At $\ell_* = \pi/2$, the values are instead $B = \sqrt{2}$, $B' = -1/\sqrt{2}$ and $B'' = -\sqrt{2}/4$. The amplitude and its derivatives must come from the same function and the same base separation. They cannot be assigned separately.

For a chosen $\ell_*, v$, the stationary charge is $R = B/(1+v^2)$, $I = Bv/(1+v^2)$, and the required forces are $F = \eta_X v + BI$ and $U' = B'(R - B)$. With $K = U''$, $v_\ell = -B'I/\eta_X$ and $v_I = -B/\eta_X$, the phase-fixed Jacobian on $(\delta\ell, \delta R, \delta I)$ is

$$J = \begin{pmatrix} [-K + B''(R-B) - B'^2]/\eta_\ell & B'/\eta_\ell & 0 \\ B' - Iv_\ell & -1 & -v - Iv_I \\ Rv_\ell & v & -1 + Rv_I \end{pmatrix}. \tag{S39}$$

Its characteristic coefficients are $a_1 = -\mathrm{tr}J$, $a_2$ the sum of the principal second-order minors and $a_3 = -\mathrm{det}J$. A Hopf candidate satisfies $a_1 a_2 = a_3$ with all three positive and nonzero transversality. Because the source is nonlinear in width, this example differs from the linear-source model. It is still a one-Fourier-component projection and does not establish a Hopf boundary of the complete local-exchange pair, whose full-wake exclusion is proved in S7.3. We retain the consistent equations and candidate condition; the inconsistent source-amplitude tuple in the supplied open report is not used as a quantitative onset example. No material frequency or stable periodic branch is inferred from this construction.

## S10. Material coefficients, mechanical restrictions and analytical evaluation

### S10.1. $BaTiO_3$ convention and parameter set

The principal $BaTiO_3$ example uses the invariant expansion and coefficients of Marton, Rychetsky and Hlinka [11], including their specified mechanical convention. Temperature is in kelvin and polarization in C m$^{-2}$. Along the scalar [001] path, $a = 2\alpha_1$, $b = 4\alpha_{11}$ and $c_6 = 6\alpha_{111}$.

| Coefficient | Value or law | SI units |
|---|---|---|
| $\alpha_1$ | $3.34 \times 10^5(T - 381)$ | J m C$^{-2}$ |
| $\alpha_{11}$ | $4.69 \times 10^6(T - 393) - 2.02 \times 10^8$ | J m$^5$ C$^{-4}$ |
| $\alpha_{111}$ | $-5.52 \times 10^7(T - 393) + 2.76 \times 10^9$ | J m$^9$ C$^{-6}$ |
| $\alpha_{12}$ | $3.230 \times 10^8$ | J m$^5$ C$^{-4}$ |
| $\alpha_{112}, \alpha_{123}$ | $4.470 \times 10^9,\ 4.910 \times 10^9$ | J m$^9$ C$^{-6}$ |
| $g_y = G_{44}$ | $2 \times 10^{-11}$ | J m$^3$ C$^{-2}$ |
| $C_{11}, C_{12}, C_{44}$ | $2.750 \times 10^{11},\ 1.790 \times 10^{11},\ 5.43 \times 10^{10}$ | J m$^{-3}$ |
| $q_{11}, q_{12}, q_{44}$ | $14.20 \times 10^9,\ -0.74 \times 10^9,\ 1.57 \times 10^9$ | J m C$^{-2}$ |

Table S1. Material coefficients used in Fig. 11. The shear convention is that of the cited energy; tensor and engineering shear components are not interchangeable without their factors. The stability of the tetragonal phase against other variants limits the temperature interval of a scalar interpretation.

Let $C_N$ be the cubic normal-strain stiffness matrix with diagonal $C_{11}$ and off-diagonal $C_{12}$, and $\mathbf{q} = (q_{12}, q_{12}, q_{11})$. For a coherent [010]-normal wall, relaxation of the normal strain at zero normal stress while retaining far-field tangential strains gives

$$A_{\mathrm{el}} = \frac{1}{2}\mathbf{q}^T C_N^{-1}\mathbf{q} - \frac{q_{12}^2}{2C_{11}}. \tag{S40}$$

This produces $f_{\text{coh}} = f_{\text{bulk}} + A_{\text{el}}(p^2 - S)^2 + \text{constant}$ and the coefficient shifts in Section 10.1. The remote stationary polarization is unchanged because the correction and its first derivative vanish at $p^2 = S$. The core cost increases by $A_{\text{el}}S^2$ and its central curvature decreases by $4A_{\text{el}}S$. This is why the relaxed wetting limit need not survive in a coherent wall.

Coexistence of the relaxed scalar branch solves $\alpha_1 = \alpha_{11}^2/(4\alpha_{111})$, with the negative-quartic and positive-sixth-order branch selected. The polar spinodal follows from the discriminant in Eq. (41). These are algebraic evaluations of the parameterization. Field spinodals use Eq. (S5). A moving coherent wall's splitting field is not supplied by these bulk curves.

### S10.2. $BiFeO_3$ polar coefficients and structural shift

The principal $BiFeO_3$ coefficients are from Karpinsky and colleagues [79]. In that convention the octahedral coordinates are displacements, not tilt angles; their units must be retained when evaluating couplings.

| Polar coefficient | Value or law | SI units |
|---|---|---|
| $a_1^P(T)$ | $0.9 \times 10^6[800\coth(800/T) - 1300]$ | J m C$^{-2}$ |
| $a_{11}^P, a_{12}^P$ | $-1.35 \times 10^9,\ 0.50 \times 10^9$ | J m$^5$ C$^{-4}$ |
| $a_{111}^P, a_{112}^P, a_{123}^P$ | $11.2 \times 10^9,\ -3.0 \times 10^9,\ -6.0 \times 10^9$ | J m$^9$ C$^{-6}$ |
| $\zeta_{11} + 2\zeta_{12} + \zeta_{44}$ | $-2.1 \times 10^{29}$ | J m$^{-1}$ C$^{-2}$ |
| $\zeta_{44}$ | $-3.0 \times 10^{29}$ | J m$^{-1}$ C$^{-2}$ |
| $\zeta_{12}$ | $\leq 0.7 \times 10^{29}$ | J m$^{-1}$ C$^{-2}$ |
| $\eta_{11}, \eta_{12}$ | $\geq 10^{30}$ | J m$^{-1}$ C$^{-2}$ |

Table S2. The listed bounds are not assigned unique tensor values. Structural amplitudes, gradient coefficients and kinetic coefficients are not inferred from these inequalities.

For a fixed [111] structural environment, Eq. (42) gives $b_R = -2.83333 \times 10^8$ J m$^5$ C$^{-4}$ and $c_R = 3.55556 \times 10^8$ J m$^9$ C$^{-6}$. Solving the sixth-order scalar stationary equation gives $P_S^2 = \left[-b_R + \sqrt{b_R^2 - 3c_R(a_1^P + h_R)}\right]/(3c_R)$. The three landmarks in Eq. (43) follow from central curvature, equal phase energies and discriminant, respectively. They are not assigned to the actual structural transition without solving the structural variables.

For $\mathbf{\Phi} = (\varphi, \varphi, 0)$ and $\mathbf{\Psi} = (0,0,\psi)$ at $\mathbf{P} = 0$, the polar Hessian eigenvalues in the cited coupling convention are

$$K_\pm = 2a_1^P + 2(\zeta_{11} + \zeta_{12})\varphi^2 + 2\eta_{12}\psi^2 \pm \zeta_{44}\varphi^2 \tag{S41a}$$

$$K_3 = 2a_1^P + 4\zeta_{12}\varphi^2 + 2\eta_{11}\psi^2. \tag{S41b}$$

All three must be positive, and the structural sector must itself be stable. The smallest additional polar-Hessian shift must exceed $-2a_1^P(T)$, as plotted in Fig. 11d. The structural shift is displayed as an axis in Fig. 11 and assigned explicit conditional values in the motion scenarios of Fig. S6. A joint wall between polar and nonpolar structural phases additionally requires its interphase path and energy.

For fields $\mathbf{q} = (\mathbf{P}, \mathbf{\Phi}, \mathbf{\Psi}, \ldots)$ with constant positive kinetic and gradient tensors, multiplying each local traveling equation by $q_i'$ and summing gives $v\int \mathbf{q}'^T\mathbf{\Gamma}\mathbf{q}'d\xi = g_R - g_L$. This determines the sign of an isolated interphase transformation once the correct phase energies are known. Dynamic screening adds electrical work, and the identity does not determine the speed without the profile and kinetic integral.

### S10.3. What was evaluated and how to reproduce the figures

The figures retain the distinction between analytical evaluation and time-dependent field calculation. Figure 1 is a geometric schematic. Figure 5 evaluates exact scalar profiles at matched spontaneous polarization and far-domain curvature. Figure 7a gives the local corrugation boundary; panels 7b,c evaluate the coupled displacement–rotation spectrum, while 7d is the finite-depth field spectrum. Figure 9a,b gives analytical temporal and finite-amplitude structural diagrams, and 9c,d shows nonlinear periodic solutions of the declared reduction. Figure 10 evaluates the separate reductions detailed in S14. Figures 2, 3, 4c, 6, 8, and S6 contain the stated field or structural calculations. Figures S1–S5 retain the complete earlier comparisons and diagnostics.

Figure 11 evaluates Eqs. (41)–(43), (S40) and the Hessian condition from (S41) using Tables S1–S2. These are parameterized analytical slices. No surface diffusivity, charge relaxation time or MIEC composition is fitted by these plots. If a dimensional velocity is later displayed, $\Gamma$ must be given with provenance; if a wavelength or frequency is displayed, all screening and projected-mode coefficients must also be specified. Numerical sampling of a closed expression is distinguished from solving the coupled nonlinear field problem.

## S11. Prior coverage and the status of the present results

### Comparison with earlier physical cases

Table S3 summarizes the correspondence between earlier treatments and the present physical cases. Table S4 then records the more detailed scope of the analytical and numerical results. The classification distinguishes a related physical mechanism from a calculation of the same boundary-value problem.

Table S3. Earlier treatments and their physical correspondence.

| Physical problem or outcome | Representative earlier treatments | Scope of the correspondence |
|---|---|---|
| Exposed-surface ionic equilibrium | Stephenson–Highland [35] | Equilibrium chemical compensation and constrained ensembles; no finite-rate deformable-front spectrum |
| Surface-charge relaxation and spreading | Kalinin–Bonnell; Gaponenko and colleagues [23, 24, 40] | Experimental evidence and phenomenological kinetics; the specific coupled law still has to be stated |
| Prescribed surface charge | Kessler–Balke; Cao–Kalinin [55, 56] | Deformation or switching with fixed/prescribed charge; distinct from a mobile conserved sheet |
| Capacitor screening-retarded velocity | Eliseev and colleagues [3] | Analytical velocity laws with specified screening relaxation and local/averaged force closures |

| Physical problem or outcome | Representative earlier treatments | Scope of the correspondence |
|---|---|---|
| Local electrochemical exchange beneath an electrode | Morozovska and colleagues [36, 58] | Ferroionic states, hysteresis and domain evolution; does not supply every isolated-front stability result |
| Contact-supplied lateral transport | McGilly, Yudin, Tagantsev and colleagues [6, 7, 39] | Transport-limited propagation with contacts; different endpoint balance from Case III |
| Shape instability from delayed screening | Udalov, Korzhenevskii and Shur [8, 46] | Direct shape-instability precedents; full mode-by-mode equivalence requires the same screening law |
| Temporal oscillations with circuit feedback | Udalov and colleagues [47, 48] | Circuit-coupled velocity dynamics, not an isolated passive surface result |
| Conserved defects in the ferroelectric volume | Genenko–Lupascu; Vorotiahin–Morozovska–Genenko [64, 65] | Aging and domain reconstruction with bulk charge; no relocation of species to an external MIEC |
| Interphase cores and vector walls | Ishibashi–Suzuki; Dorfman and colleagues; Salje, Hlinka, Yudin and colleagues [11, 66, 68, 71, 74] | Structural mechanisms and specific material constraints; screening-controlled moving thresholds require their own calculation |

Two independent labels are needed. The first concerns correspondence with earlier work: the same physical case, a restricted limit, a related model with different boundary conditions, or an original treatment not yet verified in detail. The second concerns the present result: an exact balance or solution, a controlled asymptote, a prescribed-profile result, a conditional mode reduction, or a deferred field calculation. Combining these labels prevents a known structural mechanism from being called a solved moving-wall problem in every geometry.

Table S4. Coverage of the physical cases and status of the present results.

| Case/outcome | Earlier coverage | Present treatment and limits |
|---|---|---|
| O-I equilibrium and finite-rate motion | Chemical equilibrium [35]; static surface-wall structure [10] | Exact kernel and charge response; rigid force law; local/exposed-tanh monotonicity; finite-depth deformable continuation and sampled scalar spectra in S12 |
| o-II exchange plus lateral transport | Charge spreading/passivation evidence [40] | Whole-wall drag integral, weak-exchange asymptote and local finite-width weakening criterion; finite-depth spectrum computed for the stated numerical references |

| Case/outcome | Earlier coverage | Present treatment and limits |
| --- | --- | --- |
| O-III conserved mobile sheet | Equilibrium inventory comparisons [35]; related TGS observations [42, 43, 44] | Exact jump constraint, long-wave impedance, sharp-interface similarity and finite-inventory asymptotes; closed finite-depth evolution and selected perturbations computed in S12 |
| T with historical relaxation | Capacitor velocity calculation [3] | Traveling screening solution and local/averaged/projected force distinction re-derived; complete bending branch not inferred |
| T-I electrochemical exchange | Ferroionic equilibrium and dynamics [36, 58] | Shared causal reduction; exact local monotonicity; pair exclusion under its stated assumptions |
| T-II mixed transport | Contacted propagation is related [6, 7, 39] | Finite-depth positive-slope branch, corrugation band, and resolved-pair force and matrix; large-field scalar and vector evolution in S13; contact boundary conditions remain distinct. |
| T-III conserved mobile sheet | Fixed-charge equilibrium is related [35] | Species jump constraint and inventory-dependent equal-background pulse family; general closed transient not solved |
| Blocking external MIEC | Diffuse-charge/MIEC transport and semiconductor screening [59, 60, 61, 62, 63] | Exact homogeneous linear admittance and long-distance charge-conservation constraint; nonlinear wall-specific response remains open. |
| Defects inside the ferroelectric | Genenko–Lupascu and Vorotiahin et al. [64, 65] | Kept as a distinct comparison; no claim that external-MIEC or surface-sheet relocation preserves their solution |
| Corrugation | Udalov et al. [8, 46] | Local graph spectrum and finite-depth field spectrum; nonlinear T-II evolution in S13. The 2017 period-dependent criterion is a direct predecessor; details of the 2018 full spectrum remain unverified. |
| Temporal oscillation | Circuit-coupled velocity oscillations [47, 48] | Passive-equilibrium exclusions and conditional driven-mode Hopf boundaries; supercritical periodic branch computed for the declared cubic finite-mode extension |

| Case/outcome | Earlier coverage | Present treatment and limits |
|---|---|---|
| First-order wall structure | Ishibashi–Suzuki and Dorfman et al. [66, 67, 68]; Lajzerowicz [69] is a historical lead | Exact scalar profile and interphase front; charge-dependent translation/width distinction; original Lajzerowicz full model not used for an equation-level claim |
| Vector wall structure | Salje, Hlinka, Yudin and colleagues [11, 71, 72, 73, 74, 75, 76] | Exact illustrative rotation modes and conditional coupled spectra; reference vector branches and material motion in S12; vector xy evolution in S13. Coupled material-specific structural and elastic dynamics remain open. |
| Electrical wall forces and arrays | Mokrý–Tagantsev–Fousek; Kopal et al. [5, 9] | Force and restoring-force precedents retained; a domain array is not identified with an isolated charge-conserving front |

"Not verified" indicates an access or evidence limit, not an absent result. The comparison makes no universal priority claim from the audited set.

The original first-order scalar wetting and non-Ising-wall literature is cited for the corresponding structural mechanism. The separate contribution of the present synthesis is to place those mechanisms alongside the actual charge balance and the correct moving or transient reference state. Frozen ions, mobile conserved ions, contacted electronic transport, electrode charge and mobile bulk defects remain explicitly different objects. The same distinction is used for the empirical interpretation in Section 2.

## S12. Numerical models, results and reproducibility

### S12.1. Units and the finite-depth field problem

Calculations N1, N2, the bounded-population part of N3 and the scalar stability part of N6 solve polarization in the finite $y, z$ cross-section, electrostatics in the film and exterior, and the surface charge equation. The reference units set $P_s = 1$, the quadratic coefficient magnitude to 1, $\Gamma = 1$ and $g_y = 0.5$, so the equilibrium quartic wall is $p = -\tanh y$ and its width parameter is 1. Numerical permittivities include the factor $\epsilon_0$ and the chosen energy scale. They are reduced coefficients, not relative dielectric constants. The numerical bias $e$ enters as $-ep$ in the polarization free energy. Both electrodes, where present, are held at zero potential after this thermodynamic bias has been specified. The charge reservoir has zero electrochemical offset. Consequently the bias acts on polarization alone. An experimentally imposed electrode voltage would also change the charge electrochemical potential and requires the electrical protocol of Eqs. (6) or (15).

For a straight moving base we solve the stationary equations in $\xi = y - vt$, with unknown speed $v$ and the phase condition that the depth-averaged polarization at $\xi = 0$ is zero. At the lateral ends the supplied cases approach their reservoir-equilibrated far-domain polarization and charge. Their electrochemical potential is zero. Polarization has its natural normal derivative at the upper and lower interfaces, the bottom potential is zero, and the upper electrostatic response is the exposed or finite-gap admittance. The exterior is eliminated exactly for each eigenmode of the discrete lateral Laplacian. The wall is isolated along $y$; no periodic wall pair is introduced.

The finite-depth polarization equations are $\partial_t p = -f_L{}'(p) + e + \sum_j g_j \partial_j^2 p - \partial_z \phi$. The surface law is $\partial_t c = -(\Lambda - M\nabla_\parallel^2)[f_s{}'(c) - \phi_s]$. For the unbounded linear reference $f_s{}' = 0$; this is the complete-equilibrium-compensation limit, not a finite-site adsorption model. For bounded adsorption, $f_s{}' = \theta \operatorname{atanh}(c/c_{\max})$. This signed coordinate represents a finite interval of net compensation; two-species neutral population modes have been eliminated. The steady far states are solved from $f_L{}'(p) + G_0(p-c)/h = e$ and $f_s{}'(c) = G_0(p-c)$, including the adsorption nonlinearity.

We use centered second differences in $y$, cell-centered polarization in depth, and potential on the depth faces. Let $B$ be the discrete polarization-divergence source, $C$ restrict potential to the surface, and $K(k)$ be the positive Poisson matrix including the exterior admittance at $Q = \sqrt{q^2 + k^2}$. For deviations from the far-state interpolation, $K\phi = Bp - C^T c$. The polarization electric field is $E\phi$ with $E = -B^T/\Delta z$. This adjoint pairing preserves the electrostatic work in the discrete equations and includes the bound charge generated by depth bending. The exterior admittance is $\epsilon_e Q$ or $\epsilon_d Q \coth(Qd)$, with the latter set to $\epsilon_d/d$ at zero wave number. Newton iteration uses the analytic coupled Jacobian and line search. All reported branches satisfy the complete field residual, rather than only a projected force equation.

**S12.2. Linear spectra and time evolution**

For a perturbation proportional to $\exp(st + ikx)$, the dynamic variables are the complete polarization and surface-charge profiles. The potential is an algebraic variable. Writing $R_k = -\Lambda + M(\partial_y^2 - k^2)$, the discretized eigenproblem is

$$s\begin{pmatrix}\delta p\\ \delta c\\ 0\end{pmatrix} = \begin{pmatrix} L_p - g_x k^2 - f_L{}'' + vD_y & 0 & E\\ 0 & R_k f_s{}'' + vD_y & -R_k C\\ -B & C^T & K(k)\end{pmatrix}\begin{pmatrix}\delta p\\ \delta c\\ \delta\phi\end{pmatrix}. \quad \text{(S42)}$$

The last row enforces Poisson's equation. Sparse shift-invert evaluation returns the eigenvalues near the stated spectral shift; residuals are checked in the original generalized eigenproblem. At representative coarse-grid points we also eliminate potential and calculate all dynamic eigenvalues. This independent full-spectrum check detects growing modes outside the neighborhood used for sparse continuation. The complete coarse-grid matrices have 1045 dynamic variables. The analytic steady Jacobian agrees with central finite differences to a relative error of approximately $3.2 \times 10^{-11}$. Translational eigenvalues within discretization error of zero are tracked separately. A sample of wave numbers and fields does not prove global stability over the entire continuous parameter space.

N2 uses a closed finite surface with zero lateral flux of charge and natural lateral polarization conditions. The discrete lateral electrostatic boundary is reflecting as well. We first relax an equilibrium wall with its compensation and then apply a bias step. A stabilized semi-implicit update advances polarization and surface charge while enforcing Poisson's equation at every step. The charge equation is conservative to solver precision. Perturbations at specified transverse wave numbers are advanced with the linearization of the same update on the evolving base. The reported position amplitude is the projection $-\langle \delta p, \partial_y p\rangle / \langle \partial_y p, \partial_y p\rangle$. The $k = 0$ charge perturbation has zero integral. For $k > 0$ its sinusoidal variation along the front has zero transverse average. The measured decay is for these prescribed displacement perturbations; an optimal transient amplification over all possible initial perturbations has not been computed.

### S12.3. Parameters and convergence of the surface calculations

N1 uses $\epsilon_f = 10$, $\epsilon_e = 1$, $\Lambda = 0.05$, $M = 0$, $\chi = 0$, $g_x = g_y = g_z = 0.5$, $a = -1$, $b = 1$ and $c_6 = 0$. The $h = 5$ branch uses 191 lateral points, 20 depth cells and length 40; the $h = 20$ branch initially uses 191 points, 24 cells and length 60. Eight bias values span 0–0.10. Spectra are sampled at $k = 0,0.08,0.20,0.50$. At $e = 0.06$, the refined thick-film calculation gives $v = 0.0727545$ and top-to-bottom displacement 0.9848. Increasing the domain from 60 to 80 while refining the lateral grid changes its speed by less than $5 \times 10^{-6}$ relatively. The thin-film calculation gives $v = 0.0404370$ with length 48 and 255 lateral points. Its depth displacement is approximately 0.0951. Residual norms are typically below $10^{-9}$. In the $f_s'' = 0$, local-exchange limit, high-wave-number charge-only modes become slowly relaxing; no mesh-independent spectral gap is claimed.

N2 uses $L = 40$, $h = 1$, $M = 10$, $\Lambda = 0$, $e = 0.01$, 127 lateral points and four depth cells, with time step 0.2 through $t = 5000$. The refined comparison uses 255 points, eight depth cells and time step 0.1 through $t = 2500$. At this common time the two displacements are 1.99689 and 1.99859, a difference of 0.085%. The far polar amplitudes remain near their zero-field values; this is not a near-spinodal test. The leading arrest estimate follows from the uniform residual field: $\langle p \rangle - \langle c \rangle = \epsilon_f e$ and $\langle p \rangle \simeq 2P_s X/L$, giving $X_\infty = \epsilon_f eL/(2P_s)$. The largest charge-integral error is $1.2 \times 10^{-11}$. A separate wider-domain run uses $L = 200$, 767 lateral points, four depth cells and the same $\Delta t = 0.2$ through $t = 3000$; it gives $X = 4.9943$, still below its eventual inventory estimate of 10, with charge error below $4.8 \times 10^{-11}$. The finite sample has only a limited interval between intrinsic polarization relaxation and boundary-controlled arrest; comparison with the infinite-domain square-root asymptote is therefore shown without fitting it as an exact finite-sample law.

For the bounded-population electrode branch in N3, the inputs are $L = 40$, $h = 5$, $d = 0.5$, $\epsilon_f = 10$, $\epsilon_d = 1$, $\Lambda = 0.05$, $M = 0.3$, $\theta = 0.015$ and $c_{\max} = 1.3$. The original branch uses 191 lateral points and 16 depth cells. N6 changes only $g_x$ to 0.005 and evaluates the same base at $e = 0.06$. The fastest mode is obtained by scalar maximization of the leading eigenvalue; the high-wave-number cutoff is obtained by root finding. Refinement from $(n_y, n_z, L) = (95,10,32)$ to $(255,24,48)$ changes the growth rate at $k = 0.75$ from 0.00291050 to 0.00292084. The computed maximum is $s = 0.00292286$ at $k = 0.74166$ and the cutoff is 1.32620. These numbers describe initial exponential growth. Nonlinear evolution of the selected branch is reported in S13; a unique converged asymptotic pattern remains undetermined.

### S12.4. Evaluation of the Eliseev velocity relations

The historical relaxation law and electrostatic field are re-derived in S4. We evaluate its Fourier quadrature for a tanh wall, using the rapidly decaying form factor $f_w(q)$ to control the large-wave-number tail. The junction relation uses the field at $\xi = 0, z = 0$; the comparison replaces its depth weight by $1/h$. The electric-field drive is $E_{\text{app}} = v/\mu + |E_{\text{ret}}|$. The prescribed parameter set is $P_s = w = \mu = 1$, $\tau = 2000$, $\epsilon_f = 10$ and $\epsilon_d = 1$. The velocity grid spans $10^{-6}$–$10^{-1}$. Derivatives identify candidate sign changes, which are refined by scalar root finding.

At $d = 1$, the folds are $(0.0021804,0.0560704)$ and $(0.0076366,0.0538906)$ for $h = 20$, and $(0.0017579,0.0538263)$ and $(0.0128452,0.0448725)$ for $h = 100$. The $h = 5$ branch is monotone over the sampled interval. At fixed $h = 100$, changing the gap changes both the retarding-field magnitude and its velocity dependence; a negative-slope interval is present for several intermediate gaps and absent for $d = 10$ in this slice. An exposed comparison is obtained by taking the exterior admittance $\epsilon_e|q|$ while retaining the prescribed relaxation law. This comparison tests the field kernel and does not equate the historical law with passive local electrochemical exchange.

### S12.5. Scalar and vector structural calculations

The structural time evolutions solve a one-dimensional long-wave capacitor limit on a closed interval of length 80, with $g = 0.5$, electrostatic coefficient $K = 0.3$, charge diffusivity $D = 0.5$ and unit polarization mobility. The equations are $p_t = gp_{yy} - \partial_p f - K(p - c)$, $c_t = D(c - p)_{yy}$ and, when present, $(a_\perp)_t = g(a_\perp)_{yy} - \partial_{a_\perp} f$. Both lateral ends are reflecting. The initial profile is $p = -P_s[\tanh(y + 4) + \tanh(y - 4)]/2$ with $c = p$. The rotated-domain calculation adds $a_\perp = \sqrt{1 - \delta}[\tanh(y + 4) - \tanh(y - 4)]/2$. Reflection symmetry fixes the center while allowing the core to expand or contract. No external time-dependent force is applied.

The scalar coefficients $(a, b, c_6)$ are $(-1,1,0)$, $(0.94, -4,3)$ and $(1.03, -4,3)$. The latter two have coexistence at $a = 1$. At $a = 0.94$ the central local minimum has higher energy and a bound pair forms. At $a = 1.03$ the central phase has lower energy and grows into metastable polar domains. The rotated-domain potential is specified in Section 7.3, with $C = 2$ and $\delta = -0.06$. Its in-plane variant is favored. The time discretization uses 801 cells and $\Delta t = 0.025$; the refined run uses 1601 cells and $\Delta t = 0.0125$. At $t = 300$, the coarse/refined half-amplitude widths are 1.1003/1.0989, 2.2220/2.2213, 12.4786/12.5461 and 15.5024/15.6628, respectively. All qualitative transformations persist. The largest refined charge-integral error is approximately $1.1 \times 10^{-11}$. This convergence supports the structural outcomes and roughly percent-level late-time expansion widths.

Equilibrium vector branches are calculated separately for $f = f_L\left(\sqrt{p^2 + a_\perp^2}\right) + \kappa_\perp a_\perp^2/2$, with complete compensation and a charge-free transverse component. For the sixth-order examples $c_6 = 1.5/(1 + \rho)$, $b = 2c_6(\rho - 2)/3$ and $a = -b - c_6$, which fix $P_s = 1$ and far-domain curvature 2. The interval is 120, with 1201 interior points and $g = 0.5$. Energy minimization is followed by phase-fixed Newton refinement. The complete two-component Hessian gives the low eigenvalues and stability of the relaxed branches. Residuals are below $8 \times 10^{-10}$. The quartic linear rotation threshold 0.49986 agrees with the exact 0.5. At $\rho = 0.3$ and 0.1 the thresholds are 0.04317 and 0.005746. Finite-amplitude rotated branches also exist where the Ising Hessian is positive. Energy differences change sign along those branches, demonstrating a first-order structural competition in these reference potentials. The last retained large-amplitude point before a relaxation jump is not identified as a precisely continued saddle-node.

### S12.6. Nonlinear Hopf calculation

The nonlinear extension used in Fig. S5 is

$$A' = -pA + GR - A^3 \tag{S43a}$$

$$R' = A - R - UI \tag{S43b}$$

$$I' = UR - I. \tag{S43c}$$

Its linear characteristic polynomial is Eq. (37). At $p = 2, G = 10, U = \sqrt{6}$, it factors as $(z + 4)(z^2 + 1)$. The complex pair crosses transversely as $U$ changes, with $d\mathrm{Re}z/dU = -0.2881753$. The first Lyapunov coefficient is evaluated with unit Euclidean norm for the right eigenvector $q$ and adjoint normalization $p_H^\dagger q = 1$. With the standard convention $l_1 = (2\omega_H)^{-1}\mathrm{Re}\langle p_H, C(q, q, \bar{q})\rangle$, the only nonzero cubic derivative is $C_{111} = -6$, and $l_1 = -0.1470588$. There is no quadratic contribution. The negative value gives a supercritical local periodic branch on the side where the equilibrium loses stability.

Numerical integration initializes periodic solutions, and a shooting method refines the state and period with a phase condition. Integration of the variational equations over one period gives the monodromy matrix and Floquet multipliers. At $U = 2.40$, the shooting residual is below $10^{-12}$, the period is 5.88723, and the nontrivial multipliers are 0.85587 and $7.9 \times 10^{-12}$. Nearby stable cycles are also continued at $U = 2.445, 2.43, 2.35$ and 2.30. The model has an imposed translating reference state; it demonstrates a possible nonlinear screening–structural oscillation rather than a material-specific field prediction.

The earlier linear example $G = 12, p = 2, U = 3$ remains a valid nonzero-frequency crossing when $U$ is varied. Varying $p$ at that fixed $G, U$ is tangent to the Hopf locus and does not establish transversality. For the simple cubic extension used above, that particular point also gives a degenerate first Lyapunov coefficient within numerical precision. The nonlinear demonstration therefore uses the nondegenerate $G = 10$ point. The distinct width-source construction in S9.5 retains derivatives consistent with its source function; no incompatible width-Hopf frequency from a prior analysis is transferred to the manuscript.

**S12.7. Dimensional material scenarios and limitations**

For each material path let $K_{\mathrm{act}} = f''(P_s)$ include the stated local coherent or structural constraint and define $K_* = K_{\mathrm{act}}/2$, $l_* = \sqrt{2g/K_*}$, $E_* = K_* P_s$ and $t_* = 1$ ns. The computational variables are $p = P/P_s$, $y/l_*$, $t/t_*$, $e = E_{\mathrm{eq}}/E_*$ and $v_* = l_*/t_*$. Thus $\Gamma = K_* t_*$ is an assumed kinetic coefficient. These units differ by fixed factors from the analytical normalization in Section 8.4 and are stated explicitly to prevent a change of convention from being mistaken for a physical effect. Reduced permittivities are $\epsilon_0 \epsilon_b K_*$. We use $h = 20$ nm, $d = 0.5$ nm and $\epsilon_d = 5$, with background relative permittivity 7 for $BaTiO_3$ and 10 for $BiFeO_3$. The nominal long-wave coefficients are $r_0 = t_*/(1\,\mu s) = 10^{-3}$ and $D_* = 10^{-12} t_*/l_*^2$, implemented as $\Lambda = r_0/G_0$ and $M = D_*/G_0$. The surface chemical curvature is zero in these material scenarios.

The $BaTiO_3$ scalar coherent correction is Eq. (S40). For a compatible uniform Bloch perturbation along [100] of a [010]-normal wall, set $\mathbf{q}_z = (q_{12}, q_{12}, q_{11})$, $\mathbf{q}_x = (q_{11}, q_{12}, q_{12})$ and $Q_N = C_N^{-1} - \hat{y}\hat{y}^T/C_{11}$. The extra coefficient of $a_\perp^2$ is $(\mathbf{q}_x^T Q_N \mathbf{q}_z)(p^2 - P_s^2) + q_{44}^2 p^2/(2C_{44})$. The shear convention is $f_{xz} = C_{44} e_5^2/2 - q_{44} e_5 P_x P_z$, with engineering shear $e_5 = 2e_{xz}$. Tangential shear is held at its far value for this planar compatible limit. The resulting transverse curvature is used with the same scalar base; it is not replaced by a fully relaxed bulk curvature. Nonlocal elastic relaxation near a free surface remains a separate extension.

For $BiFeO_3$ the fixed isotropic structural shift gives the transverse polar curvature about the [111] path

$$K_\perp(P) = 2(a_1^P + h_R) + 4a_{11}^P P^2 + \frac{10a_{111}^P + 4a_{112}^P - (2/3)a_{123}^P}{3} P^4. \tag{S44}$$

This expression is the diagonal-minus-off-diagonal eigenvalue of the cubic polar Hessian at $P_i = P/\sqrt{3}$. It was checked by direct differentiation. The uniform Bloch-mode operator is $L_p - K_\perp/K_* + vD_y$. A transverse mode at nonzero wave number generally produces additional bound charge and is outside this separate uniform-mode check. The fixed structural shift $h_R = 0$ or $2 \times 10^8$ J m $C^{-2}$ is a parameter slice, not an inferred octahedral amplitude. The gradient coefficient $g = 10^{-10}$ J $m^3$ $C^{-2}$ is an assumed scenario. None of these calculations claims a self-consistent $BiFeO_3$ polar–octahedral–magnetic wall.

Material branches use 191 lateral points and 32 depth cells over a reduced interval of length 40, at $e = 0, 10^{-4}, 3 \times 10^{-4}, 10^{-3}, 3 \times 10^{-3}$. The Newton tolerance is $2 \times 10^{-11}$ because the small charge rates require a stricter absolute residual. At 300 K and $e = 0.003$, refinement to 383 lateral points and 96 depth cells over length 50 changes the $BaTiO_3$ speed from 3.76899 to $3.76879\,\mu$m $s^{-1}$ and the $BiFeO_3$ speed from 21.1435 to $21.1408\,\mu$m $s^{-1}$. Estimated depth displacements are approximately 0.009 nm and 0.04 nm and converge less tightly than velocity. The subnanometer intrinsic reference widths also caution against treating the continuum profiles as atomistically resolved structures. Scalar spectra are sampled at reduced $k = 0.15$; the uniform transverse rates are negative for all listed material branches. These checks define the limited stability claim in Fig. S6.

### S12.8. Computational scope

The accompanying numerical archive contains the actual solver scripts, input parameters, stored profiles, spectra, time histories, continuation data and figure-generation code. NumPy and SciPy provide the sparse linear algebra, quadrature, optimization and time integration; their run versions are recorded in the archive. The calculations are deterministic apart from a fixed-seed Jacobian perturbation check. Code for the analytical figures and the material coefficient tables is included so that the figures can be regenerated. Numerical evaluation of an analytical integral, a coupled field solution and a finite-mode periodic orbit remain separately identified.

N1 establishes the specified deformable supplied walls. N2 establishes conserved transient motion and finite-size arrest, with selected perturbation decay. N3 evaluates the historical thickness/gap curves and a distinct bounded-population finite-depth branch. N5 establishes the structural transformations and reference vector branches. N6 establishes coupled scalar corrugation and a separate nonlinear reduced-model Hopf branch. N7 supplies declared dimensional material scenarios. These results do not imply that every physical case has been solved for arbitrary charge kinetics or material coefficients.

The nonlinear external-MIEC initial-value problem and the bounded multispecies conserved traveling-pulse problem remain outside the completed calculations. Further extensions include optimal transient amplification, near-spinodal conserved evolution, and fully coupled material-specific vector, elastic, and structural fields. The nonlinear corrugation and vector evolution now included in S13 address a selected second-order T-II reference. Their late states have not undergone an independent final-time convergence study. No quantitative material oscillation frequency is assigned without the corresponding mode couplings and kinetic coefficients.

## S13. Large-field scalar and vector evolution

### S13.1. Equations and preparation

The large-field calculations use the top-electrode geometry with local exchange and lateral transport (T-II). The film occupies $-h < z < 0$, a dielectric gap occupies $0 < z < d$, and the top and bottom electrodes are grounded. The driving $e = 0.06$ is a thermodynamic bias acting on polarization, not an applied electrode voltage. The scalar potential is $f(p) = -p^2/2 + p^4/4$. The two-component potential is

$$f(p, a) = -p^2/2 + p^4/4 + (\delta - 1)a^2/2 + a^4/4 + \beta p^2 a^2/2 \qquad \text{(S45)}$$

with $p = P_z/P_s$, $a = P_x/P_s$, $\delta = 0.02$, and $\beta = 2$. The third polarization component and elasticity are omitted. The vector divergence contributes to the electrostatic field. Polarization has natural boundary conditions through the thickness. Surface compensation obeys the finite-capacity electrochemical law of the continued T-II branch, with $\mu_c = \theta_c \text{atanh}(c/c_{\max}) - \phi_s$, local exchange $-\Lambda\mu_c$, and lateral flux $-M\nabla_{\parallel}\mu_c$. Here $\theta_c$ denotes the chemical coefficient, to distinguish it from the dimensionless front stiffness $\theta$ used in Fig. 10a. The signed physical free charge is $\sigma = -c$.

The common parameters are $h = 5$, $d = 0.5$, $\epsilon_f = 10$, $\epsilon_d = 1$, $g_x = 0.005$, $g_y = g_z = 0.5$, $\Lambda = 0.05$, $M = 0.3$, $\theta_c = 0.015$, and $c_{\max} = 1.3$. The lateral lengths are $L_x =$ 33.8872312064 and $L_y = 80$. The mesh has 256 periodic points along $x$, 319 interior points along $y$, and 12 depth cells. The $y$ boundaries hold the equilibrated far states; the interfacial charge is not globally conserved. The scalar time step is 0.4 and the final time is 4800. The vector time step is 0.1 and the final time is 2400. The reference scales are $P_s = 1$, $\xi = 1$, and $\tau = 1$ in the simulation convention, for which the zero-bias quartic wall is $-\tanh(y)$. These simulation units should not be confused with $\xi_P = \sqrt{g_y/K_b}$ and $\tau_P = \Gamma/K_b$ in Section 8; for the quartic reference $K_b = 2$, $g_y = 0.5$, and $\Gamma = 1$, so $\xi_P = 0.5$ and $\tau_P = 0.5$. The simulation bias multiplies $-ep$ and is scaled by $|a|P_s$; its value 0.06 corresponds to $E_d/(K_b P_s) = 0.03$ in Eq. (40). The numerical permittivities are reduced electrostatic coefficients that include the vacuum-permittivity and energy scales, not relative dielectric constants.

Both models start from the computed straight traveling wall and its charge wake. With $P = 16.9436156032$ and $k_0 = 2\pi/P$, the initial displacement is

$$s_0(x) = 0.06\cos(2k_0x) + 0.021\cos(k_0x + 0.7) + 0.012\cos(3k_0x + 1.6) \quad \text{(S46)}$$

$$s(x) = s_0(x) + 0.006\cos(2\pi x/L_x + 0.31). \quad \text{(S47)}$$

The vector seed is $a(x, y) = [0.004\cos(2\pi x/P + 0.2) + 0.0004\cos(2\pi x/L_x + 1.1)]/\cosh[y - s(x)]$. The weak new fundamental avoids exact repetition of two copies of the original smaller field. Opposite rotation signs are deliberately seeded. Their final arrangement is preparation dependent.

**S13.2. Numerical method and verification**

The production solver uses a Fourier representation along $x$, a sine representation for deviations from the prescribed $y$ boundary states, and a finite-depth discretization. A second-order semi-implicit backward-differentiation scheme advances polarization and charge. The self-consistent electrostatic operator retains the full finite-depth stack response. A moving coordinate follows the wall using explicit advection and a feedback velocity updated every 10 time units; no remeshing or interpolation is applied during evolution. Checkpoints retain both time levels and the nonlinear history required for exact restart. Snapshots and metrics are recorded every 20 time units, and checkpoints every 200 time units.

The compact production implementation agrees with the preceding scalar and vector operators to below $3 \times 10^{-15}$ in the physical fields over 30 updates. Restart tests reproduce uninterrupted evolution to machine precision. Earlier refinements gave roughness differences of 0.17% for a halved scalar time step at $t = 400$, 0.05% for a halved vector time step at $t = 200$, and 0.40% for a finer vector mesh at $t = 400$. These checks verify the implementation and the earlier stages of evolution; they do not establish convergence of the present late-time morphologies.

The finite charge capacity is checked at every update. The maximum absolute compensation is approximately 1.186 in the scalar run and 1.280 in the vector run, both below 1.3. A sampled integral balance includes reservoir exchange, lateral boundary flux, and moving-coordinate advection. Its root-mean-square discrepancy relative to the exchange contribution is approximately 0.027% and 0.051%, respectively. The balance is evaluated from saved records and is a diagnostic of these runs, not an assertion of exact conservation in the exchange model.

**S13.3. Morphology and interpretation**

The surface front position is obtained by interpolating the descending zero crossing of $P_z$ at each $x$. Roughness is its standard deviation after removing the mean position. Its discrete Fourier amplitudes resolve the available periodic wavelengths. In the scalar trajectory, mode 2 first exceeds mode 4 at the saved time 4260. The endpoint retains several spatial harmonics. The vector endpoint differs substantially between the two half-fields and contains a broad rotated region; the final median magnitude in the low-$P_z$ mask is approximately 0.905. Each final snapshot has one descending front crossing in each sampled column, and no detached positive or negative domain containing at least eight image cells under the declared threshold test $|P_z| > 0.3P_s$. This diagnostic does not exclude subgrid features.

The archived trajectories contain the upper-cell and depth-averaged polarization, compensation, surface potential, modal endpoint, and coordinate-origin record. Movies, full snapshots, and the executable Colab accompany the previously supplied simulation archive. Figure 8 selects the observations needed for the manuscript. The scalar and vector durations differ, and their later patterns are not independently converged asymptotic states. These two T-II trajectories do not locate first-order, exposed-surface, or MIEC instability boundaries.

## S14. Evaluation of the reduced-variable summary

### S14.1. Mixed transport and strictly conserved motion

Figure 10a evaluates $f(u) = \epsilon u + u/\sqrt{1+u^2}$ at $\epsilon = 0.1$ and maps $\theta_c(u) = u^2/\left[2(1+u^2)^{3/2}\right]$ onto the driving axis. The derivative of $f$ is positive, so the conversion is single valued. The maximum threshold is $1/\left(3\sqrt{3}\right)$. Here $B = H_0 \Delta P^2$, $v_s = 2\sqrt{Dr_0}$, $\epsilon = \eta v_s/B$, and $\theta = T\sqrt{r_0/D}/B$. Varying one physical rate generally changes more than one reduced coordinate; a vertical or horizontal cut holds the other declared reduced parameters fixed. The map applies to the local unresolved-front reduction, not to the complete nonlinear T-II simulation.

Figure 10b solves $f = \left(\sqrt{\pi}/2\right)\lambda \exp(\lambda^2)\operatorname{erfc}(\lambda)\operatorname{erfc}(-\lambda)$ for $0.005 \le f \le 0.42$ and evaluates $X/\sqrt{Dt_*} = 2\lambda\sqrt{t/t_*}$. Scaled complementary error functions avoid overflow. The maximum inversion residual is below $5 \times 10^{-13}$. This is the charge-dominated finite-film long-wave similarity problem with negligible boundary influence. It is not an interpolation through the crossover to finite-size arrest. The initial transient and arrest depend on wall friction, conserved inventory, specimen size, and preparation.

### S14.2. Central-state and rotational boundaries

For Fig. 10c, write $f'(p) = ap + bp^3 + cp^5$ with $a = \nu - 1/2$, $b = 1/2 - 2\nu$, and $c = \nu$. A central minimum exists at zero bias only for $a > 0$. Its loss at positive bias is determined by the smaller positive root $q$ of $a + 3bq + 5cq^2 = 0$, giving $p_{\rm sp} = \sqrt{q}$ and $e_{\rm sp} = ap_{\rm sp} + bp_{\rm sp}^3 + cp_{\rm sp}^5$. The negative-bias boundary follows by symmetry. This tests a scalar homogeneous minimum. It does not test interface binding, transverse stability, compatible strain, or additional structural order parameters. The line $\nu = 3/4$ marks equal scalar phase energies only at zero bias.

For Fig. 10d, the mode stiffness is $r_A + c_A k^2 + d_A k^4$, with $d_A > 0$. Choose any fixed reference wave number $k_*$ and define $q = k/k_*$, $\bar{r}_A = r_A/(d_A k_*^4)$, and $\bar{c}_A = c_A/(d_A k_*^2)$. Uniform instability occurs for $\bar{r}_A < 0$. For $\bar{c}_A < 0$, a finite-wave-number mode is unstable when $0 < \bar{r}_A < \bar{c}_A^2/4$; the first mode has $q^2 = -\bar{c}_A/2$. Below zero stiffness, uniform and modulated modes may both grow. The colors identify unstable linear sectors, not their final nonlinear attractors. A negative effective gradient coefficient must be calculated from coupling or a controlled reduction, not inserted as a property of the uncoupled polarization gradient energy.

### S14.3. Width bistability and the blocking MIEC

Figure 10e evaluates the prescribed-speed separated-front law $\dot{s} \propto -\delta_w + g(s)$, where

$$g(s) = e^{-s} + b_d\left[e^{-s/l_b} - e^{-s/l_f}\right] \tag{S48a}$$

$$l_b = 50 \tag{S48b}$$

$$l_f = 10. \tag{S48c}$$

The two fold curves satisfy $g'(s) = 0$ and $\delta_w = g(s)$. Roots with $g'(s) < 0$ are stable within this one-coordinate equation. At $b_d = 0.01$ and $\delta_w = 0.00485382$, the roots are 6.70664, 11.95181, and 32.17701. The first and third are stable. Here $s = \ell/\xi_0$, $\delta_w = \Delta_0/A$, and $b_d = G_0 P_s^2 v/\left(2hA\sqrt{v^2 + 4Dr_0}\right)$ for the complete-compensation local-capacitance outer model. The positive scale $A$ is the short-range interface-repulsion pressure and $\Delta_0$ is the central-phase cost. The wake lengths are $\left(\sqrt{v^2 + 4Dr_0} \pm v\right)/(2r_0\xi_0)$. Fixing both wake lengths and speed defines an illustrative parameter slice; at fixed applied force, the center and separation equations must be solved together. The reduction requires resolved elementary interfaces and a separation evolution slow compared with relaxation of the retained screening wake.

Figure 10f evaluates the long-distance Ohmic term of Eq. (25). For an arbitrary potential scale $\Phi_*$, the dimensionless required current is $j_* = \Delta P v/(\pi\sigma_m\Phi_*)$ and the potential drop is $j_*\ln(R/a_{\text{in}})$. The curves are evaluations of this outer asymptote, not nonlinear MIEC simulations. Their divergence demonstrates why a fixed local charge cloud cannot describe the strict isolated moving junction. Nonlinear screening or unequal diffusivities change the inner response but do not remove the conservation requirement for the outer transport geometry.

## S15. Additional figures and dimensional motion scenarios

The figures in this section retain the detailed comparisons and numerical diagnostics removed from the main narrative. They are complementary to the consolidated main-text figures. Analytical reductions, full-field calculations, and material scenarios remain identified separately in their captions.

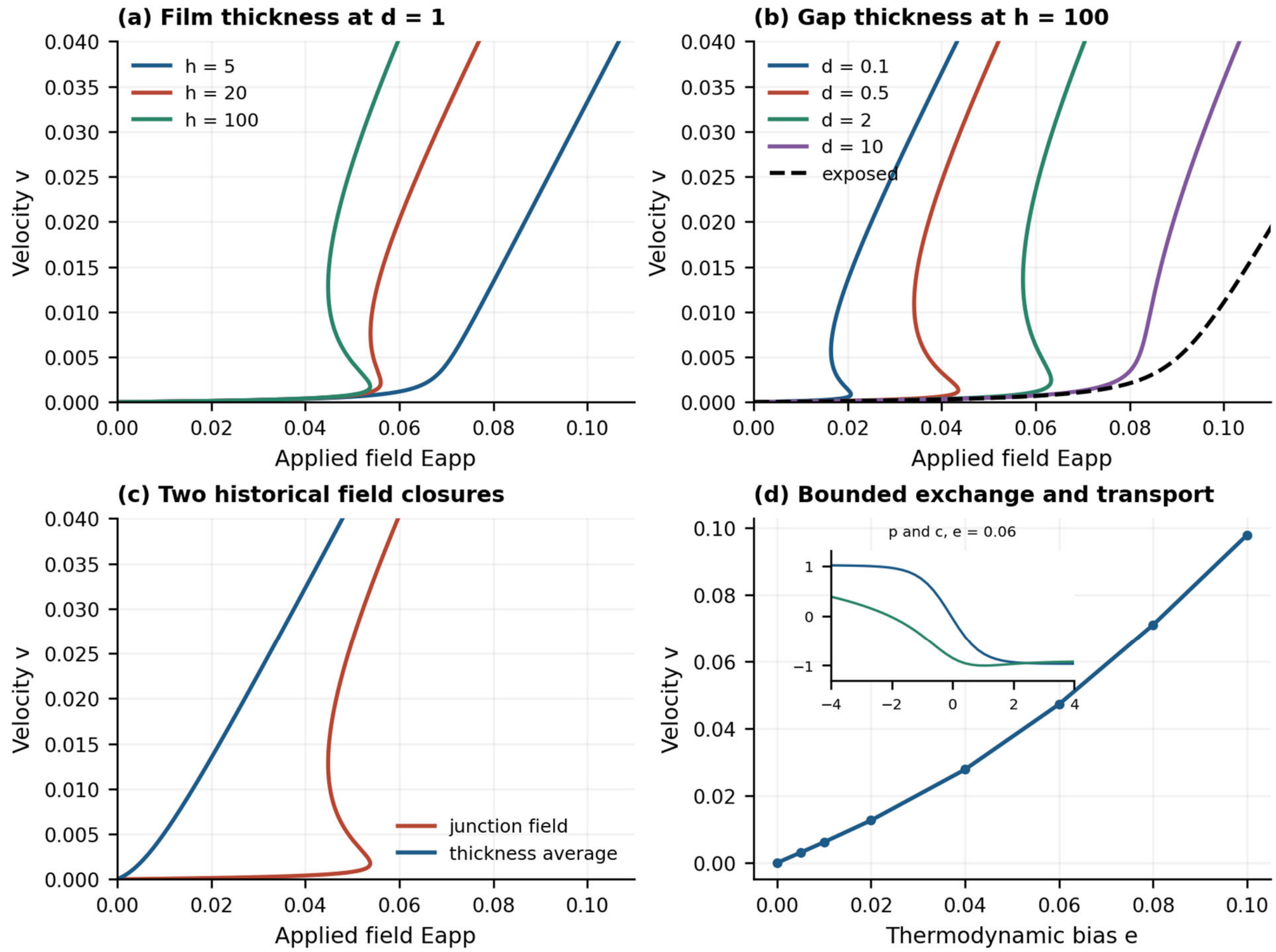

Figure S1. Electrode-controlled screening. (a) Eliseev local-junction velocity curves for different film thicknesses at $d = 1$. (b) Gap dependence at $h = 100$, with an exposed-surface comparison using the same prescribed relaxation law. (c) Junction and thickness-averaged closures at $h = 100, d = 1$. (d) Coupled finite-depth branch with bounded adsorption, together with its surface polarization and charge in the inset. Panels (a)–(c) use the historical relaxation law and applied-field closure; panel (d) solves the different passive exchange-plus-transport model with a thermodynamic bias.

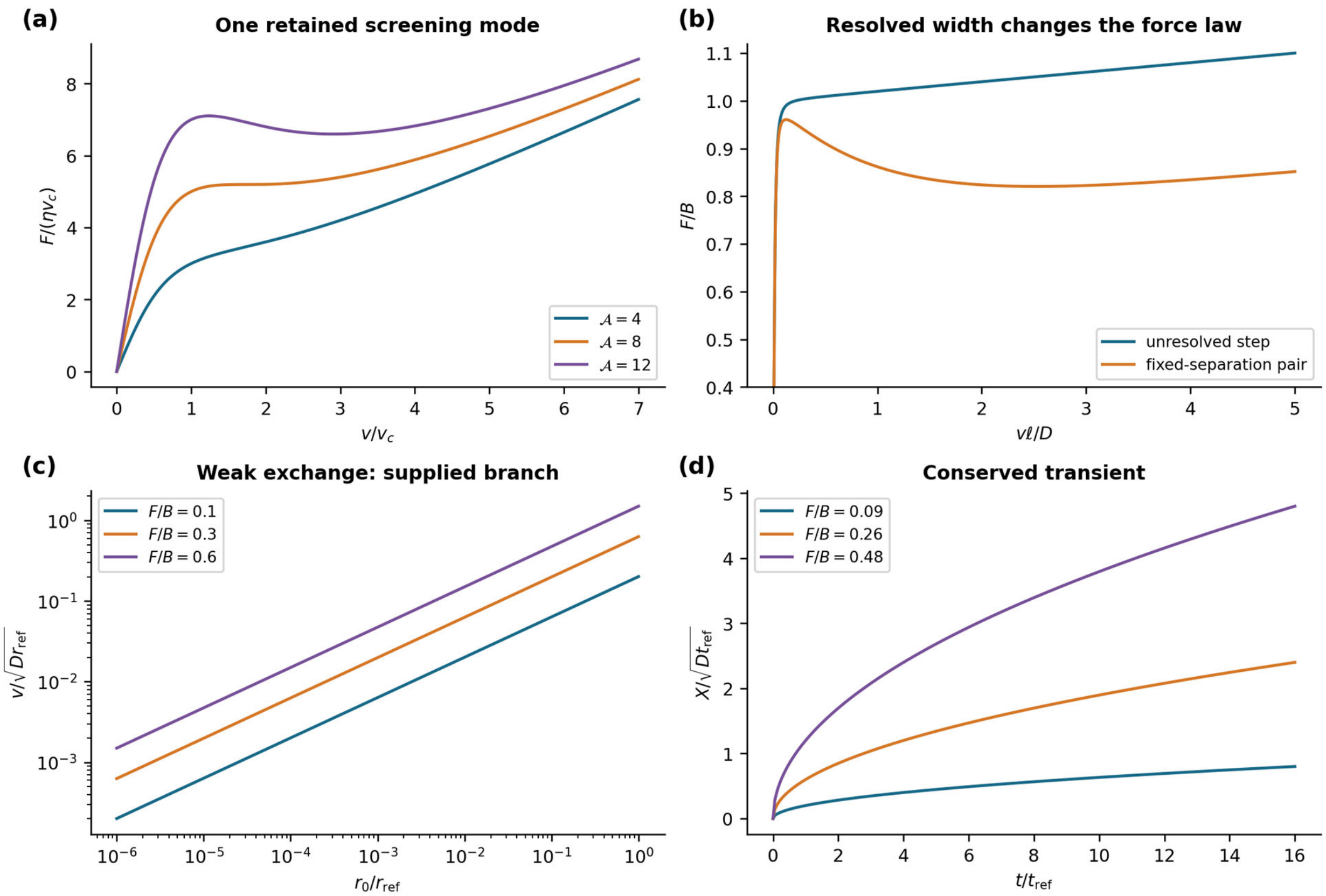

Figure S2. Analytical motion comparisons. (a) Conditional one-mode force laws, Eq. (27), with the fold threshold $\mathcal{A} = 8$. (b) An unresolved step and a fixed-separation pair under mixed exchange and transport, with $r_0\ell^2/D = 10^{-4}$ and $\eta_X D/(B\ell) = 0.02$. The pair can have a negative-slope interval; its separation is held fixed in this panel. (c) The supplied weak-exchange branch has $v \propto \sqrt{r_0}$ at fixed reduced force. (d) Conserved sharp-interface displacement for specified similarity parameters, Eq. (13). Panels (a)–(d) are distinct stated reductions; they do not constitute one universal velocity diagram.

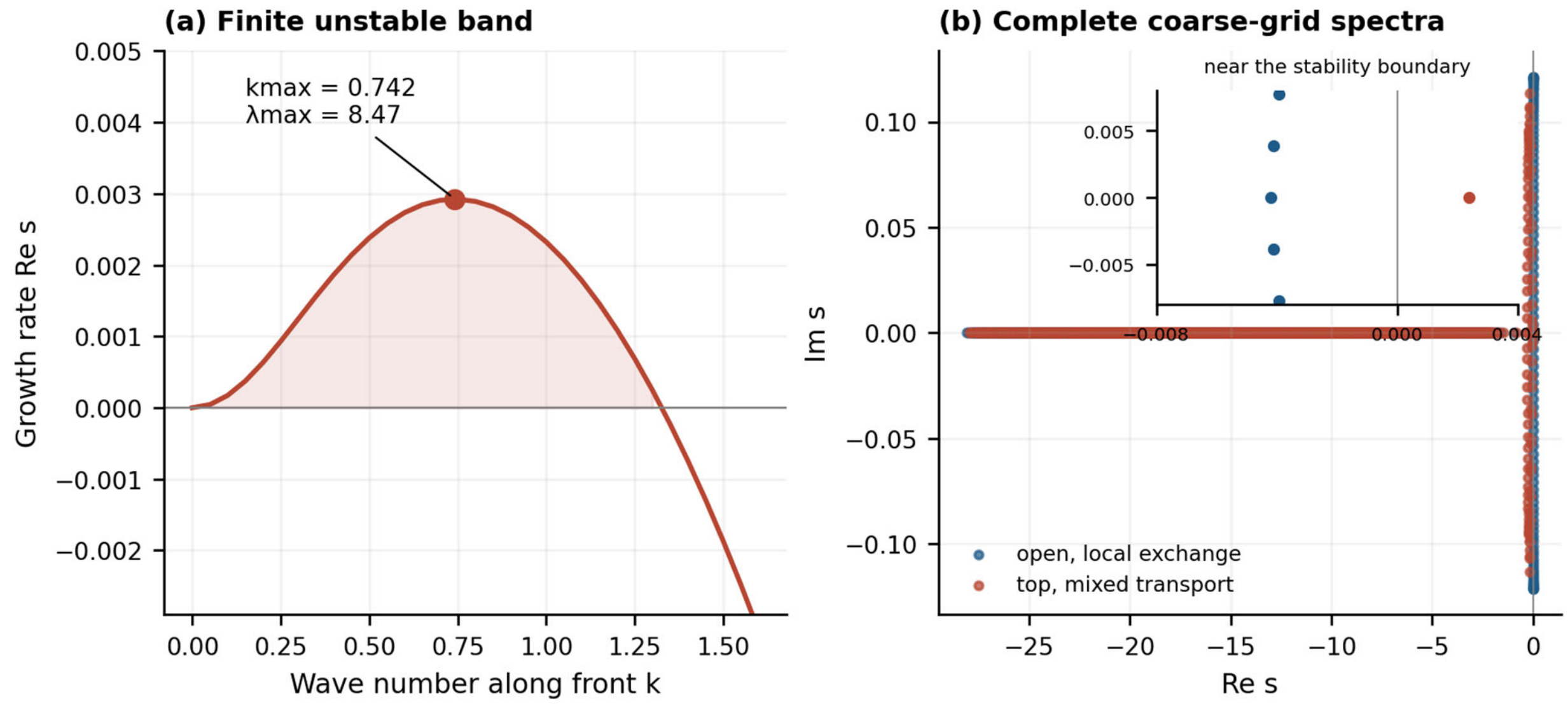

Figure S3. Numerical spatial stability. (a) Leading growth rate along the continued moving branch at $e = 0.06$ for a soft wall-front gradient coefficient. (b) Complete coarse-grid eigenvalue spectra at $k = 0.5$ for the exposed local-exchange reference and the electrode/mixed-transport example. The inset resolves the growing real eigenvalue. The curve in (a) is a coupled field eigenvalue calculation, not the sharp-step analytical approximation.

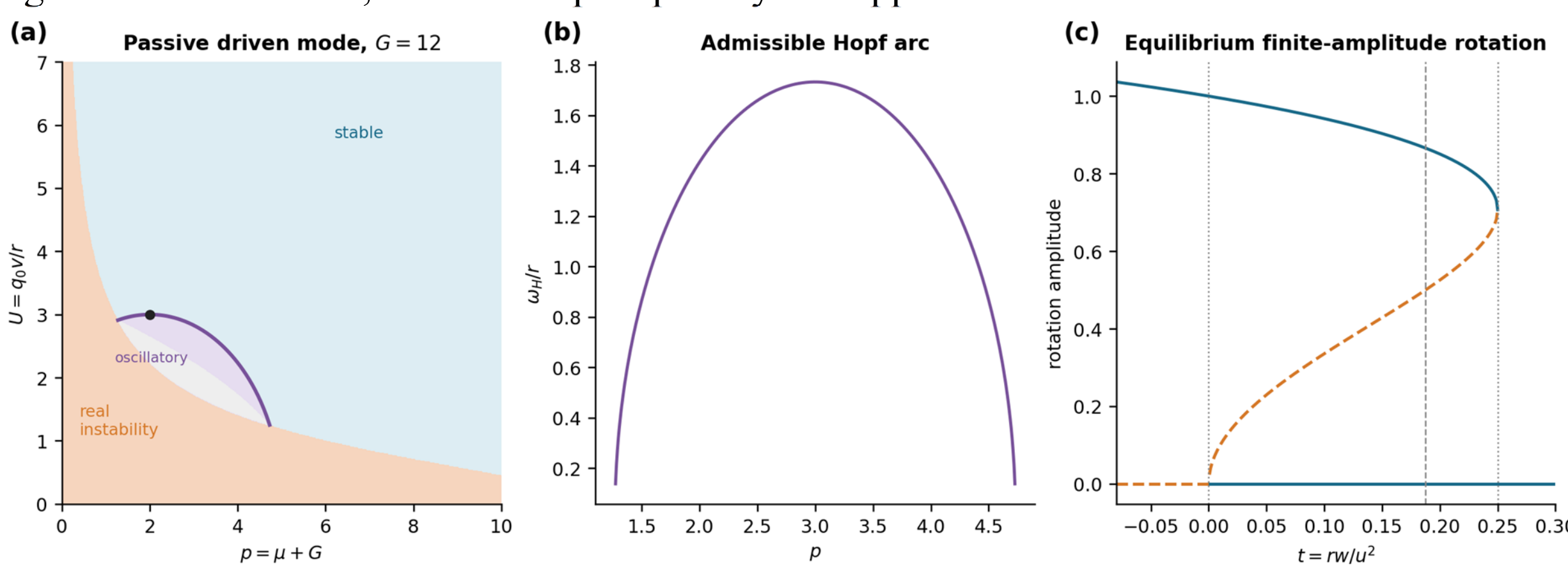

Figure S4. Conditional analytical temporal and structural diagrams. (a) The Routh–Hurwitz map of Eq. (37) at $G = 12$. Stable, stationary-unstable and oscillatory-unstable regions refer to this finite-mode model. (b) Frequency on the admissible Hopf arc. (c) A negative-quartic equilibrium rotation amplitude, showing appearance, equal-energy and stability-loss points. None of the unstable regions is labeled as a demonstrated stable oscillatory phase.

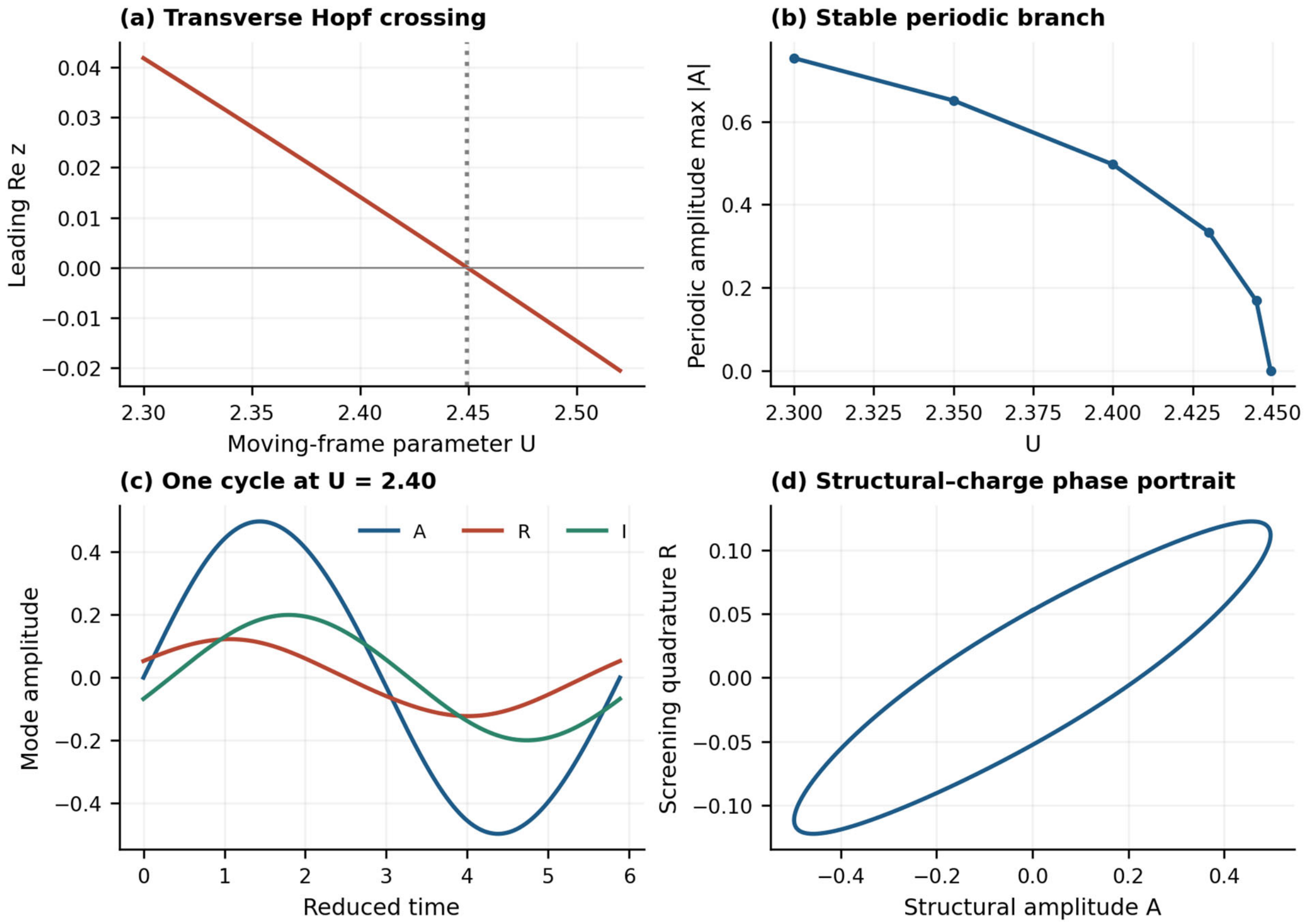

Figure S5. Stable oscillation in the declared driven structural–screening model. (a) Leading eigenvalues through $U_H = \sqrt{6}$. (b) Continued orbit amplitude. (c) One computed periodic cycle at $U = 2.40$. (d) The same orbit in structural and screening coordinates. The forcing is steady translation of the reference state; no periodically varying external parameter is imposed.

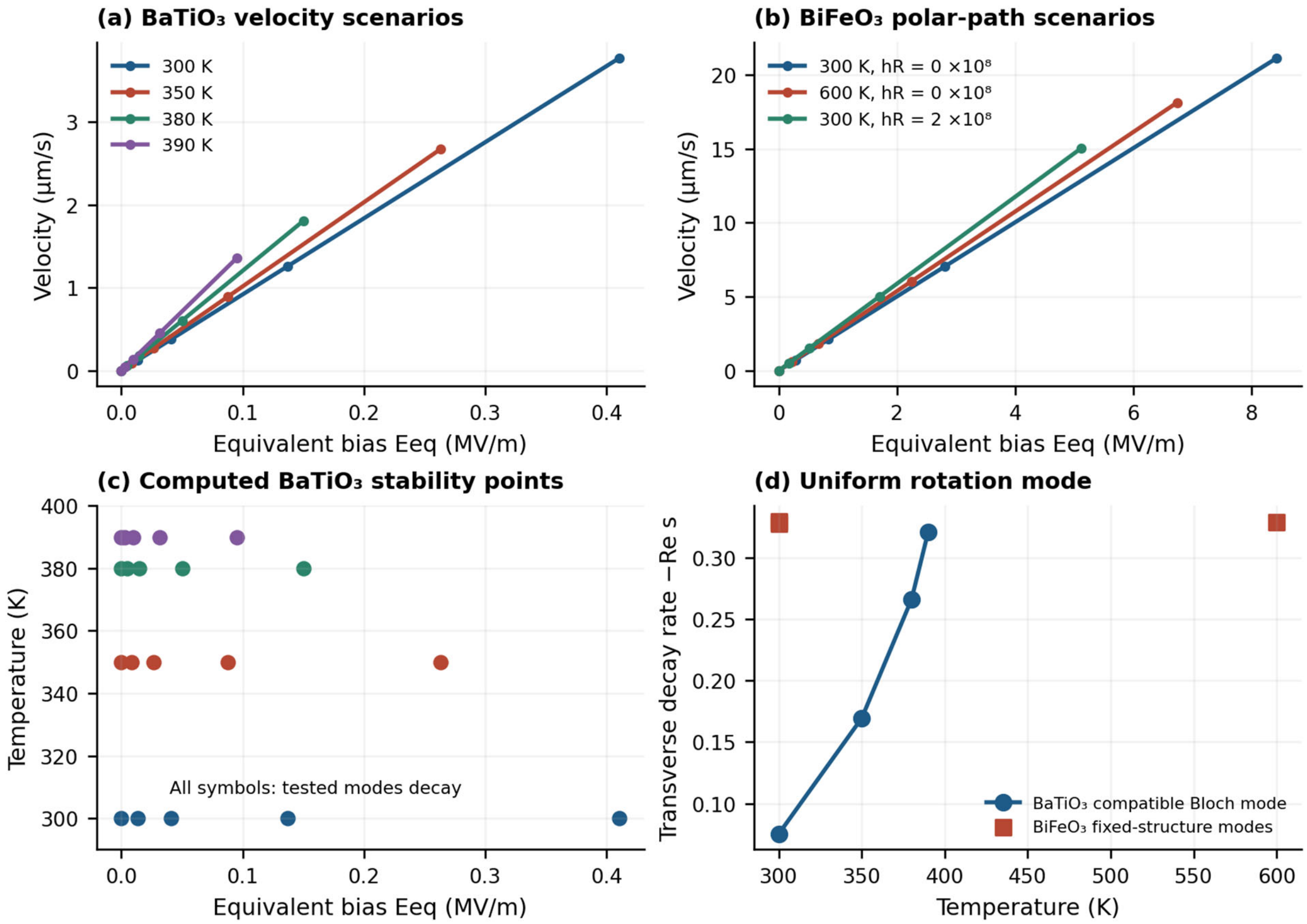

Figure S6. Conditional dimensional motion calculations. (a) $BaTiO_3$ velocity versus field-equivalent thermodynamic bias for four temperatures. (b) $BiFeO_3$ fixed-structure scenarios. (c) $BaTiO_3$ sampled stability points in field–temperature coordinates; symbols indicate calculated states, not a filled universal phase region. (d) Uniform transverse-polarization decay rates on the terminal computed branches. The structural constraint, assumed kinetics and gradient parameters are given in S10 and S12. No material Hopf frequency or uncomputed instability boundary is assigned.

## S15.1. Dimensional velocity and stability calculations

Material calculations N7. We solved finite-depth scalar traveling walls for $BaTiO_3$ at 300, 350, 380 and 390 K and for the fixed-structure $BiFeO_3$ polar path at 300 and 600 K (Fig. S6). The common electrical scenario is a 20 nm film, a 0.5 nm dielectric gap with relative permittivity 5, a nominal 1 $\mu$s long-wave exchange time and a nominal $10^{-12}$ m$^2$ s$^{-1}$ surface diffusivity. We set the polarization time unit to 1 ns. These kinetic values are declared scenarios, not fits to measured velocities. The $BiFeO_3$ gradient coefficient $10^{-10}$ J m$^3$ C$^{-2}$ is also a scenario input; the $BaTiO_3$ coefficient and coherent planar correction follow S10.

The abscissa is the field-equivalent thermodynamic bias $E_{\mathrm{eq}}$ acting on polarization alone. It is suitable for comparing domain free-energy differences and is not identified with an electrode voltage. At reduced bias $e = 0.003$, the calculated $BaTiO_3$ velocity decreases from 3.77 $\mu$m s$^{-1}$ at 300 K to 1.35 $\mu$m s$^{-1}$ at 390 K in this kinetic scenario. For fixed-structure $BiFeO_3$ at 300 K and $h_R = 0$, it is 21.1 $\mu$m s$^{-1}$; setting $h_R = 2 \times 10^8$ J m C$^{-2}$ reduces it to 15.0 $\mu$m s$^{-1}$. These comparisons hold the declared time and transport conventions fixed and therefore do not establish an intrinsic mobility ordering between real materials.

All sampled nonzero-wave-number scalar eigenvalues and the separately evaluated uniform Bloch-mode eigenvalues are decaying in these slices. The $BiFeO_3$ calculation holds its structural coordinates fixed; a fully coupled polar–octahedral spectrum has not been obtained. The diagrams therefore identify the computed moving branches and their tested stability, alongside the thermodynamic boundaries in Fig. 11. They do not extrapolate a wall-instability boundary into uncomputed fields. Representative mesh refinements change the velocities by less than 0.02%; the much smaller depth displacements carry several-percent discretization uncertainty and are reported only as estimates.

### S15.2. Additional numerical values for the main-text comparisons

Finite-depth calculation N1. We solved the coupled polarization, Poisson and local-exchange equations for an isolated scalar wall, allowing the profile to bend in depth (Fig. 2). The reduced reference has $f_L = -p^2/2 + p^4/4$, $g_y = g_z = g_x = 0.5$, $\Gamma = 1$, $\epsilon_f = 10$, $\epsilon_e = 1$, $\Lambda = 0.05$ and $\chi = M = 0$. The drive $e$ is an imposed thermodynamic polarization bias; it does not directly drive the surface carriers. At $e = 0.06$, increasing $h$ from 5 to 20 raises the velocity from 0.04044 to 0.07275 and the top-to-bottom junction displacement from 0.095 to 0.985. The thicker wall can advance more readily in the bulk and bends appreciably near its screening-limited upper junction.

Conserved transient calculation N2. Starting from an equilibrated compensated wall, we applied a constant bias $e = 0.01$ on a closed surface of length $L = 40$, with $h = 1$, $\epsilon_f = 10$, $M = 10$, $\Lambda = 0$ and the same quartic polarization potential. The long-wave charge diffusivity is $D = MG_0 = 1$. The wall moves initially, leaves a spreading charge distribution, and approaches a finite displacement (Fig. 3). The calculated final position is $X = 2.00075$, compared with the leading finite-inventory estimate $X_\infty = e\epsilon_f L/(2P_s) = 2.00000$. The charge integral changes by less than $1.2 \times 10^{-11}$ in reduced units. Refining both coordinates and halving the time step gives $X(2500) = 1.99859$.

Electrode calculation N3. Figure 4 evaluates the re-derived Eliseev junction-field relation, $E_{\text{app}} = v/\mu + |E_{\text{ret}}(0,0)|$, using $P_s = w = \mu = 1$, $\tau = 2000$, $\epsilon_f = 10$ and $\epsilon_d = 1$. At $h = 100, d = 1$ it has folds at $(v, E) = (0.001758, 0.05383)$ and $(0.01285, 0.04487)$. Reducing the film thickness to $h = 20$ narrows the negative-slope interval; at $h = 5$ the displayed branch is monotone. Increasing the gap to $d = 10$ also removes the folds in this parameter slice. These values are new evaluations of the historical equations with declared reduced parameters, rather than a reproduction of a published dimensional data set [3]. The thickness-averaged field closure gives a different curve.

We also continued the coupled finite-depth equations with a bounded surface population, $\mu = \theta \operatorname{atanh}(c/c_{\max}) - \phi_s$, using $h = 5$, $d = 0.5$, $\Lambda = 0.05$, $M = 0.3$, $\theta = 0.015$ and $c_{\max} = 1.3$. The branch remains monotone over $0 \le e \le 0.10$; at $e = 0.06$, $v = 0.04727$, the junction displacement through depth is 0.097, and $\max|c|/c_{\max} = 0.771$. This computation uses the thermodynamic bias protocol of N1 and a passive electrochemical law. Its comparison with the historical curves keeps both differences explicit. Section 8 shows that this positive-slope branch nevertheless corrugates when its stiffness along the front is reduced.

Structural calculations N5. Figure 6 answers the structural question within three explicit models. In a closed long-wave capacitor model, a prepared broad scalar core contracts to the ordinary quartic wall. For the first-order potential $f_L = ap^2/2 - p^4 + p^6/2$, a central layer is bound at $a = 0.94$ and expands at $a = 1.03$, on opposite sides of coexistence $a_t = 1$. The screening population is conserved in all these evolutions. Symmetric expansion consumes equal parts of the two oppositely polarized domains, so it requires redistribution but no change of total charge. At $t = 300$ the widths between $p = \pm P_s/2$ are approximately 1.10, 2.22 and 12.5 for the quartic, bound first-order and expanding first-order cases, respectively.

Computed corrugation N6. The coupled finite-depth branch of N3 provides a field-level example of corrugation on a positive-slope velocity curve (Fig. S3). At $e = 0.06$, reducing only the gradient coefficient along the front from $g_x = 0.5$ to $0.005$ leaves the straight base and its velocity unchanged. Its leading growth rate becomes positive over $0 < k < 1.326$. The fastest mode is at $k = 0.742$, with $\mathrm{Re}s = 0.002923$, corresponding to an initial wavelength $2\pi/k = 8.47$ in units of the reference wall width and an e-folding time of 342 reduced time units. The unstable eigenvalue is real. The high-wave-number gradient penalty closes the band.

Nonlinear oscillatory calculation N6. To determine whether a stable periodic state can actually emerge in the stated finite-mode model, we added a positive cubic structural restoring term and solved $A' = -pA + GR - A^3$, $R' = A - R - UI$, $I' = UR - I$. Here the prime denotes time in units of the charge relaxation time, and $R + iI$ is the rescaled screening mode. At $G = 10, p = 2$, the Hopf point is $U_H = \sqrt{6}$, with angular frequency 1. The eigenvalue crossing is transverse, $d\mathrm{Re}z/dU = -0.28818$, and the first Lyapunov coefficient is $-0.14706$ under the normalization specified in S12. The bifurcation is supercritical.

Shooting continuation and integration of the variational equations give a stable periodic orbit at $U = 2.40$, with amplitude $\max|A| = 0.4970$, period 5.8872 and nontrivial Floquet multipliers 0.8559 and $7.9 \times 10^{-12}$ (Fig. S5). The unit multiplier corresponds to a shift of time along the orbit. This is a demonstrated stable oscillation in the prescribed moving-frame reduction. Translation supplies the energy through the imposed moving base; the result is not a Hopf prediction for the unregularized isolated MIEC or for either calculated material. The full field calculations and the later xy evolution are documented in S12–S13; the temporal reduction is a separate model.

## S16. Small-signal response and impedance

We derive the response functions underlying Section 11 of the main text and retain the observables omitted from its displacement-focused presentation. The analysis includes polarization within domains, screening charge, surface potential, electrode current, wall displacement, interphase separation and transverse polarization. Each response is defined by its perturbation, measured coordinate and reference state. Local charge exchange, exchange with lateral transport, strictly conserved redistribution and frozen charge are treated separately under the exposed-surface, top-electrode and blocking-MIEC boundary conditions.

The derivations give explicit frequency dependence for the stated uniform-domain and collective-coordinate reductions. A general deformable wall requires the linearized field operator about its actual base profile. Transient and periodic reference states require two-time and Floquet response functions, respectively. These distinctions preserve the conservation laws when an ordinary stationary impedance does not exist. Figures S7–S19 collect the additional spectra. The results separate changes in the mobility of an existing wall from changes in its internal structure and from direct charging of the electrodes.

### S16.1. Definitions of response and impedance

We use $e^{i\omega t}$ throughout and write $s = i\omega + 0^{+}$. For a force $f$ and its conjugate coordinate $a$, define

$$\mathcal{X}_{af}(s) = \frac{\delta a}{\delta f}. \tag{S49a}$$

$$Z_a(s) = \frac{\delta f}{\delta \dot{a}} = \frac{1}{s\mathcal{X}_{af}(s)}. \tag{S49b}$$

$$|\mathcal{X}| = \sqrt{(\mathrm{Re}\mathcal{X})^2 + (\mathrm{Im}\mathcal{X})^2}. \tag{S49c}$$

$$\varphi_{\mathcal{X}} = \mathrm{atan2}(\mathrm{Im}\mathcal{X}, \mathrm{Re}\mathcal{X}). \tag{S49d}$$

For an ordinary passive polarization susceptibility, the loss is $-\mathrm{Im}\mathcal{X} \geq 0$ and a negative phase means lag. Electrical impedance is $Z_V = \delta V/\delta I$. The wall impedance $Z_X = \delta F/\delta \dot{X}$ has mechanical units and must not be presented as ohms. A modal interfacial impedance relates potential to the rate of change of a specified charge mode. Nyquist plots use $-\mathrm{Im}Z$ vertically; passive capacitive branches therefore lie above the real axis. All plotted impedance magnitudes and phases are dimensionless unless a physical conversion is supplied. The electrical definition follows ferroelectric impedance spectroscopy [13, 14], while the need to retain blocking boundary conditions and charge conservation is established in diffuse-charge and MIEC response theory [60, 61]. A dielectric contribution of a stationary wall [15] or local ac carrier conduction [16] need not imply motion of the wall center.

The observable may be the volume-averaged polarization, a wall position, a surface potential at a specified probe height, or terminal charge. A finite probe weights many spatial modes. Consequently, the location of a loss maximum is not generally the bare exchange rate or a universal material relaxation time. High frequency below means high relative to the retained relaxational rates while remaining within the overdamped, quasistatic-electrostatic model; optical phonons, inertia, electromagnetic propagation and external instrument bandwidth are outside that limit.

Table S5. Definitions of the small-signal observables.

| Quantity | Input and output | Meaning |
|---|---|---|
| Polarization response | Polarization divided by prescribed field | Intrinsic and screening-modified dielectric response |

| Quantity | Input and output | Meaning |
|---|---|---|
| Wall response | Displacement divided by force per junction length | Motion and deformation of an existing front |
| Modal charge impedance | Potential divided by charge-rate amplitude | Spatially resolved generalized electrical response |
| Terminal impedance | Voltage divided by electrode current | Requires a specified electrode and reservoir arrangement |

### S16.2. Surface electrostatics and the coupled polarization and charge response

Let $q$ be the wave number across the wall, $k$ the wave number along it, and $Q = \sqrt{q^2 + k^2}$. In a uniform-domain calculation $Q$ denotes the imposed lateral mode. For a depth-independent polarization and a surface sheet with physical free charge $\sigma = -c$, the induced surface potential is $G_Q\big(p_Q - c_Q\big)$. The kernels retained from the manuscript are

$$G_O(Q) = [\epsilon_0 \bar{\epsilon} Q \coth(\alpha Q h) + \epsilon_0 \epsilon_e Q]^{-1}. \tag{S50a}$$

$$G_T(Q) = [\epsilon_0 \bar{\epsilon} Q \coth(\alpha Q h) + \epsilon_0 \epsilon_d Q \coth(Qd)]^{-1}. \tag{S50b}$$

$$\alpha = \sqrt{\epsilon_\parallel / \epsilon_z}. \tag{S50c}$$

$$\bar{\epsilon} = \sqrt{\epsilon_\parallel \epsilon_z}. \tag{S50d}$$

The finite-film zero modes are $G_O(0) = h/(\epsilon_0 \epsilon_z)$ and $G_T(0) = \big(C_f + C_d\big)^{-1}$, where $C_f = \epsilon_0 \epsilon_z / h$ and $C_d = \epsilon_0 \epsilon_d / d$. We denote the chemical curvature by $\chi_c = f_s''(c_0)$, equal to χ in the quadratic surface energy of the main text, and reserve $\mathcal{X}$ for response. A smooth nonlinear adsorption law gives the same local linearization; saturation and interactions alter the curvature and, in a multispecies description, the chemical Hessian [35, 36].

$$A_Q = G_Q + \chi_c. \tag{S51a}$$

$$L_Q = \Lambda + M Q^2. \tag{S51b}$$

$$r_Q = L_Q A_Q. \tag{S51c}$$

$$a_Q = G_Q / A_Q. \tag{S51d}$$

$$H_Q = G_Q^2 / A_Q. \tag{S51e}$$

For uniform reference coefficients, $K_Q = f_L''(p_0) + g_\parallel Q^2$. The forced linear equations are

$$\big(K_Q + G_Q/h + \Gamma s\big)\delta p_Q - \frac{G_Q}{h}\delta c_Q = \delta E_{b,Q}. \tag{S52a}$$

$$\big(s + r_Q\big)\delta c_Q - L_Q G_Q \delta p_Q = L_Q \delta\psi_Q. \tag{S52b}$$

Here $E_b$ is a prescribed polarization-driving field, and $\psi$ is the sum of the imposed electrochemical offset and the externally produced surface potential entering the charge equation. These are independent generalized controls; an actual voltage experiment specifies a relation between them. Eliminating charge gives

$$\mathcal{X}_{PE}(Q, s) = \left[K_Q + \Gamma s + \frac{G_Q}{h}\frac{s + L_Q \chi_c}{s + r_Q}\right]^{-1}. \tag{S53a}$$

$$\delta p_Q = \mathcal{X}_{PE}\left[\delta E_{b,Q} + \frac{G_Q L_Q}{h(s + r_Q)}\delta\psi_Q\right]. \tag{S53b}$$

$$\delta c_Q = \frac{L_Q}{s+r_Q}\left(G_Q \delta p_Q + \delta\psi_Q\right). \quad \text{(S53c)}$$

The spatially uniform conserved mode is special. With $\Lambda = 0$ and $Q = 0$, the perturbation of the conserved inventory is zero. Equation (S53a) must then be evaluated as $\delta c_0 = 0$, giving the frozen-charge response for that mode. One must not replace the resulting $0/0$ at $s = 0$ by reservoir equilibrium. For a real wall, $f_L''(p_0)$ and a nonlinear $f_s''(c_0)$ vary in space, so Equation (S52) becomes an operator system, as given in Section S16.11.

### S16.3. Real and imaginary polarization response and frequency limits

Write $\mathcal{X}_{PE} = 1/(A_\omega + iB_\omega)$. For fixed reservoir offset and $L_Q > 0$,

$$A_\omega = K_Q + \frac{G_Q}{h} - \frac{L_Q G_Q^2 r_Q}{h\left(r_Q^2+\omega^2\right)}. \quad \text{(S54a)}$$

$$B_\omega = \omega\Gamma + \frac{L_Q G_Q^2 \omega}{h\left(r_Q^2+\omega^2\right)}. \quad \text{(S54b)}$$

$$\mathrm{Re}\mathcal{X}_{PE} = \frac{A_\omega}{A_\omega^2+B_\omega^2}. \quad \text{(S54c)}$$

$$-\mathrm{Im}\mathcal{X}_{PE} = \frac{B_\omega}{A_\omega^2+B_\omega^2}. \quad \text{(S54d)}$$

$$|\mathcal{X}_{PE}| = (A_\omega^2 + B_\omega^2)^{-1/2}. \quad \text{(S54e)}$$

$$\varphi_{PE} = -\mathrm{atan2}(B_\omega, A_\omega). \quad \text{(S54f)}$$

The equilibrated and frozen-charge stiffnesses are $K_0 = K_Q + G_Q\chi_c/\left(hA_Q\right)$ and $K_\infty = K_Q + G_Q/h$. On a stable branch, the low-frequency result is

$$\mathcal{X}_{PE}(s) = \frac{1}{K_0} - \frac{\Gamma_{\mathrm{eff}}}{K_0^2}s + O(s^2). \quad \text{(S55a)}$$

$$\Gamma_{\mathrm{eff}} = \Gamma + \frac{a_Q^2}{hL_Q}. \quad \text{(S55b)}$$

Thus the real part approaches $1/K_0$, the loss is $\omega\Gamma_{\mathrm{eff}}/K_0^2$, and the phase approaches zero from below. If screening is much slower than polarization, the intermediate interval $r_Q \ll \omega \ll K_\infty/\Gamma$ approaches the frozen-charge plateau $1/K_\infty$. At the true high-frequency limit of the overdamped model,

$$\mathrm{Re}\mathcal{X}_{PE} \sim \frac{K_\infty}{\Gamma^2\omega^2}. \quad \text{(S56a)}$$

$$-\mathrm{Im}\mathcal{X}_{PE} \sim \frac{1}{\Gamma\omega}. \quad \text{(S56b)}$$

The amplitude then scales as $\omega^{-1}$ and the phase approaches $-90°$. The coupled poles are two relaxation rates, rather than independent unmodified charge and polarization rates:

$$\lambda_\pm = \frac{r_Q + K_\infty/\Gamma \pm \sqrt{\left(r_Q - K_\infty/\Gamma\right)^2 + 4L_Q G_Q^2/(h\Gamma)}}{2}. \quad \text{(S57a)}$$

$$\mathcal{X}_{PE}(s) = \frac{s+r_Q}{\Gamma(s+\lambda_+)(s+\lambda_-)}. \quad \text{(S57b)}$$

For positive mobilities and a stable constrained equilibrium, both rates are positive. The discriminant is nonnegative, so this passive two-variable equilibrium model has relaxational poles and no oscillatory eigenmode. A loss maximum is not evidence of a Hopf instability. A vanishing $K_0$ gives a static softening and invalidates a finite static susceptibility at the instability itself.

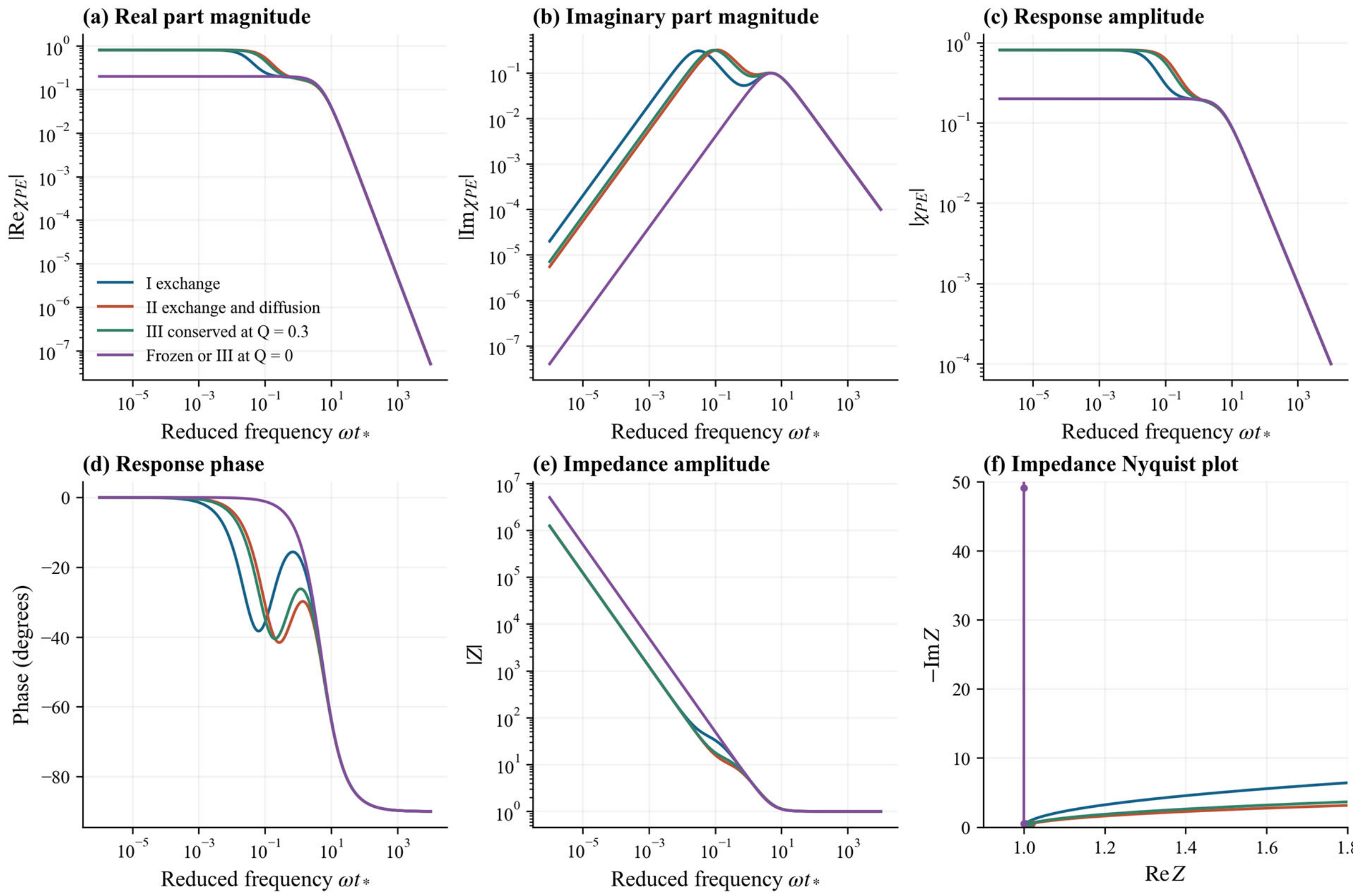

**Figure S7.** Uniform-domain response at fixed $K_Q = 1$, $\Gamma = h = 1$, $G_Q = 4$ and $\chi_c = 0.25$. The modal mobilities are $L_Q = 0.03$ for exchange, 0.12 for exchange plus diffusion, 0.09 for the conserved nonzero mode, and zero for frozen charge or the uniform conserved mode. Panels show the real part, loss, amplitude and phase of $\mathcal{X}_{PE}$, and the generalized polarization impedance $Z_P = 1/(s\mathcal{X}_{PE})$. This impedance is not a terminal resistance. The Nyquist panel displays a restricted window near its finite real intercept; the low-frequency capacitive branches extend beyond it.

### S16.4. Wall displacement and screening memory

Screening retardation couples charge redistribution to wall motion [3]. For a wall at rest, write its displacement as $X(x,t)$ and define $W(q)$ as the Fourier transform of the derivative of its equilibrium polarization profile. The intrinsic friction is $\eta = h\Gamma \int (p_0{}')^2 dy$. Eliminating the charge perturbation and varying both the source and the force projection gives

$$C(s,k) = K_{\text{eq}}(k) + s[\eta + \mathcal{M}(s,k)]. \tag{S58a}$$

$$\mathcal{M}(s,k) = \int_{-\infty}^{\infty} \frac{|W(q)|^2 H_Q}{s + r_Q} \frac{dq}{2\pi}. \tag{S58b}$$

$$K_{\text{eq}}(k) = Tk^2 + \int_{-\infty}^{\infty} |W(q)|^2 \, [K_e(Q) - K_e(|q|)] \frac{dq}{2\pi}. \tag{S58c}$$

$$K_e(Q) = \frac{G_Q \chi_c}{G_Q + \chi_c}. \tag{S58d}$$

$$\mathcal{X}_{XF}(s,k) = C(s,k)^{-1}. \tag{S58e}$$

$$Z_X(s,k) = C(s,k)/s. \tag{S58f}$$

At $k = 0$, an equilibrated wall with exchange has $K_{\text{eq}}(0) = 0$ because translation of the wall and its screening distribution costs no static energy. For a tanh wall of width parameter $w$ and polarization jump $\Delta P$,

$$|W(q)| = \Delta P \frac{\pi q w/2}{\sinh(\pi q w/2)}. \tag{S59}$$

The finite width provides the ultraviolet cutoff. At stable equilibrium, each term in the memory is a passive relaxation. Define $Z_X = R - iX_Z$ at $s = i\omega$. Then

$$R = \eta + \int \frac{|W|^2 H_Q r_Q}{r_Q^2+\omega^2} \frac{dq}{2\pi}. \tag{S60a}$$

$$X_Z = \frac{K_{\text{eq}}(k)}{\omega} + \int \frac{|W|^2 H_Q \omega}{r_Q^2+\omega^2} \frac{dq}{2\pi}. \tag{S60b}$$

$$\text{Re}\mathcal{X}_{XF} = \frac{X_Z}{\omega(R^2+X_Z^2)}. \tag{S60c}$$

$$-\text{Im}\mathcal{X}_{XF} = \frac{R}{\omega(R^2+X_Z^2)}. \tag{S60d}$$

These expressions give the real part, loss, amplitude and phase for each surface case without fitting an equivalent circuit. For an electrode measurement they still require a voltage-to-force conversion and a charge readout. With a purely thermodynamic uniform field acting on a single wall in a lateral interval $L_y$, the rigid-wall polarization contribution is $\delta\overline{P} = (\Delta P/L_y)\delta X$ and $\delta F = h\Delta P\delta E$. Hence $\mathcal{X}_{PE}^{\text{wall}} = h(\Delta P)^2\mathcal{X}_{XF}/L_y$ for that explicitly specified field protocol.

### S16.4.1. Local exchange and exchange with lateral transport

For Case I, $r_Q = \Lambda(G_Q + \chi_c)$. Even with purely local exchange, a nonlocal electrostatic kernel produces a distribution of rates. In the strict local-capacitance approximation, $G_Q = G_0$, all these charge modes have the same rate $r_0$. Put $I = \int (p_0')^2 dy$ and $\zeta = I a_0^2/\Lambda$. At $k = 0$,

$$Z_X(s) = \eta + \frac{\zeta r_0}{s+r_0}. \tag{S61a}$$

$$\mathcal{X}_{XF}(s) = \frac{s+r_0}{s[\eta s + r_0(\eta+\zeta)]}. \tag{S61b}$$

$$(\text{Re}Z_X - \eta - \zeta/2)^2 + (\text{Im}Z_X)^2 = (\zeta/2)^2. \tag{S61c}$$

The impedance traces a semicircle between $\eta + \zeta$ at low frequency and $\eta$ at high frequency, with maximum imaginary magnitude at $\omega = r_0$. The displacement has amplitude $1/[\omega(\eta+\zeta)]$ and phase $-90°$ at low frequency. The divergent displacement susceptibility is the response of an unconfined wall, not a finite static dielectric constant. A weak restoring force would regularize it. At high frequency the finite-width wall sees the prepared screening pattern as fixed, so $C \sim \eta s + K_{\text{fr}}$ and $\mathcal{X} \sim 1/(\eta s)$.

For Case II, exchange removes the conserved zero-mode obstruction, while diffusion spreads the relaxation rates. Under local capacitance define $D = MA_0$ and $B = (\Delta P)^2 H_0$. The sharp-wall integral can be evaluated exactly:

$$Z_X(s) = \eta + \frac{B}{2\sqrt{D}\sqrt{s+r_0}}. \tag{S62a}$$

$$Z_X(0) = \eta + \frac{B}{2\sqrt{Dr_0}}. \tag{S62b}$$

For $\omega \ll r_0$, the charge response gives finite friction. For $r_0 \ll \omega$ and a diffusion length still larger than the wall width, $Z_X - \eta$ scales as $(i\omega)^{-1/2}$. Intrinsic friction can mask that range. At frequencies above $D/w^2$, a sharp-wall extrapolation is inappropriate: the finite-width integral instead has $\mathcal{M} \sim K_{\text{fr}}/s$. Both electrical geometries require their own full kernel outside the local-capacitance limit. In particular, the exposed kernel contains a nonanalytic long-wave correction, and a single fitted exchange time does not capture all geometric dispersion.

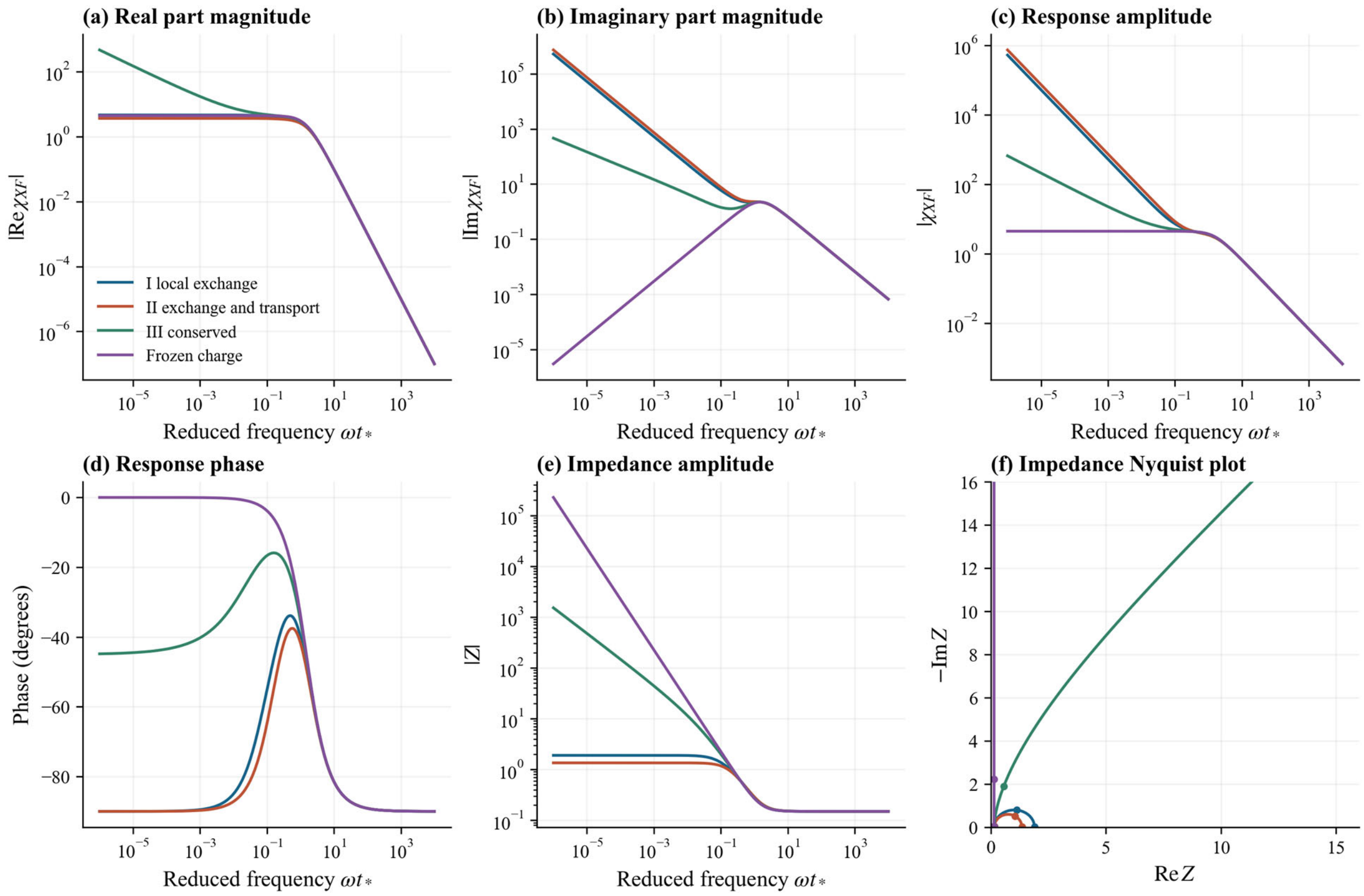

**Figure S8.** Exposed-surface wall response for local exchange, exchange plus diffusion, conserved diffusion and frozen charge. The full $G_O(Q)$ and finite tanh profile are used. Reduced parameters are $h = 10$, $\epsilon_f = 10$, $\epsilon_e = 1$, $\alpha = 1$, $\chi_c = 0.2$, $\Delta P = 2$, $w = 1$ and $\eta = 0.15$. Nonzero kinetic coefficients are $\Lambda = 0.2$ and $M = 1$. Here $\epsilon_f$ and $\epsilon_e$ include the unit conversion. The response is $\mathcal{X}_{XF}$ and impedance is force divided by wall velocity. The conserved low-frequency branch approaches a 45° response, whereas frozen charge gives a finite static displacement. Nyquist dots mark frequencies $10^{-3}$, $10^{-1}$ and 10; curves outside the displayed window are retained in the data.

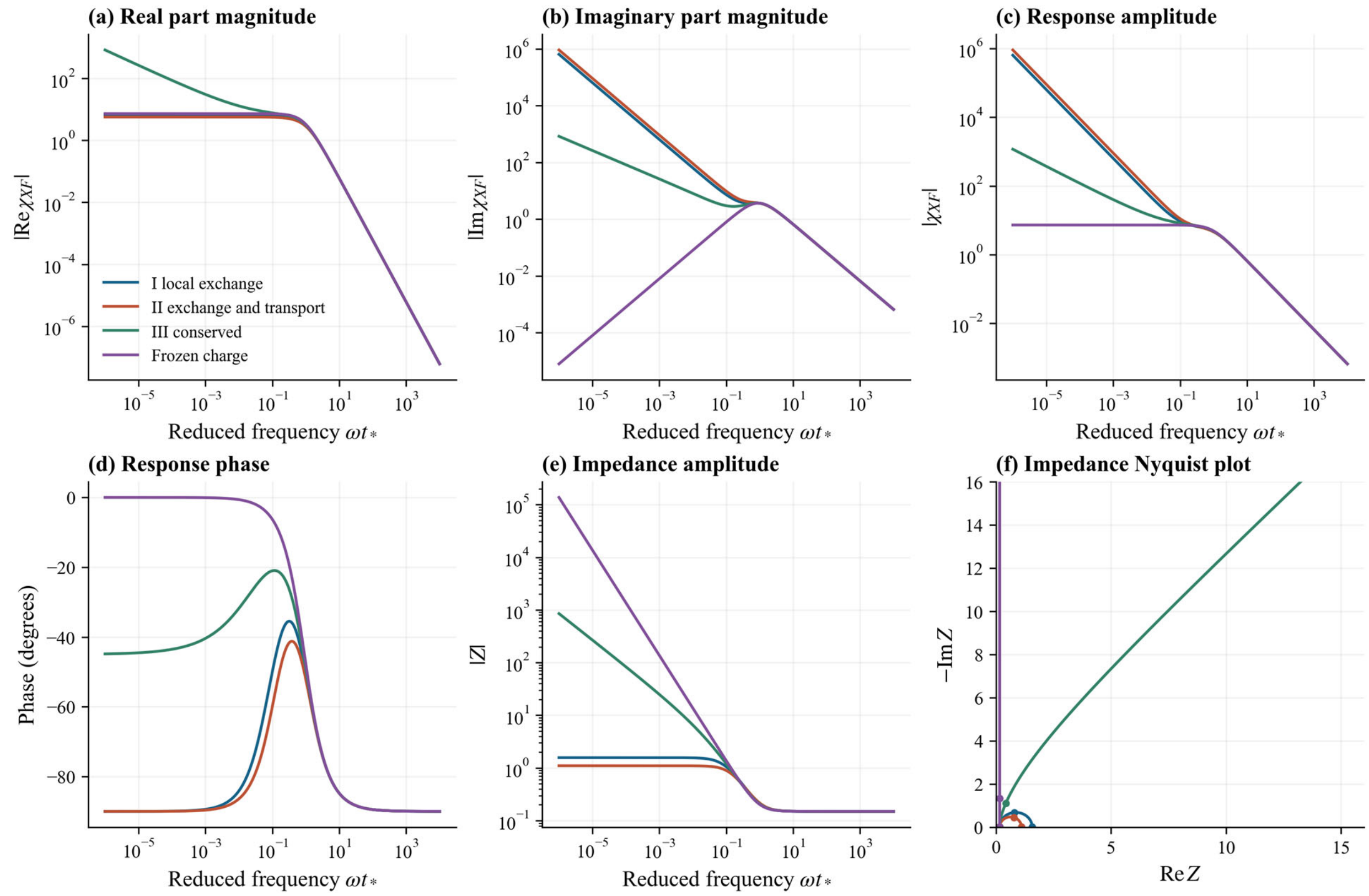

**Figure S9.** The same displacement and impedance calculations for the electrode-covered surface using $G_T(Q)$, gap thickness $d = 2$ and reduced gap permittivity $\epsilon_d = 2$. All other inputs match Figure S8. The ionic kinetic law is unchanged; the electrode changes the field generated by charge mismatch. These are wall spectra under force control, not electrode terminal spectra.

### S16.4.2. Strict conservation and the finite-size crossover

For Case III, $\Lambda = 0$. In a finite film at long wavelength, $r_q = Dq^2$. The uniform translation source has $W(0) = \Delta P \neq 0$, so the low-frequency memory diverges. On an infinite interval,

$$Z_X(s) = \eta + \frac{B}{2\sqrt{Ds}}. \tag{S63a}$$

$$\mathcal{X}_{XF}(s) = \left[\eta s + \frac{B}{2\sqrt{D}}\sqrt{s}\right]^{-1}. \tag{S63b}$$

$$\mathcal{X}_{XF}(i\omega) \sim \frac{\sqrt{2D}}{B\sqrt{\omega}}(1 - i). \tag{S63c}$$

The displacement amplitude scales as $\omega^{-1/2}$, with equal real part and loss and phase $-45°$. The wall impedance has a Warburg form, shifted by $\eta$, and its Nyquist branch approaches a straight line of slope one. This is a consequence of conserved redistribution and is compatible with the absence of a steadily translating screened front. The result assumes a stable stationary reference and fixed far-domain polarization; domain compressibility renormalizes the outer diffusion, as discussed below.

For a closed interval $-L/2 < y < L/2$ with reflecting charge boundaries and a sharp wall initially at its center, the Neumann Green function at the wall is known exactly. The uniform mode and nonuniform diffusion modes give

$$\mathcal{G}_N(0,0;s) = \frac{1}{Ls} + \frac{2}{L}\sum_{n=1}^{\infty}\frac{1}{s + D(2\pi n/L)^2}. \tag{S64a}$$

$$\mathcal{G}_N(0,0;s) = \frac{\coth\left[(L/2)\sqrt{s/D}\right]}{2\sqrt{Ds}}. \tag{S64b}$$

$$Z_X^{\text{closed}}(s) = \eta + B\mathcal{G}_N(0,0;s). \tag{S64c}$$

$$C^{\text{closed}}(s) = \frac{B}{L} + s\left(\eta + \frac{BL}{12D}\right) + O(s^2). \tag{S64d}$$

This is an explicit crossover function rather than a join between asymptotes. At $\omega \ll D/L^2$, the static displacement susceptibility is $L/B$ and its phase tends to zero. The impedance contains a series restoring term $B/(Ls)$ and a finite dissipative intercept $\eta + BL/(12D)$. At $D/L^2 \ll \omega \ll D/w^2$, before intrinsic friction dominates, the infinite-interval diffusion form is recovered. An off-center wall or different edge conditions changes the Green function; Equation (S64) specifically describes reflecting edges and a centered source.

If the far-domain polarization relaxes quasistatically while charge redistributes, the effective charge curvature and diffusivity are

$$A_{\text{eff}} = \chi_c + \frac{G_0 h K_b}{h K_b + G_0}. \tag{S65a}$$

$$D_{\text{eff}} = M A_{\text{eff}}. \tag{S65b}$$

The fixed-domain results use $A_0 = G_0 + \chi_c$. One must recalculate both the interfacial coupling and the restoring coefficient when domain compressibility matters; replacing only $D$ in a result derived with frozen far-domain polarization is not a complete correction. Near a bulk spinodal, retain the coupled polarization and charge operator.

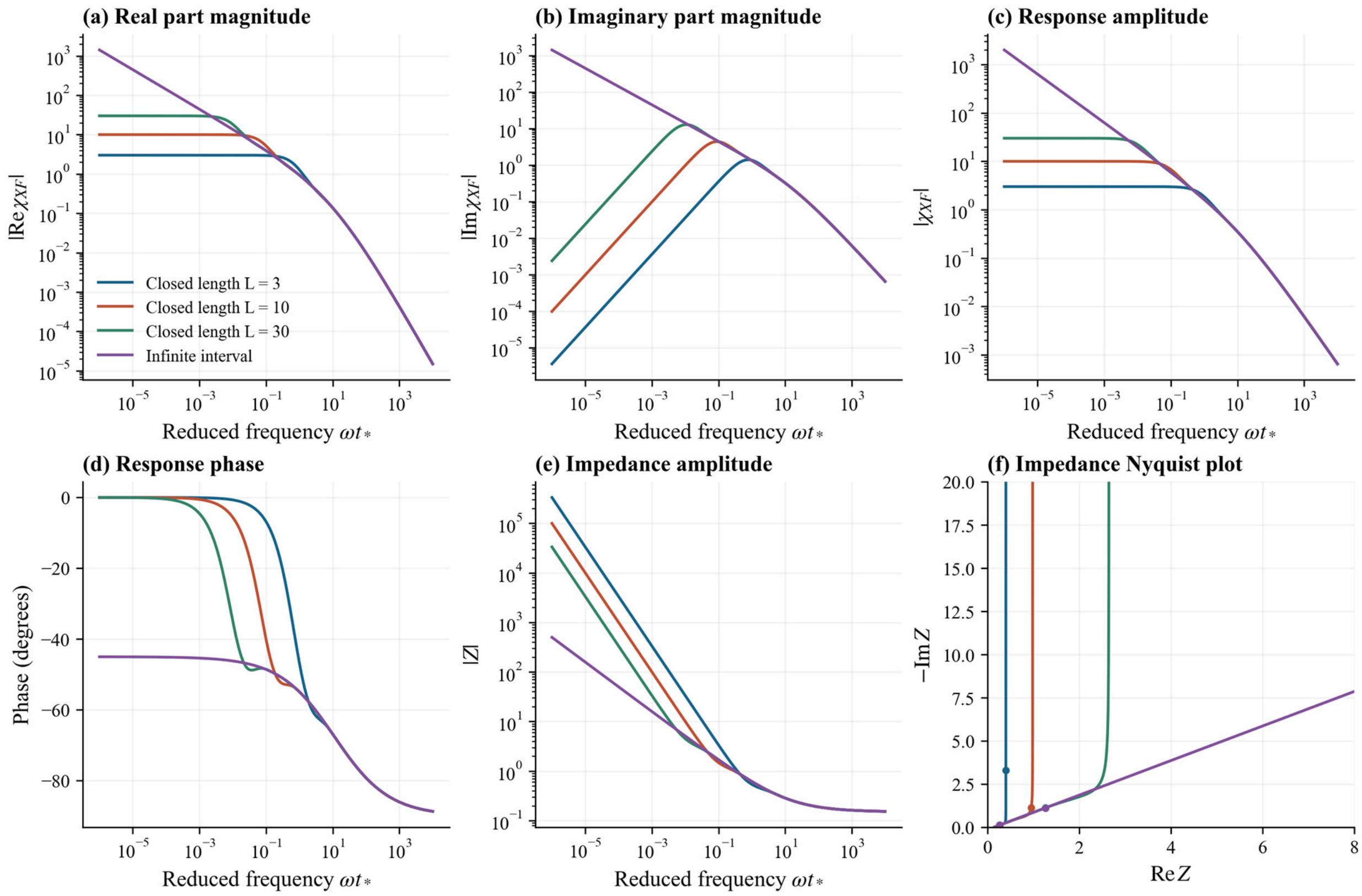

**Figure S10.** Exact centered sharp-wall response on a closed surface for $L = 3$, 10 and 30, compared with an infinite interval. Here $B = D = 1$ and $\eta = 0.15$. The real response saturates at $L/B$ and the phase returns to zero at low frequency. The infinite interval retains the fractional response. The impedance Nyquist branches of finite specimens become nearly vertical at low frequency; the displayed window omits their unbounded capacitive tails.

### S16.4.3. Frozen charge

Frozen charge means $\delta c = 0$ everywhere, not merely a conserved spatial average. For a wall prepared with the equilibrated screening pattern of its initial position, uniform displacement relative to that pattern has stiffness

$$K_{\mathrm{fr}} = \int_{-\infty}^{\infty} |W(q)|^2 \, H_q \frac{dq}{2\pi}. \tag{S66a}$$

$$\mathcal{X}_{XF}^{\mathrm{fr}}(s) = \frac{1}{K_{\mathrm{fr}} + \eta s}. \tag{S66b}$$

$$Z_X^{\mathrm{fr}}(s) = \eta + \frac{K_{\mathrm{fr}}}{s}. \tag{S66c}$$

The low-frequency real response is $1/K_{\mathrm{fr}}$, loss is $\eta\omega/K_{\mathrm{fr}}^2$, and phase is zero. At high frequency, amplitude is $1/(\eta\omega)$ and phase is $-90°$. The impedance is a vertical Nyquist line at $\mathrm{Re}Z_X = \eta$. This restoring force depends on preparation and is not an added substrate pinning potential. An arbitrary frozen charge distribution requires the Hessian about its actual stationary wall, not automatically Equation (S66a).

### S16.5. Electrical terminal impedance and electrochemical forcing

The response to a voltage must include its action on screening charge. For the top-electrode stack, take the bottom electrode at zero, the top at $-V$, and the physical reservoir offset fixed. Set $G_0 = \left(C_f + C_d\right)^{-1}$, $b = C_d/\left(C_f + C_d\right)$ and $C_g = C_f C_d/\left(C_f + C_d\right)$. The external controls in Equation (S52) are

$$\delta E_b = b\delta V/h. \tag{S67a}$$

$$\delta\psi = -b\delta V. \tag{S67b}$$

$$T_s = \frac{s+L_0\chi_c}{s+r_0}. \tag{S67c}$$

$$\frac{\delta p}{\delta V} = \frac{b}{h}\mathcal{X}_{PE}T_s. \tag{S67d}$$

Define $q_e$ as the positive charge conjugate to $V$; with the stated voltage sign it is minus the top-electrode charge. Then $q_e = C_g V + b(p-c)$. The terminal capacitance per area and area-normalized impedance are

$$C_V(s) = C_g + \frac{b^2L_0}{s+r_0} + \frac{b^2}{h}\mathcal{X}_{PE}(s)T_s^2. \tag{S68a}$$

$$Z_{V,A}(s) = \frac{1}{sC_V(s)}. \tag{S68b}$$

For electrode area $A_e$, the total device impedance is $Z_{V,A}/A_e$. At zero in-plane wave number, $L_0 = \Lambda$; lateral diffusivity does not enter a homogeneous-domain terminal spectrum. Cases I and II are therefore identical in this measurement at matched parameters. Cases III and frozen charge are also identical for the homogeneous uniform mode because neither changes its total surface charge. A wall, patterned electrode or local probe introduces nonzero spatial wave numbers and can distinguish them.

$$C_V(0) = C_g + \frac{b^2}{A_0} + \frac{b^2}{hK_0}\left(\frac{\chi_c}{A_0}\right)^2 \quad (L_0 > 0). \tag{S69a}$$

$$C_V^{\text{cons}}(s) = C_g + \frac{b^2}{h(K_\infty + \Gamma s)} \quad (Q = 0). \tag{S69b}$$

$$C_V(s) = C_g + \frac{b^2}{s}\left(L_0 + \frac{1}{h\Gamma}\right) + O(s^{-2}). \tag{S69c}$$

The $\chi_c = 0$ reservoir-equilibrated limit gives $C_V(0) = C_d$ and $\delta p/\delta V \to 0$: the slow charge response cancels the film field while the voltage falls across the gap. This is a statement about the specified fixed-reservoir protocol, not a vanishing intrinsic polarizability. At high frequency the geometric capacitance remains. Consequently the total impedance approaches $1/\left(i\omega C_g\right)$; its magnitude scales as $\omega^{-1}$ and its phase approaches $-90°$. At low frequency, a stable finite capacitance gives the same exponent but a different coefficient, with dissipative real parts between the limits. A chemical exchange pathway is a separate reservoir port; the formula reports the top-electrode current under a fixed reservoir potential.

A second well-defined response drives $\psi_Q$ while holding $E_b$ fixed. Eliminating polarization gives the modal electrochemical impedance

$$Z_{\psi c}(Q,s) = \frac{1}{L_Q} + \frac{1}{s}\left[A_Q - \frac{G_Q^2}{h(K_\infty + \Gamma s)}\right]. \tag{S70}$$

The low-frequency differential chemical capacitance is the reciprocal of the square bracket at $s = 0$. For $L_Q > 0$, the high-frequency real intercept is $1/L_Q$. For the strictly conserved uniform mode $L_0 = 0$, this charge port is blocked. A spatially varying chemical drive can still redistribute a conserved population. An exposed surface has no unique two-terminal impedance until electrodes, return paths and the reservoir control are specified; Equations (S58) and (S70) are its defined response functions.

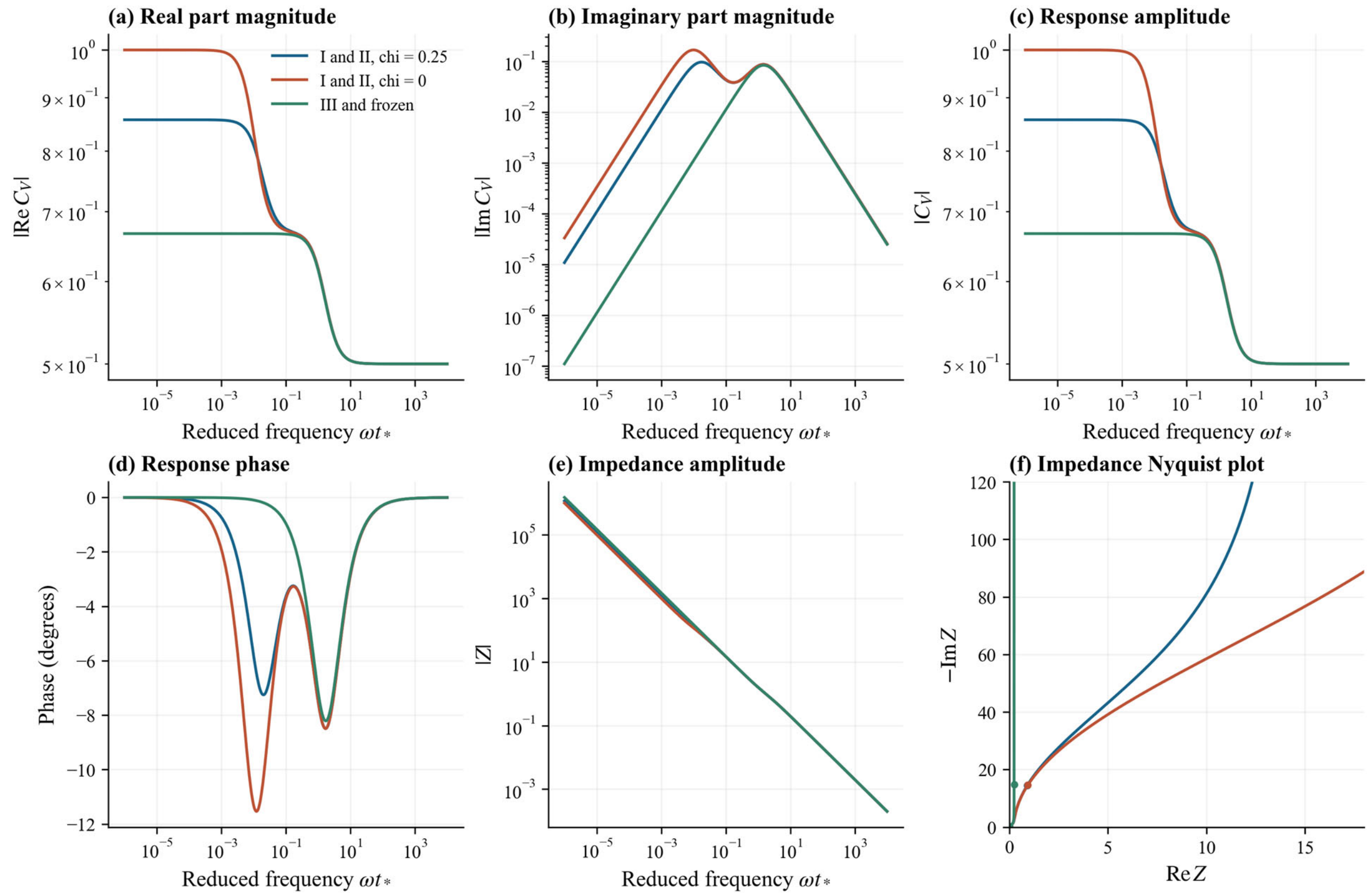

**Figure S11.** Top-electrode terminal spectra for a homogeneous domain. The plotted response is capacitance $C_V$, not $\mathcal{X}_{PE}$. Parameters are $C_f = C_d = 1$, $h = \Gamma = K_Q = 1$ and $\Lambda = 0.03$ where exchange is allowed. Curves compare $\chi_c = 0.25$, $\chi_c = 0$ and a fixed total charge. The geometric high-frequency capacitance is 0.5. Local exchange and exchange plus diffusion coincide at $Q = 0$; conserved and frozen uniform charge coincide as well. The area-normalized impedance is $1/(sC_V)$.

### S16.6. Blocking mixed ionic electronic conductor

The MIEC has volume carriers and no independent surface charge sheet. Both species have zero normal particle flux at the ferroelectric interface. First consider a homogeneous neutral reference with equal diffusivities $D_m$. This is the Debye linearization, appropriate for small background potential; small AC amplitude alone does not justify linearizing a strongly charged DC background around neutrality. Define $\Omega_D = D_m\kappa^2$ and select the square-root branch with positive real part.

$$p_m(Q,s) = \sqrt{Q^2 + \kappa^2 + s/D_m}. \tag{S71a}$$

$$Y_m(Q,s) = \epsilon_m Q p_m \frac{s+\Omega_D}{sp_m+\Omega_D Q}. \tag{S71b}$$

$$H_m(Q,s) = \left[\epsilon_f Q + Y_m(Q,s)\right]^{-1}. \tag{S71c}$$

$$\delta\phi_s = H_m \delta\sigma_b. \tag{S71d}$$

$$Z_{m,Q} = H_m/s. \tag{S71e}$$

Here $Y_m$ relates interface potential to normal displacement; it is not itself the electrical current admittance. The generalized charge-port impedance is $H_m/s$. The effective ferroelectric permittivity $\epsilon_f$ and MIEC permittivity $\epsilon_m$ are absolute permittivities. The harmonic electric-field mode and the decaying space-charge mode are both necessary for the blocking boundary condition. The original MIEC electromechanical literature likewise distinguishes the response by ionic and electronic boundary conditions [61].

### S16.6.1. Static and high-frequency limits at nonzero wave number

$$p_0 = \sqrt{Q^2 + \kappa^2}. \tag{S72a}$$

$$H_m(Q, 0) = \frac{1}{\epsilon_f Q + \epsilon_m p_0}. \tag{S72b}$$

$$H_m(Q, \infty) = \frac{1}{(\epsilon_f + \epsilon_m) Q}. \tag{S72c}$$

The potential generated by a prescribed bound-charge mode increases from the screened static value to the frozen-carrier value. Its imaginary part is positive in this convention; this does not imply active behavior because potential is the force conjugate to charge. The impedance has nonnegative real part. Writing $H_m = H_0 + sH_1 + \cdots$,

$$H_1 = -\frac{\epsilon_m p_0}{(\epsilon_f Q + \epsilon_m p_0)^2}\left[\frac{1}{2D_m p_0^2} + \frac{1 - p_0/Q}{\Omega_D}\right]. \tag{S73a}$$

$$Z_{m,Q}(s) = \frac{H_0}{s} + H_1 + O(s). \tag{S73b}$$

$$H_m(Q, s) = \frac{1}{\epsilon_\Sigma Q} - \frac{\sigma_m}{\epsilon_\Sigma^2 Q s} + O\left(s^{-3/2}\right). \tag{S73c}$$

Here $\epsilon_\Sigma = \epsilon_f + \epsilon_m$ and $\sigma_m = \epsilon_m D_m \kappa^2$. Thus $Z_{m,Q}$ is capacitive at both low and high frequency, but with different inverse capacitances. Its low-frequency real intercept is $H_1$. When $Q \ll \kappa$, the charging scale is of order $D_m \kappa Q$, slower than the Debye rate $D_m \kappa^2$. This geometry-dependent scale is analogous to the distinction between Debye and cell-charging times in blocking electrochemical systems [60].

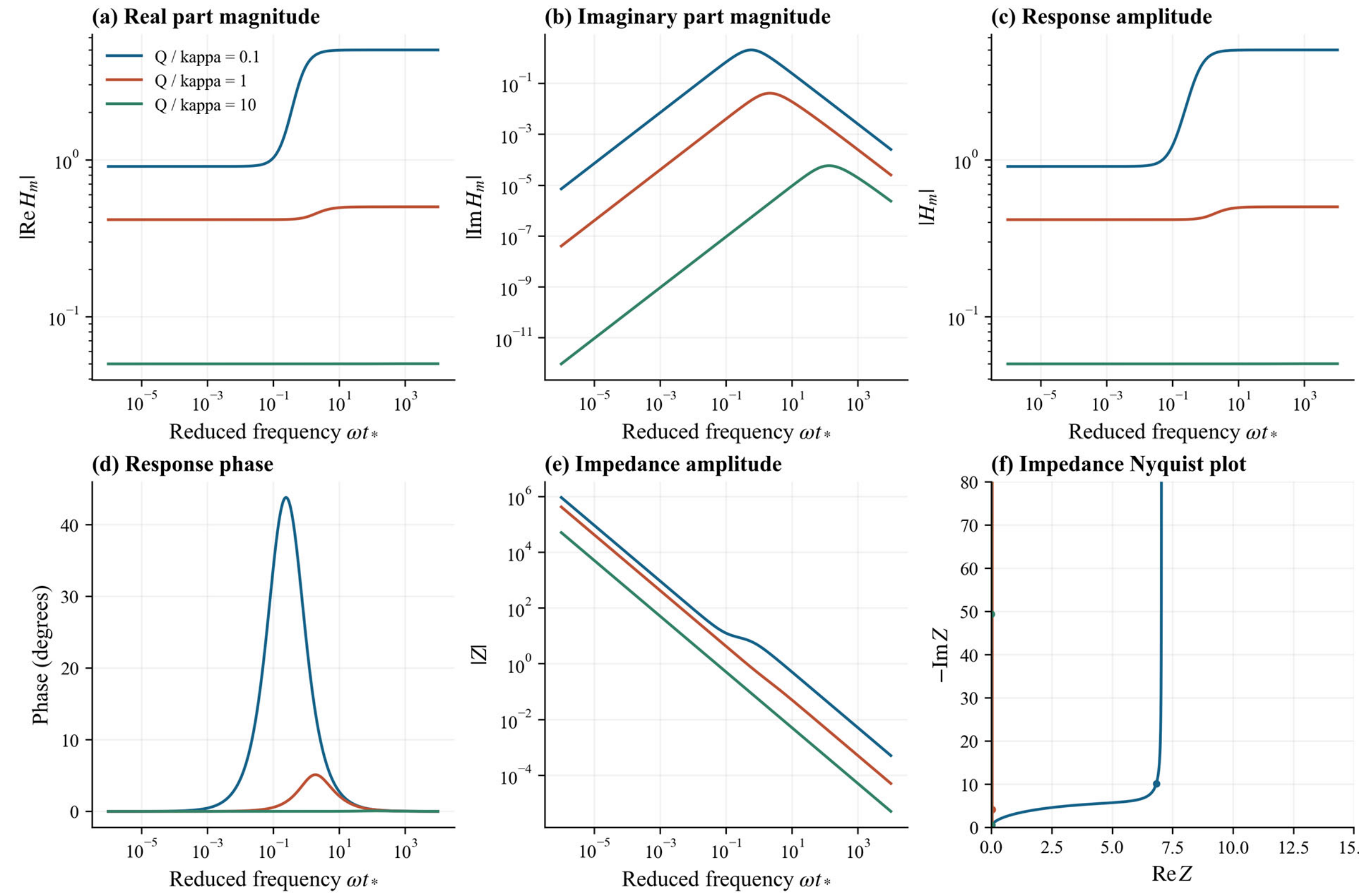

**Figure S12.** Blocking-MIEC modal response at $Q/\kappa = 0.1$, 1 and 10, with $\epsilon_m = \epsilon_f = D_m = \kappa = 1$. The response is $H_m = \delta\phi_s/\delta\sigma_b$, whose imaginary part is positive. The impedance is $Z_{m,Q} = H_m/(i\omega)$ and has negative imaginary part. Long-wavelength perturbations have a larger screening dispersion and a slower charging scale. The spectra describe nonzero modes and do not assume a uniform terminal voltage across two infinite half-spaces.

### S16.6.2. Unequal carrier diffusivities

Let $\kappa_i^2 + \kappa_e^2 = \kappa^2$, with both diffusivities positive. The exact homogeneous-coefficient response requires two charge modes and the harmonic mode. With $t = p^2 - Q^2$, the roots obey

$$(t - \kappa_i^2 - s/D_i)(t - \kappa_e^2 - s/D_e) - \kappa_i^2\kappa_e^2 = 0. \tag{S74a}$$

$$p_\pm = \sqrt{Q^2 + t_\pm}. \tag{S74b}$$

$$\phi(z) = A_0 e^{-Qz} + A_+ e^{-p_+ z} + A_- e^{-p_- z}. \tag{S74c}$$

$$A_0 + A_+ + A_- = \phi_s. \tag{S75a}$$

$$QA_0 - \sum_{l=\pm} \frac{s p_l A_l}{D_i t_l - s} = 0. \tag{S75b}$$

$$QA_0 - \sum_{l=\pm} \frac{s p_l A_l}{D_e t_l - s} = 0. \tag{S75c}$$

$$Y_m = \epsilon_m \frac{QA_0 + p_+ A_+ + p_- A_-}{\phi_s}. \tag{S75d}$$

This three-by-three algebraic system, together with Equation (S71c), is the analytical response for unequal diffusivities. Degenerate limits are obtained from the original equations or the equal-diffusivity expression, rather than dividing by vanishing denominators. Static screening is unchanged, while the dynamic spectrum gains a composition-relaxation contribution. The slow root has $t_- \sim s/D_a$, where

$$D_a = \frac{D_i D_e \kappa^2}{D_i \kappa_i^2 + D_e \kappa_e^2}. \tag{S76}$$

The high-frequency expansion uses $\sigma_m = \epsilon_m(D_i\kappa_i^2 + D_e\kappa_e^2)$. The zero-frequency and zero-wave-number limits need not commute; one should not replace the entire spectrum by one conductivity or one Debye time. Figure S18 evaluates the exact algebraic response for three diffusivity ratios.

### S16.6.3. Long-distance restriction on uniform junction motion

A localized interface charge mode has finite $Q$, but the uniform displacement of a straight isolated wall contains arbitrarily small $q$. At fixed nonzero $s$, the equal-diffusivity kernel has the limit

$$H_m(Q,s) \sim \frac{s}{Q[\sigma_m + \epsilon_\Sigma s]} \quad (Q \to 0). \tag{S77a}$$

For a prescribed interface-junction coordinate with source $W(q)$, its charge contribution to the velocity impedance at rest is

$$Z_X^m(s) - \eta = \int \frac{|W(q)|^2[H_m(|q|,s) - H_m(|q|,0)]}{s}\frac{dq}{2\pi}. \tag{S77b}$$

$$Z_{X,\text{far}}^m(s) \sim \frac{(\Delta P)^2}{\pi[\sigma_m + \epsilon_\Sigma s]} \ln\frac{q_c}{q_{\min}}. \tag{S77c}$$

Here $q_c$ is the upper edge of the applicable long-distance sector. The coefficient diverges logarithmically when $q_{\min} \to 0$. Oscillating about a fixed mean position does not remove this issue: each instantaneous uniform displacement changes the integrated bound charge of the isolated junction. A bounded force and an unbounded isolated straight geometry therefore do not produce a finite uniform harmonic displacement through this interface source model. A finite transverse wave number, a balanced domain pattern, or an explicitly specified outer return geometry removes the zero-mode obstruction. This is a restriction on the uniform junction coordinate, not a claim that every local polarization or structural response vanishes.
Figure S13a–c evaluates a declared spectral cutoff $q_{\min} = 1/R$. It isolates the sensitivity to the outer distance and is not a solution for a particular contact shape. The leading logarithm is physically robust; finite terms and the precise low-frequency spectrum depend on the actual contact boundary problem. Moreover, an infinitely deep rigid wall has divergent intrinsic volume friction. The plotted coordinate is a localized junction mode with an assigned finite projected friction. A physical depth profile must be used to predict an actual junction displacement.

### S16.6.4. A defined blocking-cell terminal impedance

To give a genuine electrical impedance for a specified MIEC geometry, consider an additional reference cell of thickness $L$ between two blocking planar electrodes. This is a separate boundary-value problem, not a replacement for the isolated ferroelectric junction. For equal diffusivities and a neutral background, the charge perturbation is proportional to $\sinh(pz)$ and the potential contains both this mode and a linear field. Enforcing zero carrier flux at both electrodes gives the exact impedance per area

$$Z_{\text{cell},A}(s) = \frac{sL + 2\Omega_D \tanh(pL/2)/p}{\epsilon_m s(s + \Omega_D)}. \tag{S78a}$$

$$p = \sqrt{\kappa^2 + s/D_m}. \tag{S78b}$$

$$C_{\text{cell}}(0) = \frac{\epsilon_m \kappa}{2\tanh(\kappa L/2)}. \tag{S78c}$$

$$Z_{\text{cell},A}(s) \sim \frac{1}{sC_{\text{cell}}(0)} + R_0 \quad (s \to 0). \tag{S78d}$$

$$R_0 = \frac{L + (L/2)\mathrm{sech}^2(\kappa L/2) - 3\tanh(\kappa L/2)/\kappa}{\epsilon_m \Omega_D}. \quad \text{(S78e)}$$

For $\kappa L \gg 1$, the dominant charging time is approximately $L/(2D_m\kappa)$ and $R_0 \simeq L/\sigma_m$. At high frequency, $Z_{\mathrm{cell},A} \sim L/(\epsilon_m s)$, the geometric dielectric impedance. Compact layers, electronic contacts that transmit carriers, or Faradaic reactions change these results and are not included. The boundary-condition dependence is consistent with established blocking-cell theory [60].

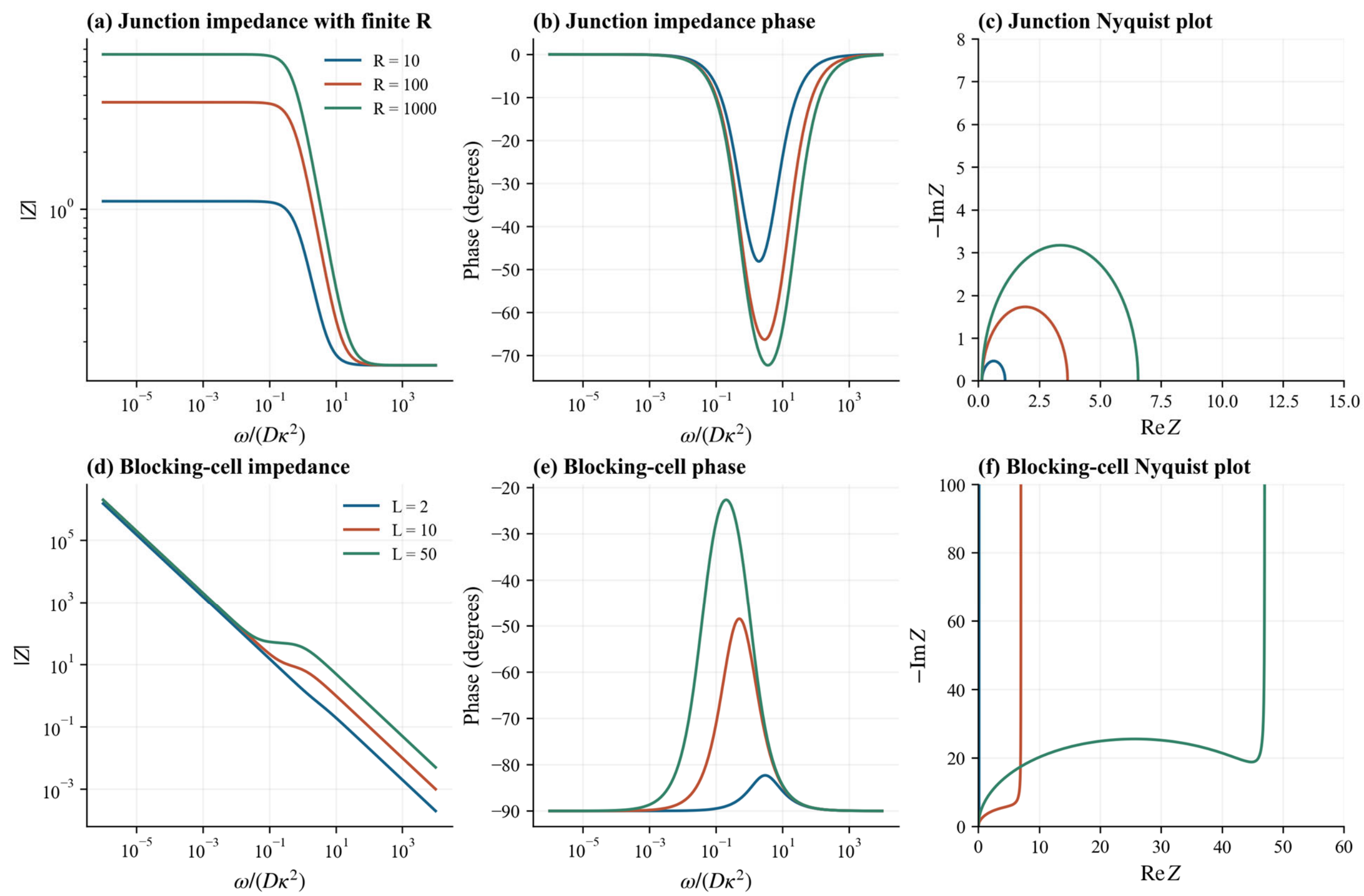

**Figure S13.** Two explicitly distinguished MIEC boundary problems. Panels a–c show the projected junction impedance with tanh source $\Delta P = 2$, $w = 1$, $\eta = 0.15$, equal unit permittivities and $D_m = \kappa = 1$, using cutoffs $R = 10$, 100 and 1000. Increasing the outer distance increases the low-frequency resistance; the strict infinite-junction limit is not a finite additional curve. Panels d–f show the genuine terminal impedance of the separate blocking cell for $L = 2$, 10 and 50. These cell spectra retain the two electrodes and cannot be assigned to the original half-space without changing its geometry.

### S16.7. First-order walls and the response of the intermediate layer

A first-order wall near coexistence has two elementary interphase boundaries. Their center $X$ and separation $\ell$ are independent coordinates. For sharp elementary jumps of magnitude $P_s$, their polarization sources are

$$S_X(q) = 2P_s\cos(q\ell/2). \quad \text{(S79a)}$$

$$S_\ell(q) = -iP_s\sin(q\ell/2). \quad \text{(S79b)}$$

At a symmetric stationary wall the cross-memory vanishes by parity. In the local-capacitance model with exchange and diffusion, let $\lambda_s = \sqrt{(s + r_0)/D}$. The two memories are

$$\mathcal{M}_X(s) = \frac{H_0 P_s^2}{\sqrt{D}\sqrt{s + r_0}}\left(1 + e^{-\ell\lambda_s}\right). \quad \text{(S80a)}$$

$$\mathcal{M}_\ell(s) = \frac{H_0 P_S^2}{4\sqrt{D}\sqrt{s+r_0}}\left(1 - e^{-\ell\lambda_s}\right). \tag{S80b}$$

$$\mathcal{X}_{XF} = [s(\eta_X + \mathcal{M}_X)]^{-1}. \tag{S80c}$$

$$\mathcal{X}_{\ell f} = [K_\ell + s(\eta_\ell + \mathcal{M}_\ell)]^{-1}. \tag{S80d}$$

Here $f$ is a force conjugate to separation and $K_\ell = U''(\ell_0)$ at the actual stable base. For identical intrinsic half-wall frictions $\eta_h$ , $\eta_X = 2\eta_h$ and $\eta_\ell = \eta_h/2$ . The sharp-interface expressions require separations larger than the elementary interface width; finite profiles regularize the high-frequency limit. With strict conservation, the center memory diverges as $s^{-1/2}$, whereas

$$\mathcal{M}_\ell(0) = \frac{H_0 P_S^2 \ell_0}{4D} = \frac{a_0^2 P_S^2 \ell_0}{4M}. \tag{S81a}$$

$$\tau_\ell = \frac{\eta_\ell + \mathcal{M}_\ell(0)}{K_\ell}. \tag{S81b}$$

The width source has zero net charge, so the layer can breathe without the long-distance charge transfer needed for translation. For the large-separation binding energy $U = 2h\gamma + h\Delta f\ell + hA_w e^{-\kappa_t \ell}$,

$$\ell_0 = \kappa_t^{-1}\ln(A_w \kappa_t/\Delta f). \tag{S82a}$$

$$K_\ell = h\kappa_t \Delta f. \tag{S82b}$$

Approaching coexistence from $\Delta f > 0$, the static separation response diverges as $1/\Delta f$ and the low-frequency relaxation becomes slow. Under conservation its friction grows logarithmically through $\ell_0$. At exact coexistence there is no finite bound-layer equilibrium in this asymptotic model, so a stationary Debye spectrum cannot be assigned to the continuing unbinding process. In a second-order scalar material the ordinary zero-polarization crossing is not an independent stable paraelectric layer; its breathing response is a deformation eigenmode of the full wall operator, not Equation (S82).

A uniform longitudinal electric field at zero bias drives common translation but, by the symmetry exchanging the two opposite domains, need not drive symmetric layer expansion at linear order. A bias, asymmetry or temperature modulation can give a nonzero separation force. The divergent $\mathcal{X}_{\ell f}$ should therefore not be identified automatically with a divergent longitudinal dielectric susceptibility. The same charge-neutral source also removes the leading MIEC infrared divergence for this structural coordinate.

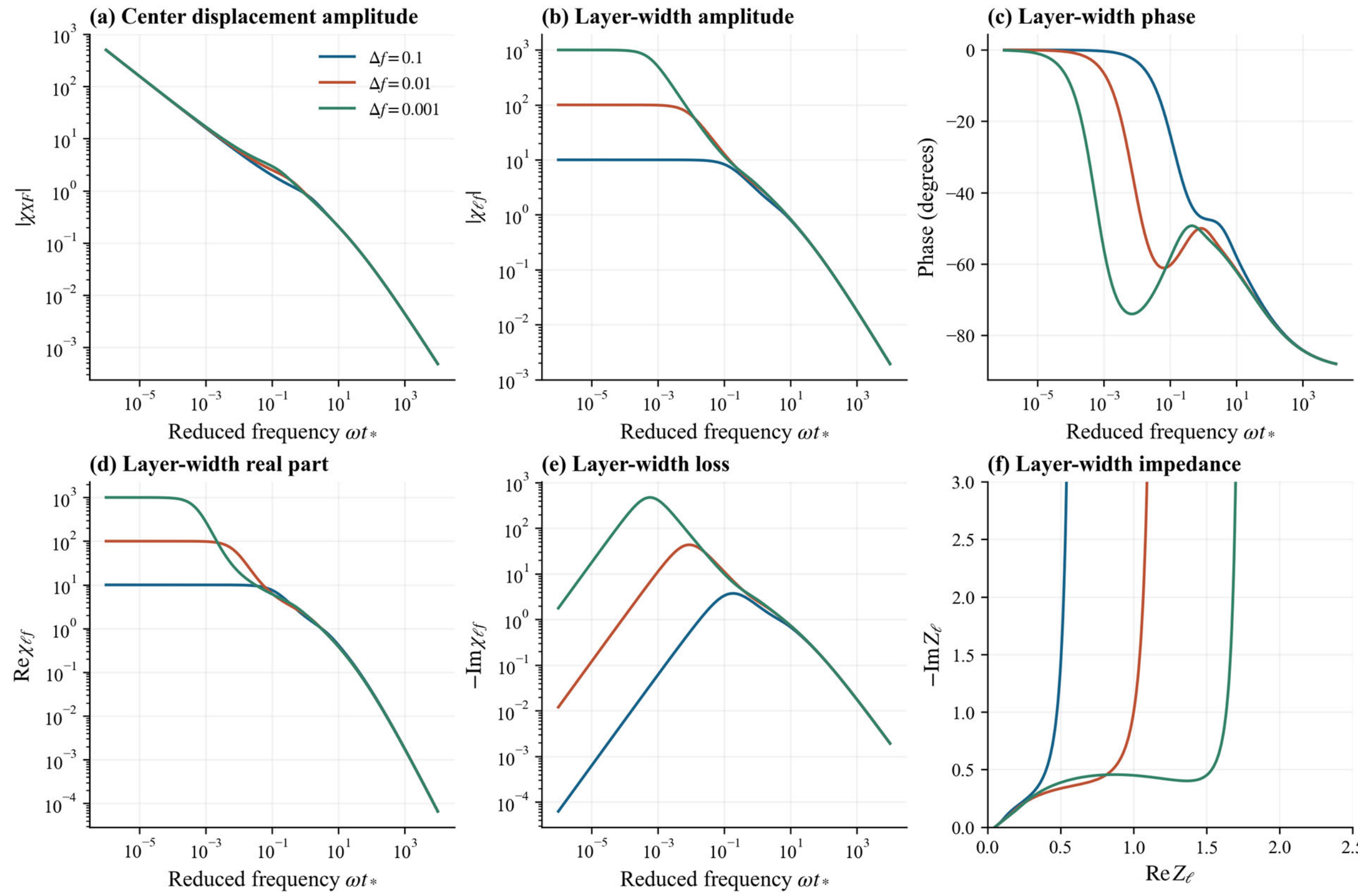

**Figure S14.** Center and separation responses for a conserved first-order pair. Parameters are $H_0 = P_s = D = h = A_w = \kappa_t = 1, \eta_X = 0.2$ and $\eta_\ell = 0.05$. The phase costs $\Delta f = 0.1$, 0.01 and 0.001 give $\ell_0 = \ln(1/\Delta f)$ and $K_\ell = \Delta f$. The width response shows increasing static amplitude and decreasing characteristic frequency near coexistence, while the center retains fractional low-frequency behavior. The input in the width panels is a conjugate separation force; electric-field visibility requires the appropriate coupling. The smallest separation is only a qualitative sharp-interface illustration.

### S16.8. Multiaxial polarization and coupled wall modes

For an isolated tangential polarization mode of amplitude $A$, the leading stable equilibrium response is $(K_A + \eta_A s)^{-1}$. A zero $K_A$ marks a rotational instability, while negative quartic terms can produce multiple stable structures before this linear stiffness vanishes. Each branch has its own small-signal spectrum; switching between branches is nonlinear and is not part of one linear susceptibility. A uniform longitudinal field can be symmetry-forbidden from exciting the odd tangential mode, whereas a transverse field couples directly.

A reciprocal displacement–rotation reduction at nonzero wave number has

$$\mathsf{D}(s,k) = \begin{pmatrix} \eta_X s + bk^2 & ick \\ -ick & \eta_A s + \mu + dk^2 \end{pmatrix}. \tag{S83a}$$

$$\mathcal{D} = (\eta_X s + bk^2)(\eta_A s + \mu + dk^2) - c^2k^2. \tag{S83b}$$

$$\mathcal{X}_{AA} = \frac{\eta_X s + bk^2}{\mathcal{D}}. \tag{S83c}$$

$$\mathcal{X}_{XA} = \frac{-ick}{\mathcal{D}}. \tag{S83d}$$

The stable equilibrium condition is $b(\mu + dk^2) > c^2$ for $k \neq 0$, with $b, d, \eta_X, \eta_A > 0$. Low-frequency responses are entries of the inverse static Hessian. At high frequency the diagonal responses scale as $1/s$; the cross-response scales as $1/s^2$. At $k = 0$ this particular gradient coupling vanishes. Surface and MIEC screening add the frequency-dependent matrix

$$\mathcal{M}_{ij}(s,k) = \int \frac{S_i(q,k)^* H_Q S_j(q,k)}{s+r_Q}\frac{dq}{2\pi}. \tag{S84a}$$

$$\mathcal{X}(s,k) = [\mathsf{K}(k) + s\eta + s\mathcal{M}(s,k)]^{-1}. \tag{S84b}$$

The entries must be derived from actual polarization and charge sources. For the MIEC, replace the surface memory by the corresponding contraction of $[H_m(Q,s) - H_m(Q,0)]/s$. A uniform tangential Bloch mode can be uncharged, but a modulated mode or depth-dependent rotation can generate charge. The measured scalar response is a forcing and detection projection of this matrix; a soft mode can be weak or invisible when that overlap vanishes.

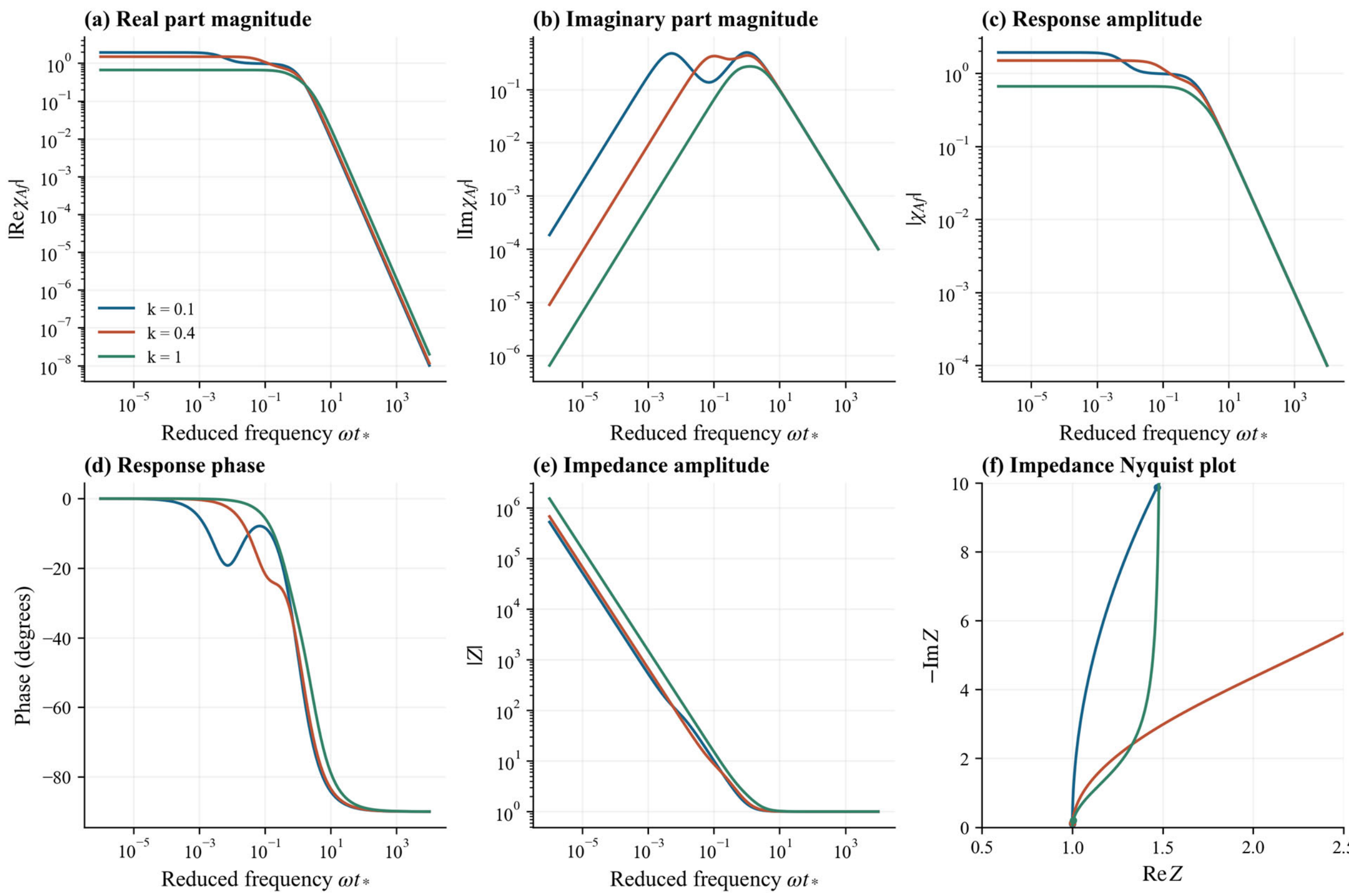

**Figure S15.** Rotation response of the reciprocal two-mode model, with $b = d = \mu = \eta_X = \eta_A = 1$, $c = 0.7$ and $k = 0.1$, 0.4 and 1. The forcing is conjugate to tangential amplitude. The plotted $\mathcal{X}_{Af} = \mathcal{X}_{AA}$ and its generalized impedance illustrate how coupling to displacement adds a slow component without an oscillatory equilibrium eigenmode. These are a declared structural projection, not material-specific BaTiO3 or BiFeO3 spectra.

### S16.9. Response about steady translation and near instability

When a stationary traveling state exists, response is defined in its moving frame. For the prescribed wall profile, define the causal residual electrostatic kernel

$$\mathcal{Z}(Q,z) = G_Q - \frac{H_Q r_Q}{z+r_Q}. \tag{S85a}$$

$$C_v(s,k) = \eta s + Tk^2 + \int \frac{|W(q)|^2}{2\pi}[\mathcal{Z}(Q, s - iqv) - \mathcal{Z}(|q|, -iqv)]dq. \tag{S85b}$$

The second term inside the brackets is the variation of the force projection and ensures translational invariance. The frequency response is $1/C_v(i\omega, k)$, with the causal continuation retained before integration. Taking an arbitrary real part inside this complex-frequency expression would destroy the response. Near $s = k = 0$,

$$C_v(s,k) = F'(v)s + T_{\text{eff}}k^2 + \cdots. \tag{S86a}$$

$$\frac{\delta v}{\delta F} \to \frac{1}{F\prime(v)} \quad (\omega \to 0,\ k = 0). \tag{S86b}$$

A small positive force slope produces a large differential mobility. A negative slope implies a growing mode in this reduction; the algebraic harmonic solution is then not a stable steady measurement. Approaching a spatial instability from the stable side, $C_v(0,k)$ becomes small and the response to a perturbation of that wavelength increases. A uniform measurement can miss it. For the local sharp-wall mixed-transport model, the integral has the exact form

$$S_v = \sqrt{v^2 + 4Dr_0}. \tag{S87a}$$

$$C_v(s,k) = \eta s + Tk^2 + B\left[\frac{s+v^2/(2D)}{\sqrt{v^2+4D(r_0+Dk^2+s)}} - \frac{v^2}{2DS_v}\right]. \tag{S87b}$$

$$T_{\text{eff}} = T - \frac{BDv^2}{S_v^3}. \tag{S87c}$$

This one-coordinate model does not have a Hopf crossing. A finite-frequency enhancement is allowed in models that also retain an internal structural mode. The driven structural–screening model used in the manuscript has, for forcing in its structural equation,

$$\mathcal{X}_A(z) = \frac{(z+1)^2+U^2}{(z+p)[(z+1)^2+U^2]-G(z+1)}. \tag{S88a}$$

$$z = s/r. \tag{S88b}$$

Put $a = 1 + U^2$ and $x = \omega/r$. The numerator has parts $n_R = a - x^2$ and $n_I = 2x$; the denominator has $d_R = pa - G - (p+2)x^2$ and $d_I = x(a + 2p - G - x^2)$. The signed response is therefore

$$\text{Re}\mathcal{X}_A = \frac{n_R d_R + n_I d_I}{d_R^2 + d_I^2}. \tag{S88c}$$

$$\text{Im}\mathcal{X}_A = \frac{n_I d_R - n_R d_I}{d_R^2 + d_I^2}. \tag{S88d}$$

Its static value is $a/(pa - G)$ on a stable branch. The first correction is $-[a^2 + G(2-a)]z/(pa-G)^2$, so the low-frequency phase follows from this derivative and need not have the passive-equilibrium sign for arbitrary driven parameters. At high frequency, $\mathcal{X}_A \sim 1/z - p/z^2$, giving an inverse-frequency amplitude and a phase approaching $-90°$. The generalized velocity impedance tends to $1/r$ in this normalization. For the traveling wall at $k = 0$, the low-frequency response instead retains the translation pole $1/[sF'(v)]$. The high-frequency displacement is $1/(\eta s)$; the sharp-wall correction to impedance is proportional to $s^{-1/2}$ until finite wall width becomes relevant.

At $G = 12$, $p = 2$, the Hopf point is $U = 3$ with $\omega_H/r = \sqrt{2}$. Figure S16 evaluates only stable points $U > 3$, where the complex poles approach the imaginary axis and the forced peak sharpens. The AC subsystem can exchange energy with the DC motion, so its differential impedance need not have nonnegative real part at every frequency. This is not a violation of the passive equilibrium result. A divergent linear response at the exact Hopf point requires nonlinear saturation, while a response about the developed periodic orbit requires the Floquet formulation below. The model is a stated projection and is not evidence of oscillations in every screening geometry.

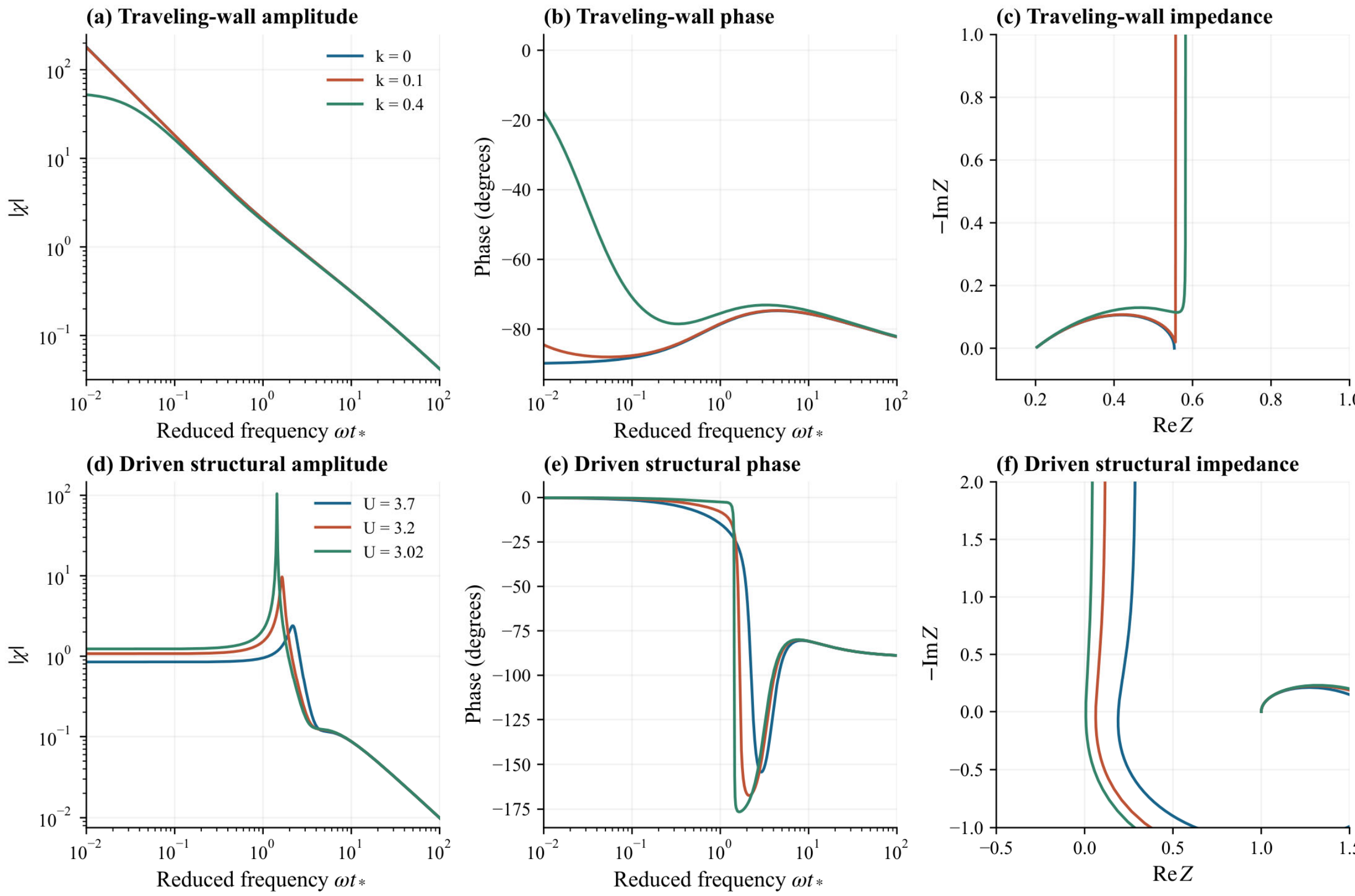

**Figure S16.** Two different driven reductions. Panels a–c show the stable sharp-wall exchange-plus-diffusion response with $B = D = 1$, $r_0 = 0.25$, $v = 1$, $\eta = 0.2$, $T = 0.4$, and $k = 0$, 0.1 and 0.4. The tension margin is $T_{\text{eff}} = 0.04645 > 0$. Panels d–f show the separate structural model at $G = 12$, $p = 2$ and $U = 3.7$, 3.2 and 3.02. Its spectral peak sharpens as $U$ approaches 3 from the stable side. All displayed structural poles were checked to have negative real parts. Negative differential impedance portions are permitted because the base is driven.

### S16.10. Low and high frequency signatures across the cases

Table S6. Low- and high-frequency limits of the response functions.

| Model and observable | Low-frequency response | High-frequency response within the model |
|---|---|---|
| Unconfined wall with exchange | Displacement proportional to 1/frequency; finite velocity mobility | Displacement proportional to 1/frequency with intrinsic friction |
| Infinite conserved surface | Displacement proportional to inverse square root of frequency; 45 degree lag | Intrinsic wall response after finite-width cutoff |
| Closed conserved surface | Finite static displacement; phase tends to zero | Diffusion regime followed by intrinsic response |
| Frozen screening pattern | Finite static displacement set by preparation | Inverse-frequency displacement and 90 degree lag |

| Model and observable | Low-frequency response | High-frequency response within the model |
|---|---|---|
| Stable layer separation or rotation | Finite static response and linear loss; enhancement near a soft mode | Inverse-frequency diagonal response |
| Top-electrode terminal impedance | Capacitive with equilibrated or constrained capacitance | Capacitive with geometric capacitance |
| Blocking MIEC at finite wave number | Screened modal capacitance plus finite dissipative part | Frozen-carrier modal capacitance |
| Uniform isolated MIEC junction | Outer-distance dependence; no finite uniform impedance in the strict limit | The long-distance restriction persists at finite frequency |

Figure S17 collects the impedance phases that complement the amplitude and Nyquist plots. In Figures S7–S12 and S15 the first two panels use logarithmic magnitudes of the real and imaginary parts. Their signs are fixed by the definitions: equilibrium polarization, capacitance, displacement and rotation responses have positive real part and negative imaginary part in the displayed examples; $H_m$ has positive real and imaginary parts. Every data file also contains the signed values. These sign conventions prevent an interfacial potential response from being mistaken for a dielectric loss spectrum.

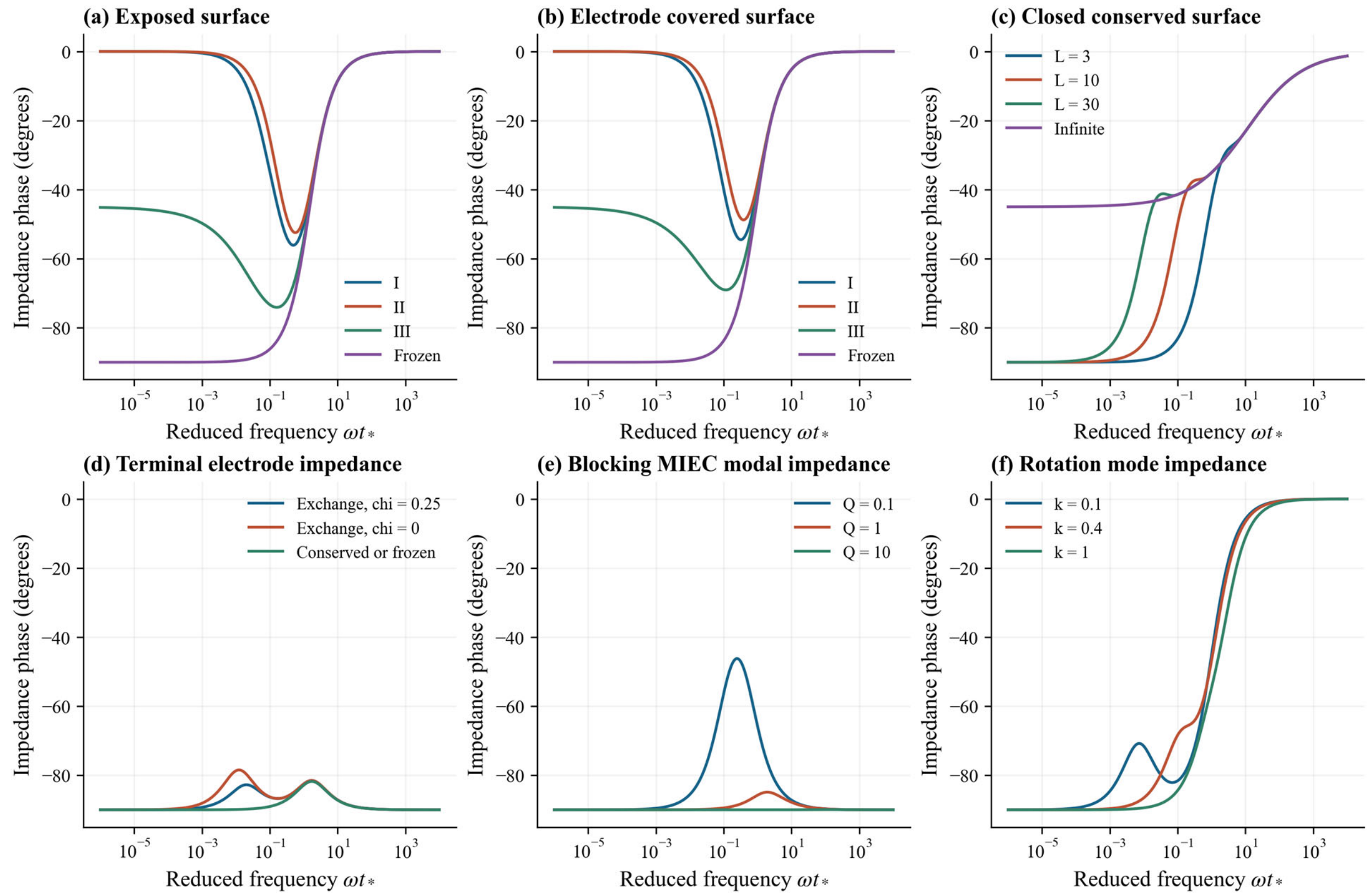

**Figure S17.** Impedance phase versus frequency for the exposed and covered wall cases, closed conserved surfaces, homogeneous top-electrode terminal response, finite-wave-number MIEC response and coupled rotation. The same parameters and curve ordering as Figures S8–S12 and S15 are used. A conserved diffusion range approaches $-45°$, whereas restoring-force and electrical-capacitance dominated impedances approach $-90°$. Wall friction with exchange approaches zero phase at both frequency extremes.

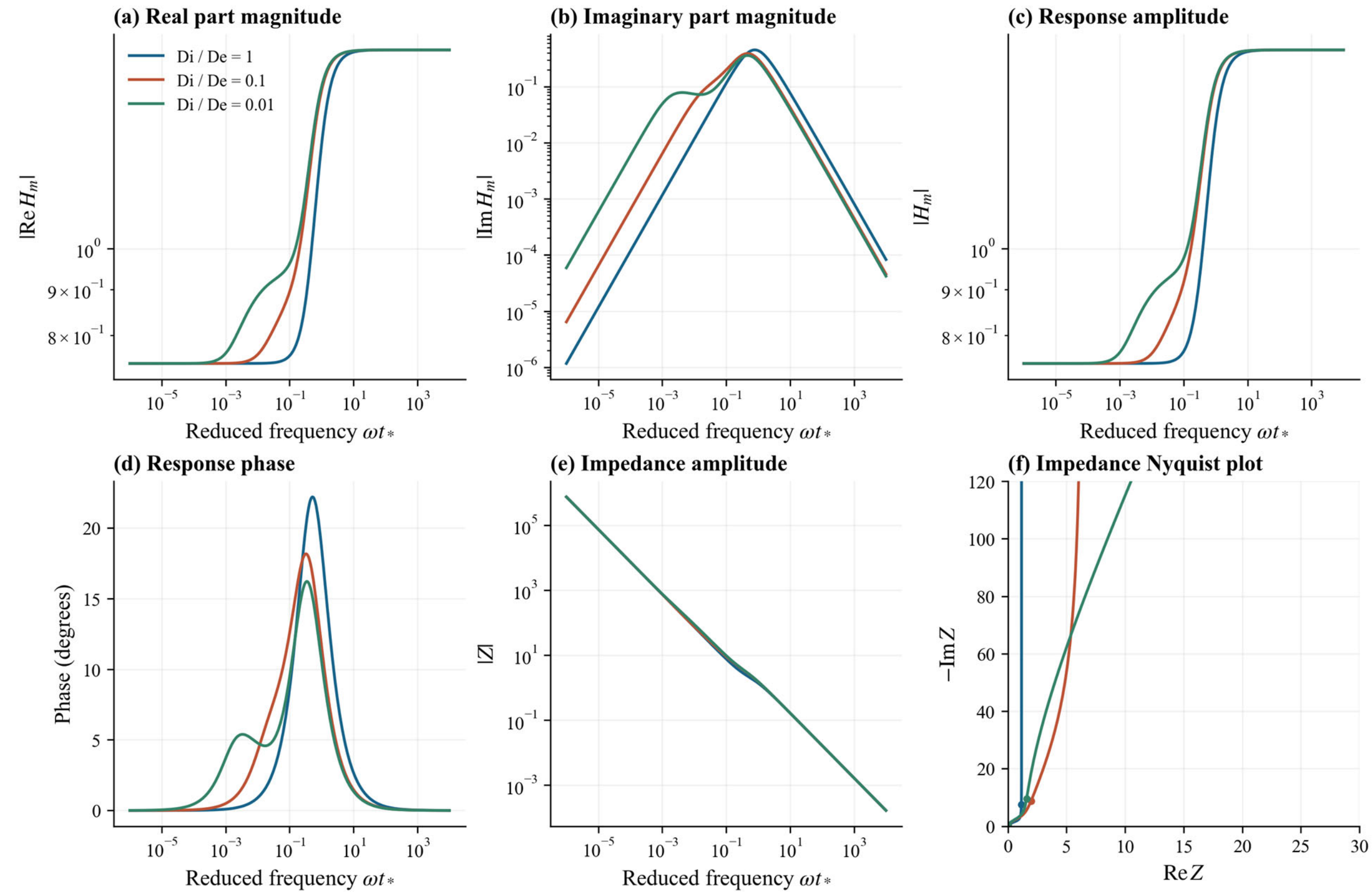

**Figure S18.** Exact two-carrier boundary algebra for unequal diffusivities. The response is $H_m$ and impedance is $H_m/s$, with $Q = 0.3$, $\epsilon_f = \epsilon_m = 1$, $\kappa_i^2 = \kappa_e^2 = 0.5$, $D_e = 1$ and $D_i/D_e = 1, 0.1$ and $0.01$. The static and frozen-carrier limits coincide across the curves, but the charge and composition dynamics produce different intermediate dispersion. No single relaxation time reproduces the whole family.

### S16.11. General deformable walls and nonstationary reference states

The preceding closed expressions retain specified collective coordinates or uniform depth profiles. The complete small-signal calculation for an already solved wall keeps its dependence on the coordinate $\xi = y - vt$ and film depth $z$. With polarization perturbation $u$, surface charge perturbation $d_c$ and potential perturbation $\varphi$, the scalar forced equations are

$$\Gamma s u = \left[g_y\,\partial_\xi^2 + g_z\,\partial_z^2 - g_x k^2 + \Gamma v\,\partial_\xi - f_L''(p_0)\right]u - \partial_z\varphi + f_P. \tag{S89a}$$

$$\epsilon_0\left[\epsilon_\parallel\left(\partial_\xi^2 - k^2\right) + \epsilon_z\,\partial_z^2\right]\varphi = \partial_z u. \tag{S89b}$$

$$\left(s - v\,\partial_\xi\right)d_c = -\left[\Lambda - M\left(\partial_\xi^2 - k^2\right)\right]\left[f_s''(c_0)d_c - \varphi_s - f_\psi\right]. \tag{S89c}$$

The electrical boundary conditions include the imposed electrode or exterior perturbation; the interface source is $u_s - d_c$. Natural polarization conditions are retained at the film surfaces. Vector polarization replaces $f_L''$ by its Hessian and includes all bound-charge components. Variable mobilities, reaction derivatives and additional species must be linearized about the same base. After eliminating the electrostatic constraint, the exact stationary transfer function can be written

$$\mathcal{X}_{ab}(s,k) = \mathsf{D}_{ab} + \mathsf{O}_a(s\mathsf{M} - \mathsf{L}_k)^{-1}\mathsf{B}_b. \tag{S90}$$

The input operator $\mathsf{B}_b$ defines the applied field, voltage, chemical potential or stress. The output $\mathsf{O}_a$ defines polarization, junction displacement, probe potential or electrode charge; $\mathsf{D}_{ab}$

contains instantaneous geometric terms. Low-frequency behavior follows from the constrained inverse of $-\mathsf{L}_k$ when it exists, and high-frequency behavior from the expansion in inverse powers of $s$. Conserved modes and translation must be treated explicitly before inversion. Equilibrium modes satisfy detailed balance and yield relaxational poles; moving bases need not. This operator formula is the analytical answer for an arbitrary deformable wall. Its numerical spectrum depends on the actual base profile; no uncomputed finite-depth or material spectrum is represented by the reduced figures.

For a conserved transient, coefficients depend on time. The response is then a two-time kernel,

$$\delta a(t) = \int_{t_0}^{t} \mathsf{O}\,(t)\mathsf{U}(t,t')\mathsf{M}^{-1}\mathsf{B}(t')f(t')\,dt'. \tag{S91}$$

Here $\mathsf{U}$ is the propagator of the linearized evolving base. A windowed susceptibility depends on the age and duration of the measurement. A frozen-time spectrum is an approximation requiring the base to evolve slowly over an excitation period and the relevant relaxation interval. One cannot assign a unique time-translation-invariant impedance to an indefinitely spreading charge wake or an unbinding layer.

For a periodic base of frequency $\Omega$, write $\mathsf{L}(t) = \sum_m \mathsf{L}_m\, e^{im\Omega t}$. A perturbation at $\omega$ generates sidebands. Their amplitudes obey

$$[i(\omega + n\Omega)\mathsf{M} - \mathsf{L}_0]u_n - \sum_{m\neq 0} \mathsf{L}_m\, u_{n-m} = \mathsf{B}_n f_\omega. \tag{S92}$$

The response is a Floquet matrix indexed by sideband, with the phase mode of the periodic orbit handled explicitly. A single Bode or Nyquist curve is insufficient unless a particular sideband and measurement protocol are selected. Equations (S91) and (S92) cover the transient and periodic cases analytically while retaining the reference-state dependence.

### S16.12. Parameter conversion and reproducibility

The figures evaluate the formulas in reduced units. If the plotted angular frequency is $\hat{\omega} = \omega t_*$, its physical cyclic frequency is $f = \hat{\omega}/(2\pi t_*)$. For a polar reference, useful material scales are $\xi_P = \sqrt{g_y/K_b}$, $\tau_P = \Gamma/K_b$ and $E_* = K_b P_s$. The independent screening groups include $G_0/(hK_b)$, $\chi_c/G_0$, $\Lambda(G_0+\chi_c)\tau_P$, $M(G_0+\chi_c)\tau_P/\xi_P^2$, and the film and gap dimensions in units of $\xi_P$. For the MIEC, retain $\kappa\xi_P$, the diffusivity ratio, susceptibility fractions and the actual outer geometry.

BaTiO3 and BiFeO3 enter through the appropriate stable branch curvature, gradient and kinetic coefficients, and mechanical or structural constraints. A first-order material also requires the central-phase cost and elementary-interface properties; BiFeO3 can require coupled octahedral and elastic fields. Static Landau coefficients alone do not fix adsorption rates, diffusivities, polarization damping or contact geometry. Consequently the reduced plots require those inputs before conversion to hertz or, for a defined electrical port, ohms. All symbolic dependencies needed for that conversion are retained here.

The supplementary figures evaluate closed expressions or one-dimensional quadratures. The accompanying response-function notebook and spectra package reproduce Figures S7–S18 and provide the signed real and imaginary parts, amplitudes, phases and impedance curves. The additional junction evaluation in Section S16.13 uses the same blocking boundary equations with unequal diffusivities. All figures are reduced-model calculations, rather than fits to experimental spectra or additional phase-field simulations.

The coupled polarization formula agrees with direct two-variable inversion to approximately $1.1 \times 10^{-15}$ relative error, and the voltage-to-terminal-capacitance expression agrees to $2.3 \times 10^{-16}$. Doubling the spatial quadrature resolution for a representative exposed mixed-transport wall changes its impedance by $4.3 \times 10^{-10}$ relatively. The closed-interval expression agrees with a truncated 39,999-mode Neumann sum to $1.8 \times 10^{-5}$; the residual is the convergent series tail. The computed equilibrium impedances have nonnegative real parts. The driven structural examples have stable poles, and the traveling-wall example has positive long-wave tension. These checks validate the analytical evaluations, not a general nonlinear material model.

### S16.13. Parameters and additional MIEC spectra for the main figure

The main-text Figure 12 assembles surface-wall displacement and two MIEC junction comparisons. For panels a–c, the electrode-covered electrostatic kernel is evaluated with the dimensionless values $h = 10$, $\epsilon_0\bar{\epsilon} = 10$, $\alpha = 1$, $d = 2$ and $\epsilon_0\epsilon_d = 2$. The wall source has $\Delta P = 2$ and tanh width parameter $w = 1$. The chemical curvature is $\chi_c = 0.2$ and the projected friction is $\eta = 0.15$. Case I uses local mobility $\Lambda = 0.2$ and no lateral transport. Case II retains the same exchange and adds surface mobility $M = 1$. Case III has the same lateral mobility and zero exchange. The frozen curve fixes the charge perturbation to zero about the initially equilibrated pattern. The full finite-width quadrature is used, including at high frequency. The associated exposed-surface comparison is Figure S8; Figure S9 gives all components for the covered surface.

These surface coefficients are expressed in a consistent set of length, time, polarization and electrostatic-kernel units, denoted $\ell_*$, $t_*$, $P_*$ and $G_*$. The force per junction length has scale $F_* = P_*^2 G_*$. Thus the numerical susceptibility is the physical susceptibility multiplied by $F_*/\ell_*$, and the numerical impedance is the physical impedance multiplied by $\ell_*/(F_* t_*)$. The dimensionless exchange mobility is $\Lambda G_* t_*$ and the dimensionless lateral mobility is $M G_* t_*/\ell_*^2$. Numerical permittivities are measured in units $\ell_*/G_*$. These scales specify a reduction and do not assign a material relaxation time.

Panel d uses the separate local-capacitance, sharp-wall expression (S64c) at $B = D = 1$ and $\eta = 0.15$, with closed lengths $L = 3$, 10 and 30 and the infinite-interval limit. This panel isolates finite-size conservation; it is not a second evaluation of the nonlocal covered-surface parameters. A symmetric wall at the interval center and reflecting charge boundaries are assumed. The diffusion response appears between the finite-size and wall-width scales when that interval exists. The smallest length is a sharp-interface illustration rather than a resolved profile calculation.

Panels e and f use the same junction projection and both impose zero normal ionic and electronic flux. Length is measured in $\kappa^{-1}$, time in $(D_e\kappa^2)^{-1}$, and the permittivity unit is chosen so that $\epsilon_m = \epsilon_f = 1$. The source has $\Delta P = 2$, $w = 1$ and finite projected friction $\eta = 0.15$. For panel e, the carrier diffusivities are equal and the outer distances are $R\kappa = 10$, 100 and 1000. For panel f, $R\kappa = 100$, the inverse-square screening lengths are $\kappa_i^2 = \kappa_e^2 = 1/2$, and $D_i/D_e = 1$, 0.1 and 0.01. The finite-cutoff displacement impedance is

$$Z_X^m(s; R) = \eta + \frac{1}{\pi s}\int_{1/R}^{\infty} |W(q)|^2 \left[H_m(q, s) - H_m(q, 0)\right] dq. \tag{S93a}$$

$$\mathcal{X}_{XF}^m(s; R) = \frac{1}{s Z_X^m(s; R)}. \tag{S93b}$$

The exact two-root boundary system (S74)–(S75) is solved at each frequency and wave number for unequal diffusivities. The equilibrium potential kernel is $H_m(q, 0) = \left[\epsilon_f q + \epsilon_m\sqrt{q^2 + \kappa^2}\right]^{-1}$ and is independent of the diffusivity ratio. The subtraction retains translational invariance of the equilibrated junction projection. Numerical integration uses 2,200 logarithmically spaced wave numbers from $1/R$ to 40; the finite tanh source makes the omitted upper tail negligible. The unequal-diffusivity spectra use 780 logarithmically spaced frequencies between $10^{-8}$ and $10^4$. The spectral cutoff specifies sensitivity to an outer distance; a contact-specific prediction requires the corresponding finite-geometry electrostatic and transport solution.

To compare the loop shapes in main-text panel f, we remove the intrinsic friction and divide by the zero-frequency screening resistance:

$$\hat{Z}_X^m(s) = \frac{Z_X^m(s)-\eta}{Z_X^m(0)-\eta}. \tag{S94}$$

Here $Z_X^m(0)$ denotes the zero-frequency limit at fixed finite $R$. For the three diffusivity ratios, the total limiting impedances in the stated units are approximately 3.663, 19.469 and 177.532, respectively. Normalization therefore removes a large change in magnitude. The equal-diffusivity curve is close to a single relaxation arc in this example, whereas unequal diffusivities add a broad slow contribution and a shoulder. The complete displacement and unnormalized impedance spectra are shown in Figure S19. The phase of the unconfined displacement tends to $-90°$ at sufficiently low frequency for fixed cutoff. This limit must not be exchanged with $R \to \infty$.

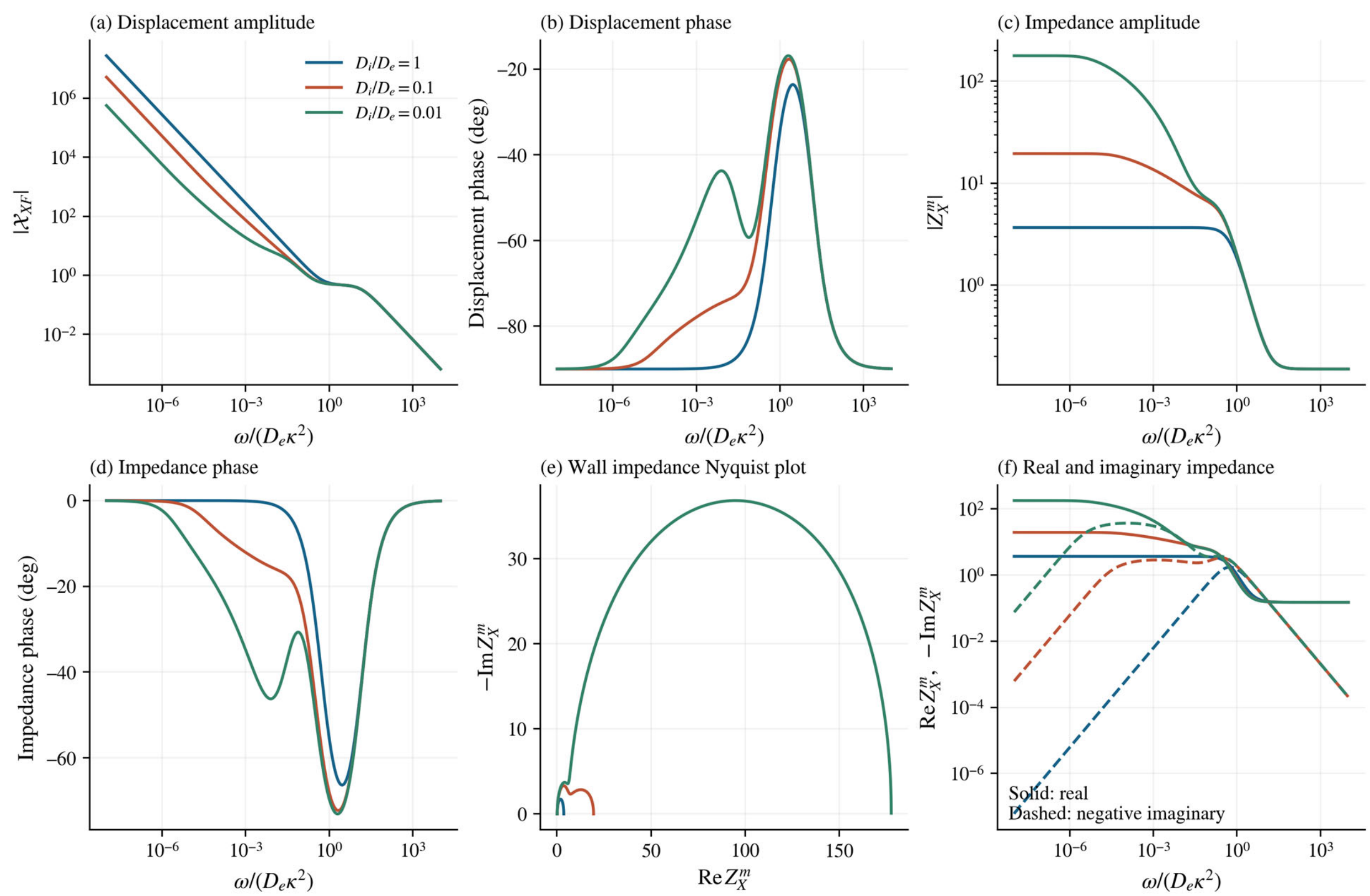

**Figure S19.** Wall-junction response of a blocking MIEC at fixed $R\kappa = 100$ for $D_i/D_e = 1$, 0.1 and 0.01. Panels give displacement amplitude and phase, impedance amplitude and phase, the unnormalized Nyquist plot, and the real and negative imaginary impedance. Solid and dashed lines in the last panel denote the real and negative imaginary parts, respectively. The electrical boundary conditions and finite junction friction are unchanged across the curves. These are projected force-to-velocity impedances, rather than terminal impedances in ohms. A specified contact geometry and wall depth profile are required for a dimensional junction prediction.

Doubling the quadrature to 4,400 wave numbers changes the unequal-diffusivity junction impedance by less than $7 \times 10^{-11}$ in relative magnitude on the validation frequency grid. The equal-diffusivity calculation agrees with the previously evaluated junction integral to the same tolerance. All displayed junction impedances have positive real part. The unequal-carrier kernel approaches the equal-carrier expression continuously and recovers the common static screening kernel. These checks establish consistency of the analytical boundary algebra and its numerical evaluation; they do not determine the remote contact geometry.

**S16.14. Surface potential and conversion to electrical observables**

A surface-potential measurement detects the joint polarization and charge response. For a surface sheet, the induced potential in a lateral Fourier mode is

$$\delta\phi_{s,Q} = G_Q\left(\delta p_Q - \delta c_Q\right) + \delta\phi_{\mathrm{ext},Q}. \tag{S95a}$$

For a polarization-driving perturbation with $\delta\psi_Q = 0$ and no direct exterior potential, the uniform-domain result gives

$$\mathcal{X}_{\phi E}(Q,s) = G_Q \frac{s+L_Q\chi_c}{s+r_Q} \mathcal{X}_{PE}(Q,s). \tag{S95b}$$

For a chemical perturbation at fixed applied electrostatic potential and $\delta E_{b,Q} = 0$,

$$\mathcal{X}_{\phi\psi}(Q,s) = \frac{G_Q L_Q}{s+r_Q}\left[\frac{G_Q}{h}\frac{s+L_Q\chi_c}{s+r_Q}\mathcal{X}_{PE}(Q,s) - 1\right]. \tag{S95c}$$

These are potential transfer functions; their imaginary parts do not have to share the sign convention of dielectric loss. For $L_Q > 0$, the low-frequency field-to-potential response is $G_Q\chi_c/\left(A_Q K_0\right)$. It vanishes at $\chi_c = 0$ because charge follows the polarization sufficiently to remove the induced potential in this particular control. At high frequency, it approaches $G_Q/(\Gamma s)$ within the overdamped model. The chemical-to-potential response has the static value $\left(G_Q/A_Q\right)\left[G_Q\chi_c/\left(hA_QK_0\right) - 1\right]$ and approaches $-G_QL_Q/s$ at high frequency. For the conserved uniform mode, the chemical perturbation cannot change the total charge, so this chemical response is zero; the field response retains the frozen-inventory polarization response.

For an exposed surface with a homogeneous exterior, a Fourier component of the induced potential at height $z$ is multiplied by $e^{-Qz}$. A probe measurement additionally averages over its lateral sensitivity function. Consequently probe height and tip geometry change the spectral weighting even when the microscopic relaxation law is fixed. A driven wall requires the spatial operator response in Section S16.11 and the derivative of the equilibrium wall and screening profiles as its sources; the uniform-domain formula alone does not describe a junction image. The covered geometry requires the transfer function inside its dielectric layer, rather than the exposed half-space propagation factor.

An electrical experiment requires a work-conjugate terminal voltage and charge. For a reciprocal stable equilibrium reduction with generalized coordinates collected in $\mathbf{a}$, the quadratic electric enthalpy and its driven response can be written as

$$\mathcal{G} = 1/2\,\mathbf{a}^T\mathbf{H}\mathbf{a} - V\mathbf{g}^T\mathbf{a} - 1/2\,C_\infty V^2. \tag{S96a}$$

$$(\mathbf{H} + s\mathbf{R})\delta\mathbf{a} = \mathbf{g}\,\delta V. \tag{S96b}$$

$$\delta Q_e = C_\infty \delta V + \mathbf{g}^T\delta\mathbf{a}. \tag{S96c}$$

$$Z_V(s) = \{s[C_\infty + \mathbf{g}^T(\mathbf{H} + s\mathbf{R})^{-1}\mathbf{g}]\}^{-1}. \tag{S96d}$$

Here $\mathbf{H}$ is the constrained free-energy Hessian, $\mathbf{R}$ is the dissipation matrix on the allowed coordinates, and $\mathbf{g}$ is the terminal coupling derived from the same geometry. Conserved zero modes must first be removed or explicitly constrained. In the absence of an independent leakage path, the current is $s\delta Q_e$. For a single wall coordinate this expression adds a term proportional to the displacement susceptibility to the terminal capacitance. For coupled charge and wall motion, direct voltage forcing of the charge must be retained; eliminating it can produce a frequency-dependent effective wall coupling. The explicit uniform-domain capacitor example is Section

S16.5. This construction explains why a force-to-velocity Nyquist curve cannot be relabeled as an electrode impedance without specifying the drive and readout.

**References**

Reference numbers are shared with the main paper; the complete list is repeated here for convenient reading.